\documentclass[a4paper,UKenglish]{lipics-v2016}

\usepackage{microtype}

\usepackage[numbers,sort]{natbib}

\usepackage{stmaryrd}

\usepackage{tabulary}
\usepackage{etex}

\newif\iflong
\longtrue  

\newif\ifdraft
\drafttrue 

\usepackage{myfloat}
\floatstyle{ruled}
\restylefloat{figure}

\newcommand{\tr}[2]{\iflong{}\S#1\else{}\cite[\S{}#2]{ext}\fi}
\newcommand{\tra}[2]{\iflong{}(\S#1)\else{}\cite[\S{}#2]{ext}\fi}

\newcommand{\nanomalies}{A}
\newcommand{\ngeneral}{B}
\newcommand{\nproofs}{C}
\newcommand{\ncompleteness}{D}

\usepackage{pifont}
\newcommand{\cross}{\ding{56}}

\renewcommand{\O}{\mathcal{O}}

\newcommand{\myparagraph}[1]{\textbf{\color{darkgray}\sffamily#1.}}

\usepackage[frame,all]{xy}
\usepackage{xspace}
\usepackage{amsmath}
\usepackage{amssymb}
\usepackage{empheq}
\usepackage{graphicx}
\usepackage{mathtools}
\usepackage{nicefrac}
\usepackage{relsize}

\usepackage{wrapfig}

\DeclareMathAlphabet{\mathpzc}{OT1}{pzc}{m}{it}

\theoremstyle{plain}
\newtheorem{proposition}[theorem]{Proposition}

\usepackage{executions}
\usepackage{execgraphs}

\newcommand{\account}{\ensuremath{\mathsf{acct}}}

\newcommand{\leaveout}[1]{}

\usepackage{paralist}
\setdefaultleftmargin{10pt}{}{}{}{}{}
\newenvironment{mywrapfigure}[3][]{
  \floatstyle{boxed}
  \restylefloat{figure}
  \wrapfigure[#1]{#2}{#3}}
  {\endwrapfigure
    \floatstyle{ruled}
    \restylefloat{figure}
    }

\newcommand{\txlaws}{{\textbf{(a)}}}
\newcommand{\deplaws}{{\textbf{(b)}}}
\newcommand{\allexeclaws}{{\textbf{(c)}}}
\newcommand{\cmexeclaws}{{\textbf{(d)}}}

\newcommand{\txId}{{\textbf{(a.1)}}}
\newcommand{\txComp}{{\textbf{(a.2)}}}

\newcommand{\txDistrR}{{\textbf{(a.3)}}}
\newcommand{\txDistrL}{{\textbf{(a.4)}}}
\newcommand{\depWRTx}{{\textbf{(b.1)}}}
\newcommand{\depWWTx}{{\textbf{(b.2)}}}
\newcommand{\depRWTx}{{\textbf{(b.3)}}}
\newcommand{\depWRIrrefl}{{\textbf{(b.4)}}}
\newcommand{\depWWIrrefl}{{\textbf{(b.5)}}}
\newcommand{\depRWIrrefl}{{\textbf{(b.6)}}}
\newcommand{\ARtrans}{{\textbf{(c.5)}}}
\newcommand{\VIStrans}{{\textbf{(c.4)}}}
\newcommand{\VISinAR}{{\textbf{(c.6)}}}
\newcommand{\ARirrefl}{{\textbf{(c.12)}}}
\newcommand{\VISnotAVIS}{{\textbf{(c.11)}}}

\newcommand{\WRinVIS}{{\textbf{(c.1)}}}
\newcommand{\WWinAR}{{\textbf{(c.2)}}}
\newcommand{\LWW}{{\textbf{(c.7)}}}
\newcommand{\RWinAVIS}{{\textbf{(c.3)}}}
\newcommand{\AVISright}{{\textbf{(c.8)}}}
\newcommand{\AVISleft}{{\textbf{(c.9)}}}
\newcommand{\AVISnotVIS}{{\textbf{(c.10)}}}

\newcommand{\Axiom}{{\textbf{(d.1)}}}
\newcommand{\CoAxiomAR}{{\textbf{(d.2)}}}
\newcommand{\CoAxiomL}{{\textbf{(d.3)}}}
\newcommand{\CoAxiomR}{{\textbf{(d.4)}}}

\newcommand{\Srf}{\text{(V1)}}
\newcommand{\Sconflict}{\text{(V3)}}
\newcommand{\SvisTrans}{\text{(V2)}}
\newcommand{\Saxiom}{\text{(V4)}}
\newcommand{\Svo}{\text{(A1)}}
\newcommand{\Svis}{\text{(A2)}}
\newcommand{\SarTrans}{\text{(A4)}}
\newcommand{\Scoaxiom}{\text{(A5)}}
\newcommand{\Sext}{\text{(A3)}}
\newcommand{\Sad}{\text{(N1)}}
\newcommand{\SavisL}{\text{(N2)}}
\newcommand{\SavisR}{\text{(N3)}}

\iflong
\title{Algebraic Laws for Weak Consistency (Extended Version)}
\titlerunning{Algebraic Laws for Weak Consistency (Extended Version)} 
\else
\title{Algebraic Laws for Weak Consistency}
\titlerunning{Algebraic Laws for Weak Consistency}
\fi

\iflong

\author[1]{Andrea Cerone}
\author[2]{Alexey Gotsman}
\author[3]{Hongseok Yang\vspace{-5pt}}
\affil[1]{Imperial College London, UK, \texttt{a.cerone@imperial.ac.uk}}
\affil[2]{IMDEA Software Institute, Madrid, Spain, \texttt{alexey.gotsman@imdea.org}}
\affil[3]{University of Oxford, UK, \texttt{hongseok.yang@cs.ox.ac.uk} \vspace{-15pt}}
\authorrunning{A. Cerone,\, A. Gotsman,\, H. Yang} 
\else

\author[1]{Andrea Cerone}
\author[2]{Alexey Gotsman}
\author[3]{Hongseok Yang}
\affil[1]{Imperial College London, UK, \texttt{a.cerone@imperial.ac.uk}}
\affil[2]{IMDEA Software Institute, Madrid, Spain, \texttt{alexey.gotsman@imdea.org}}
\affil[3]{University of Oxford, UK, \texttt{hongseok.yang@cs.ox.ac.uk}}
\authorrunning{A. Cerone,\, A. Gotsman,\, H. Yang} 
\fi

\Copyright{Andrea Cerone, Alexey Gotsman and Hongseok Yang}

\EventEditors{Roland Meyer and Uwe Nestmann}
\EventNoEds{2}
\EventLongTitle{28th International Conference on Concurrency Theory (CONCUR 2017)}
\EventShortTitle{CONCUR 2017}
\EventAcronym{CONCUR}
\EventYear{2017}
\EventDate{September 5--8, 2017}
\EventLocation{Berlin, Germany}
\EventLogo{}
\SeriesVolume{85}
\ArticleNo{22} 

\begin{document}
%


\maketitle

\begin{abstract}
Modern distributed systems often rely on so called weakly consistent databases,
which achieve scalability by weakening consistency guarantees of distributed transaction 
processing. The semantics of such databases have been formalised in two different styles, one based
on abstract executions and the other based on dependency graphs. The choice
between these styles has been made according to intended applications. The former has
been used for specifying and verifying the implementation of the databases, while
the latter for proving properties of client programs of the databases. 
In this paper, we present a set of novel algebraic laws (inequalities) that
connect these two styles of specifications. The laws relate binary relations used in
a specification based on abstract executions to those used in a specification
based on dependency graphs. We then show that this algebraic connection gives
rise to so called robustness criteria: conditions which ensure that a client program 
of a weakly consistent database does not exhibit anomalous behaviours due to
weak consistency. These criteria make it easy to reason about these client programs, 
and may become a basis for dynamic or static program analyses. 
For a certain class of consistency models specifications, we prove a full 
abstraction result that connects the two styles of specifications.

\end{abstract}

\section{Introduction}
\label{sec:introduction}

Modern distributed systems often rely on databases that achieve scalability by weakening
consistency guarantees of distributed transaction processing. These databases are said 
to implement weak consistency models. Such weakly consistent databases
allow for faster transaction processing, but exhibit anomalous behaviours, which do not arise
under a database with a strong consistency guarantee, such as serialisability. Two important 
problems for the weakly consistent databases are: (i) to find 
elegant formal specifications of their consistency models and to prove that these
specifications are correctly implemented by protocols used in the databases; (ii) to develop
effective reasoning techniques for applications running on top of such databases. 
These problems have been tackled by using two different formalisms, which model the run-time 
behaviours of weakly consistent databases differently.

When the goal is to verify the correctness of a protocol implementing 
a weak consistency model, the run-time behaviour of a distributed database is often 
described in terms of \emph{abstract executions} \cite{repldatatypes},
which abstract away low-level implementation details of the database (\S \ref{sec:abstract.executions}).
An example of abstract execution is depicted in Figure \ref{fig:example}; ignore 
the bold edges for the moment. It comprises four transactions, $T_0$, $T_1$, $T_2$, and $S$; 
transaction $T_0$ initializes the value of an object $\account$ to $0$;
transactions $T_1$ and $T_2$ increment the value of $\account$ by   
$50$ and $25$, respectively, after reading its initial value;
transaction $S$ reads the value of $\account$. 
In this abstract execution, both the updates of
$T_1$ and $T_2$ are \textbf{VIS}ible to transaction $S$, as witnessed by the two $\VIS$-labelled
edges: $T_1 \xrightarrow{\VIS} S$ 
and $T_2 \xrightarrow{\VIS} S$. 
\setlength{\intextsep}{0pt}
\begin{mywrapfigure}[11]{r}{0.65\textwidth}
\vspace{-5pt}
\begin{center}
\begin{tikzpicture}[scale=0.9, every node/.style={transform shape}, font=\small]
\node(Ra1) {$\RD\; \account: 0$};
\path(Ra1.center) + (-2.75, -1) node (Wa0) {$\WR\; \account: 0$};
\path (Ra1.center) + (2.5, 0) node (Wa1) {$\WR\; \account: 50$};
\path (Ra1.center) + (0,-2.0) node (Ra2) {$\RD\; \account: 0$};
\path (Ra2.center) + (2.5, 0) node (Wa2) {$\WR\; \account: 25$};
\path (Wa1.center) + (2.2,-1.0) node (Rf) {$\RD\; \account: 25$};
\begin{pgfonlayer}{background}
\node(t1) [background, fit=(Ra1) (Wa1), inner sep=0.2cm] {};
\node(t2) [background, fit=(Ra2) (Wa2), inner sep=0.2cm] {};
\node(t3) [background, fit=(Rf), inner sep=0.1cm] {};
\node(t0) [background, fit=(Wa0), inner sep=0.1cm]{};
\path (t3.west) + (0.2,0.1) node (h1) {};
\path (t3.west) + (0.2,-0.1) node (h2) {};
\path(t1.south) + (-1,0) node (a12s) {};
\path(t2.north) + (-1,0) node (a12e) {};
\path(t1.south) + (1,0) node (a21e) {};
\path(t2.north) + (1,0) node (a21s) {};
\path(t2.south east) + (-0.2,0) node (wr23s){};
\path(t3.south) + (1,0) node (wr23e) {};
\path[->] 
  (t0.north east) edge node[above, pos=0.4] {$\VIS$\;\;\;\;} (t1.south west)
  (t0.south east) edge node[below, pos=0.4] {$\VIS$\;\;\;\;} (t2.north west)
  (t1.south) edge node[right] {$\AR$} (t2.north)
  (t1.east) edge[bend left=30] node[above] {\;\;$\VIS$} (t3.north)
  (t2.east) edge[bend right=30] node[below] {\;\;$\VIS$} (t3.south);
\path[very thick, ->]
  (t0.north) edge[bend left=30] node[left] {\textbf{\textsf{WR}}, \textbf{\textsf{WW}}\;\;\;\;\;} (t1.west)
  (t0.south) edge[bend right=30] node[left] {\textbf{\textsf{WR}}, \textbf{\textsf{WW}}\;\;\;\;\;} (t2.west)
  (a12s.center) edge[bend right=20] node[left] {\textbf{\textsf{RW}}, \textbf{\textsf{WW}}} (a12e.center)
  (a21s.center) edge[bend right=20] node[right] {\textbf{\textsf{RW}}} (a21e.center)
  (wr23s.center) edge[bend right=40] node[below] {\textbf{\textsf{WR}}} (wr23e.center);
%
\path(t0.north west) + (0.5,0.2) node[font=\normalsize] (T0) {$T_0$};
\path(t1.north west) + (0.5,0.2) node[font=\normalsize] (T1) {$T_1$};
\path(t2.south west) + (0.5,-0.2) node[font=\normalsize] (T2) {$T_2$};
\path(t3.north east) + (-0.5,0.2) node[font=\normalsize] (T3) {$S$};
\end{pgfonlayer}
\end{tikzpicture}
\vspace{-12pt}
\end{center}
\caption{An example of abstract execution and of dependency graph.}
\label{fig:example}
\end{mywrapfigure}
\setlength{\intextsep}{10pt plus 6pt minus 3pt}
On the other hand, the update of $T_1$ is not visible to $T_2$, and 
vice versa, as indicated by the absence of an edge labelled with $\VIS$ between these transactions. 
Intuitively, the absence of such an edge means that
$T_1$ and $T_2$ are executed concurrently.
Because $S$ sees $T_1$ and $T_2$, as indicated by $\VIS$-labelled edges from $T_1$ and $T_2$ to $S$, 
the result of 
reading the value of $\account$ in $S$ must be one of the values written by $T_1$ and $T_2$. However, because these 
transactions are concurrent, there is a race, or \emph{conflict}, between them.
The $\AR$-labelled edge connecting $T_1$ to $T_2$, is used to \textbf{AR}bitrate the conflict: 
it states  that the update of $T_1$ is older than the one of $T_2$, hence the query of $\account$ 
in $S$ returns the value written by the latter.

The style of specifications of consistency models in terms of abstract executions can be given by imposing 
constraints over the relations $\VIS, \AR$ (\S \ref{sec:specification}).
A set of transactions $\T = \{T_1, T_2, \cdots\}$, 
 called a \emph{history}, is allowed by a consistency model specification
if it is possible to exhibit two witness relations $\VIS, \AR$ over $\T$ such that the resulting abstract execution 
satisfies the constraints imposed by the specification. For example, \emph{serialisability} can be 
specified by requiring that the relation $\VIS$ should be a strict total order. The set of transactions $\{T_0, T_1, T_2, S\}$ 
from Figure \ref{fig:example} is not serialisable: it is not possible to choose a relation $\VIS$ such that the resulting abstract execution relates the transactions $T_1, T_2$ and the results of read operations are consistent with 
visible updates.

Specifications of consistency models using abstract executions have been used in
the work on proving the correctness of protocols implementing weak consistency models, as well
as on justifying operational, implementation-dependent descriptions of these 
models~\cite{framework-concur,repldatatypes,ev_principles,ev_transactions}.  

The second formalism used to define weak consistency models is based on the notion 
of \emph{dependency graphs} \cite{adya}, and it has been used for proving properties of client programs 
running on top of a weakly consistent database.
Dependency graphs capture the data dependencies of transactions at run-time (\S \ref{sec:dgraphs});  
the transactions $\{T_0, T_1, T_2, S\}$ depicted above, together with the bold edges but without
normal edges, constitute an example of dependency graph.
The edge $T_2 \xrightarrow{\RF(\account)} S$\footnote{For simplicity, references to the object 
$\account$ have been removed from the dependencies of 
Figure \ref{fig:example}.} denotes a \emph{write-read dependency}. It means that the read of
$\account$ in transaction $S$ returns the value written by transaction $T_2$,
and the edges $T_0 \xrightarrow{\RF(\account)} T_1$ and $T_0 \xrightarrow{\RF(\account)} T_2$
mean something similar. The edge $T_1 \xrightarrow{\VO(\account)} T_2$ 
denotes a \emph{write-write dependency}, and says that the write to $\account$ in $T_2$ supersedes
the write to the same object in $T_1$. The remaining edges $T_1 \xrightarrow{\AD(\account)} T_2$ 
and $T_2 \xrightarrow{\AD(\account)} T_1$ express \emph{anti-dependencies}. The former
means that $T_1$ reads a value for object $\account$ which is older  
than the value written by $T_2$.

When using dependency graphs, consistency models are specified as sets of transactions 
for which there exist $\RF, \VO, \AD$ relations that satisfy certain properties, 
 usually stated as particular relations being acyclic \cite{SIanalysis,giovanni_concur16}; for 
 example, serialisability can be specified by requiring that dependency graphs are acyclic.
Because dependencies of transactions can be over-approximated at the compilation time,
specifications of consistency models in terms of dependency graphs have been widely used for manually or automatically reasoning about properties of client programs of weakly consistent
databases~\cite{fekete-tods,chopping}. 
They have also been used in the complexity 
and undecidability results for verifying implementations of consistency models~\cite{bouajjani_popl17}.

Our ultimate aim is to reveal a deep connection between these two styles of specifying
weak consistency models, which was hinted at for specific consistent models in 
the literature. Such a connection would, for instance, give us a systematic way to derive
a specification of a weak consistency model based on dependency graphs from
the specification based on abstract executions, while ensuring that the original
and the derived specifications are equivalent in a sense. In doing so, it would enable us
to prove properties about client programs of a weakly consistent database
using techniques based on dependency graphs~\cite{SIanalysis,psi-chopping,bouajjani_popl17}
even when the consistency model of the database is specified in terms of abstract executions.

In this paper, we present our first step towards this ultimate aim. 
First, we observe that each abstract execution determines an underlying dependency graph. 
Then we study the connection between these two structures at an algebraic level. 
We propose a set of algebraic laws, parametric in the 
specification of a consistency model to which the original abstract execution belongs (\S \ref{sec:laws}). 
These laws can be used to derive properties of the form $R_{\mathsf{G}} \subseteq R_{\mathsf{A}}$: 
here $R_{\mathsf{G}}$ is an expression from the Kleene Algebra with Tests \cite{kat} 
whose ground terms are run-time dependencies of transactions, 
and tests are properties over transactions. The relation $R_{\mathsf{A}}$ is one 
of the fundamental relations of abstract executions: $\VIS$, $\AR$, or 
a novel relation $\overline{\VIS^{-1}}$ that we call \emph{anti-visibility}, defined as $\overline{\VIS^{-1}} = \{(T,S) \mid \neg(S \xrightarrow{\VIS} T)\}$. 
Some of the algebraic laws that we propose show that there is a direct 
connection between each kind of dependencies and the relations of abstract 
executions: $\RF \subseteq \VIS, \VO \subseteq \AR$, and $\AD \subseteq \overline{\VIS^{-1}}$. 
The other laws capture the connection between the relations of abstract 
executions $\VIS, \AR$, and $\overline{\VIS^{-1}}$. The exact nature of 
this connection depends on the specification of the consistency model 
of the considered abstract execution.

We are particularly interested in deriving properties of the form 
$R_{\mathsf{G}} \subseteq \AR$. Properties of this form 
give rise to so called
robustness criteria for client programs, conditions ensuring that a program 
only exhibits serialisable behaviours even when it runs under a weak consistency 
model~\cite{fekete-tods,giovanni_concur16,vechev_popl17}.
Because $\AR$ is a total order, this implies that $R_{\mathsf{G}}$ must be acyclic, 
hence all cycles must be in the complement of $R_{\mathsf{G}}$. We can then 
check for the absence of such \emph{critical} cycles at compile time: 
because dependency graphs of serialisable databases are always acyclic, this  
ensures that said application only exhibits serialisable behaviours.

As another contribution we show that, for a relevant class of consistency models, 
our algebraic laws can be used to derive properties which are not only necessary, 
but also sufficient, for dependency graphs in such models (\S \ref{sec:completeness}).

\section{Abstract Executions}
\label{sec:abstract.executions}

We consider a database storing objects in $\Obj = \{x,y,\cdots\}$, 
which for simplicity we assume to be integer-valued.
Client programs can interact with the database by 
executing operations from a set $\Op$, grouped inside \emph{transactions}. 
We leave the set $\Op$ unspecified, apart from requiring that 
it contains read and write operations over objects: 
$\{\WR(x,n), \RD(x,n) \mid x \in \Obj, n \in \mathbb{N} \} \subseteq \Op$. \vspace{0.2\baselineskip}

\myparagraph{Histories}
To specify a consistency model, we first define the set of 
all client-database interactions allowed by the model. We start by introducing 
(run-time) \emph{transactions} and \emph{histories}, which record such 
interactions in a single computation.
Transactions are elements from a set $\TrSet = \{T, S, \cdots \}$; 
the operations executed by transactions are given by   
a function $\behav : \TrSet \rightarrow 2^{\Op}$, which maps a transaction 
$T$ to a set of operations that are performed by the transaction
and can be observed by other transactions. 
We often abuse notations and just write $o \in T$ (or $T \ni o$) instead of 
$o \in \behav(T)$. We adopt similar conventions for 
$\O \subseteq \behav(T)$ and $\O = \behav(T)$ where $\O$ is a subset of operations.

We assume that transactions enjoy 
\emph{atomic visibility}
: for each object $x$,
(i) a transaction $S$ never observes two different writes to $x$ from a single transaction $T$ 
and (ii) it never reads two different values of $x$. Formally, the requirements are that
if $T \ni (\WR\; x: n)$ and $T \ni (\WR\; x: m)$, or $T \ni (\RD\;x: n)$ and $T \ni (\RD\;x: m)$, 
then $n = m$. Our treatment of atomic visibility is taken from our previous work 
on transactional consistency models~\cite{framework-concur}. Atomic visibility 
is guaranteed by many consistency models \cite{ramp,PSI,fekete-tods}. 
We point out that
although we focus on transactions in distributed systems in the paper,
our results apply to weak shared-memory models \cite{cat}; there a transaction $T$ 
is the singleton set of a read operation ($T = \{\RD\; x: n\}$), 
that of a write operation ($T = \{\WR\; x: n\}$), or the set of read and write
representing a \emph{compare and set} operation ($T = \{\RD\; x: n,\; \WR\; x: m\}$).

For each object $x$, we let 
$\WTr_x := \{T \mid \exists n.\;(\WR\;x:n) \in T\}$ and 
$\RTr_x := \{T \mid \exists n,\;(\RD\;x:n) \in \T\}$ be the 
sets of transactions that write to and read from $x$, respectively.

\begin{definition}
\label{def:history}
A history $\hist$ is a finite set of transactions $\{T_1, T_2, \cdots, T_n\}$.
\end{definition}

\myparagraph{Consistency Models}
A consistency model $\Gamma$ is a set of histories 
that may arise when client programs interact with the database.
To define $\Gamma$ formally,
 we augment histories with two relations, called \emph{visibility} and \emph{arbitration}.

\begin{definition}
\label{def:aexec}
An \emph{abstract execution} $\aexec$ is a tuple $(\T, \VIS, \AR)$ 
where $\T$ is a history and 
$\VIS,\AR \subseteq (\T \times \T)$ are relations on transactions such that
$\VIS \subseteq \AR$
and $\AR$ is a strict total order\footnote{A relation 
$R \subseteq \T \times \T$ is a strict (partial) order if it is transitive and irreflexive; it is total 
if for any $T, S \in \T$, either $T = S$, $(T,S) \in R$ or $(S,T) \in R$.}. 
\end{definition}
We often write $T \xrightarrow{\VIS} S$ for $(T,S) \in \VIS$, and similarly for 
other relations. For each abstract execution $\aexec = (\T, \VIS, \AR)$, 
we let $\T_{\aexec} := \T$,  $\VIS_{\aexec} := \VIS$, 
and $\AR_{\aexec} := \AR$.

In an abstract execution $\aexec$, $T \xrightarrow{\VIS_{\aexec}} S$ means 
that the read operations in $S$ may depend on the updates 
of $T$, while $T \xrightarrow{\AR_{\aexec}} S$ means that the 
update operations of $S$ supersede those performed by $T$. 
Naturally, one would expect that 
the value fetched by read operations in a transaction $T$ is 
the most up-to-date one among all the values written by transactions visible to $T$.
For simplicity, we assume that such a transaction always exists.
\begin{definition}
\label{def:constraints}
An abstract execution $\aexec = (\T, \VIS, \AR)$ respects the \emph{Last Write Win} (\textbf{LWW}) policy, 
if for all $T \in \T$  such that $T \ni (\RD\; x: n)$, 
the set $\T' := \left(\VIS^{-1}(T) \cap \WTr_x \right)$ is not empty, 
and $\max_{\AR}(\T') \ni (\WR\; x: n)$, where $\max_{\AR}(\T')$ is the $\AR$-supremum of $\T'$.
\end{definition}

\begin{definition}
\label{def:assumption}
An abstract execution $\aexec = (\T, \VIS, \AR)$ \emph{respects 
causality} if $\VIS$ is transitive. Any abstract execution 
that respects both causality and the LWW policy is said 
to be \emph{valid}.
\end{definition}

We always assume an abstract execution to be valid, unless otherwise stated.
Causality is respected by all abstract executions allowed
by several interesting consistency models. They also simplify the mathematical 
development of our results. In \tra{\ref{app:general}}{\ngeneral}, we explain how our results 
can be generalised for consistency models that do not respect causality. We also 
discuss how the model can be generalised to account for sessions and 
session guarantees \cite{terry1994session}.

We can specify a consistency model using abstract executions in two steps. 
First, we identify 
properties on abstract executions, or \emph{axioms},  that formally express an informal consistency guarantee, and
form a set with the abstract executions satisfying the properties. Next, we project abstract 
executions in this set to underlying histories, and define a consistency model 
$\cm$ to be the set of resulting histories.

Abstract executions hide low-level operational details of the interaction between client programs 
and weakly consistent databases. This benefit has been exploited for proving
that such databases implement intended consistency models~\cite{repldatatypes,crepldatatypes,ev_principles,ev_transactions,framework-concur}.

\subsection{Specification of Weak Consistency Models} 
\label{sec:specification}

In this section we introduce a simple framework for specifying 
consistency models using the style of specification discussed above. 
In our framework, axioms of consistency models relate the visibility and 
arbitration relations via inequalities of the form 
$R_1 \rcomp \AR_{\aexec} \rcomp R_2 \subseteq \VIS_{\aexec}$, 
where $R_1$ and $R_2$ are particular relations over transactions, and 
$\aexec$ is an abstract execution.
As we will explain later, axioms of this form establish a necessary condition for two transactions 
in an abstract execution $\aexec$ to be related by $\VIS_{\aexec}$, 
i.e. they cannot be executed concurrently.
Despite its simplicity, the framework is 
expressive enough to capture several consistency models for 
distributed databases \cite{framework-concur,redblue}; as we will show in 
\S \ref{sec:laws}, one of the benefits of this simplicity is that we can 
infer robustness criteria of consistency models in a systematic way.

As we will see, the relations 
$R_1, R_2$ in axioms of the form above, may depend on the visibility relation of the abstract execution $\aexec$. 
To define such relations, we introduce the notion of \emph{specification function}.

\begin{definition}
\label{def:spec.function}
\label{def:cons.guarantee}
A function $\rho: 2^{(\TrSet \times \TrSet)} \rightarrow
2^{(\TrSet \times \TrSet)}$ is a \emph{specification function} if
for every history $\T$ and relation $R \subseteq \T \times \T$,
then $\rho(R) = \rho(\T \times \T) \cap R?$. 
Here $R?$ is the reflexive closure of $R$.
A \emph{consistency guarantee}, or simply \emph{guarantee}, is a pair of specification functions $(\rho, \pi)$.
\end{definition}
Definition \ref{def:spec.function} ensures that specification functions are 
defined locally: for any $R_1, R_2 \subseteq \T \times \T$, $\rho(R_1 \cup R_2) = 
\rho(R_1) \cup \rho(R_2)$, and in particular for any $R \subseteq \T \times \T$, 
$\rho(R) = \left( \bigcup_{T,S \in \T} \rho(\{(T,S)\}) \right) \cap R?$. 
The reflexive closure in Definition \ref{def:spec.function} is needed because we will always apply 
specification functions to irreflexive relations 
(namely, the visibility relation of abstract 
executions), although the result of this application need 
not be irreflexive. For example, $\rho_{\Id}(R) := \Id$, where $\Id$ is the identity function, 
is a valid specification function.

Each consistency guarantee $(\rho, \pi)$ defines, for each abstract execution $\aexec$, 
an axiom of the form $\rho(\VIS_{\aexec}) \rcomp \AR_{\aexec} \rcomp \pi(\VIS_{\aexec}) 
\subseteq \VIS_{\aexec}$: if this axiom is satisfied by $\aexec$, we say that 
$\aexec$ satisfies the consistency guarantee $(\rho, \pi)$.
Consistency guarantees impose a 
condition on when two transactions $T,S$ in an abstract 
execution $\aexec$ are not allowed to 
execute concurrently,
i.e. they must be related by a $\VIS_{\aexec}$ edge. 
By definition, in abstract executions visibility edges 
cannot contradict arbitration edges, 
hence it is only natural 
that the order in which the transactions $T, S$ above are 
executed is determined by the arbitration order: 
in fact, the definition of specification function ensures that 
$\rho(\VIS_{\aexec}) \subseteq \VIS_{\aexec}?$ and
$\pi(\VIS_{\aexec}) \subseteq \VIS_{\aexec}?$,
so that
$(\rho(\VIS_{\aexec}) \rcomp \AR_{\aexec} \rcomp 
\pi(\VIS_{\aexec})) \subseteq \AR_{\aexec}$
for all abstract executions $\aexec$. 

\begin{definition}
\label{def:cm.spec}
A \emph{consistency model specification $\Sigma$} or \emph{x-specification} 
is a set of consistency guarantees $\{(\rho_i, \pi_i)\}_{i \in I}$ for some 
index set $I$.

We define $\aeset(\Sigma)$ to be the set of valid abstract executions that satisfy all
the consistency guarantees of $\Sigma$. We let $\modelof(\Sigma) := \{\hist_{\aexec} 
\mid \aexec \in \aeset(\Sigma)\}$.
\end{definition}\vspace{0.2\baselineskip}

\myparagraph{Examples of Consistency Model Specifications}
Figure \ref{fig:models} shows several examples of specification functions
and consistency guarantees. 
In the figure 
we use the relations $\circled{\T} := \{(T, T) \mid T \in \T\}$
and $\circled{o} := \{(T, T) \mid T \ni o\}$  for $\T \subseteq \TrSet$ and $o \in \Op$. 
The guarantees in the figure can be composed together to specify, 
among others, several of the consistency models considered in \cite{framework-concur}: 
we give some examples of them below. 
Each of these consistency models 
allows different kinds of anomalies: 
\setlength{\intextsep}{0pt}
\begin{wrapfigure}[12]{r}{0.48\textwidth}
\vspace{-5pt}
{\small
\[
\begin{array}{| lcl | l l l | }
\hline
\text{Function} & & \text{Definition} \\ 
\hline
\rho_{\Id}(R) &$=$& \Id \\
\rho_{\SI}(R) &=&  R \setminus \Id \\
\rho_{x}(R)  &=& \circled{\WTr_x} \\
\rho_S(R) &=& \circled{\SerTX}\\
\hline
\text{Guarantee} & \multicolumn{2}{c |}{\text{Associated Axiom}}\\
\hline
(\rho_{\Id}, \rho_{\Id}) & \multicolumn{2}{r |}{\AR \subseteq \VIS}\\
 (\rho_{\Id}, \rho_{\SI}) & \multicolumn{2}{r |}{\AR \rcomp \VIS \subseteq \VIS}\\
(\rho_{x}, \rho_{x}) & \multicolumn{2}{r |}{\circled{\WTr_x} \rcomp \AR \rcomp \circled{\WTr_x} \subseteq \VIS}\\ 
 (\rho_{S}, \rho_{S}) & \multicolumn{2}{r |}{\circled{\SerTX} \rcomp \AR \rcomp \circled{\SerTX} \subseteq \VIS}\\
 \hline
\end{array}
\]
}
\vspace{-15pt}
\caption{Some Specification Functions and Consistency Guarantees}
\label{fig:models}
\end{wrapfigure}

\setlength{\intextsep}{10pt plus 6pt minus 3pt}
due to lack of space, these 
are illustrated in \tra{\ref{app:anomalies}}{\nanomalies}.

\noindent\textbf{Causal Consistency \cite{cops}: }
This is the weakest consistency model
we consider. It is specified by $\Sigma_{\CC} = \emptyset$. 
In this case, all abstract executions in $\aeset(\Sigma_{\CC})$ respect causality.
The execution in Figure \ref{fig:example} is an example in $\aeset(\Sigma_{\CC})$.

\noindent\textbf{Red-Blue Consistency \cite{redblue}:} 
This model extends causal consistency 
by marking a subset of transactions as serialisable, 
and ensuring that no two such transactions appear to execute concurrently.
We model red-blue consistency via 
the x-specification $\Sigma_{\CCSER} = \{(\rho_S, \rho_S)\}$. In the definition of $\rho_S$,
an element $\SerTX \in \Op$ is used to mark transactions as serialisable, and
the specification requires that
in every execution $\aexec \in \aeset(\Sigma_{\CCSER})$, 
any two transactions $T,S \ni \SerTX$  in $\aexec$
be compared by $\VIS_{\aexec}$. The abstract execution from Figure \ref{fig:example} 
is included in $\aeset(\Sigma_{\CCSER})$, but if it were modified so that transactions 
$T_1, T_2$ were marked as serialisable, then the result would not belong to 
$\aeset(\Sigma_{\CCSER})$. 

\noindent\textbf{Parallel Snapshot Isolation (PSI) \cite{PSI,NMSI}: } This model 
strengthens causal consistency by enforcing the \emph{Write Conflict Detection} 
property:  transactions writing to one same object do not execute concurrently.
We let  
$\Sigma_{\PSI} = \{(\rho_{x}, \rho_{x})\}_{x \in \Obj}$:  
every execution $\aexec \in \aeset(\Sigma_{\PSI})$ satisfies 
the inequality 
$(\circled{\WTr_x} \rcomp \AR_{\aexec} \rcomp \circled{\WTr_x}) \subseteq \VIS_{\aexec}$, for all $x \in \Obj$.

\noindent\textbf{Snapshot Isolation (SI) \cite{si}: } This consistency model 
strengthens PSI by requiring that, in executions, the set of transactions visible 
to any transaction $T$ is a prefix of the arbitration relation. 
Formally, we let $\Sigma_{\SI} = \Sigma_{\PSI} \cup \{(\rho_{\Id}, \rho_{\SI})\}$. 
The consistency guarantee $(\rho_{\Id}, \rho_{\SI})$ ensures that any abstract 
execution $\aexec \in \aeset(\SI)$ satisfies the property $(\AR_{\aexec} \rcomp \VIS_{\aexec})
\subseteq \VIS_{\aexec}$\footnote{To be precise, the 
property induced by the guarantee $(\rho_\Id, \rho_{\SI})$ is 
$(\AR_{\aexec} \rcomp (\VIS_{\aexec} \setminus \Id)) \subseteq \AR_{\aexec}$. 
However, since $\VIS_{\aexec}$ is an irreflexive relation, $\VIS_{\aexec} \setminus \Id = \VIS_{\aexec}$. 
Also, note that $\rho(R) = R$ is not a specification function, so we cannot replace the guarantee 
$(\rho_{\Id}, \rho_{\SI})$ with $(\rho_{\Id}, \rho)$.}. 

Similarly to what we did to specify Red-Blue consistency, 
we can strengthen SI by allowing the possibility to mark transactions 
as serialisable. The resulting x-specification is $\Sigma_{\SI+\SER} = 
\Sigma_{\SI} \cup \{(\rho_S, \rho_S)\}$. This x-specification captures a fragment of Microsoft SQL server,
which allows the user to select the consistency model at which 
a transaction should run \cite{selectisolevel}.

\noindent\textbf{Serialisability: } Executions in this consistency model require 
the visibility relation to be total. This can be formalised via the x-specification 
$\Sigma_{\SER} := \{(\rho_{\Id}, \rho_{\Id})\}$. 
Any $\aexec \in \aeset(\Sigma_{\SER})$  is such that 
$\AR_{\aexec} \subseteq \VIS_{\aexec}$, thus 
enforcing $\VIS_{\aexec}$ to be a strict total order.\vspace{1.2\baselineskip}

\section{Dependency Graphs}
\label{sec:dgraphs}

We present another style of specification for consistency models 
based on dependency graphs, 
introduced in \cite{adya}. These are structures that capture the 
data-dependencies between transactions accessing one same object. 
Such dependencies can be over approximated at compilation time. For 
this reason, they have found use in 
static analysis 
\cite{fekete-tods,SIanalysis,psi-chopping,giovanni_concur16} for programs 
running under a weak consistency model. 

\begin{definition}
\label{def:rf.exec}
A \textbf{\em dependency graph} is a tuple $\G = (\T, \RF, \VO, \AD)$,
where $\T$  is a history and
\begin{enumerate}
\item\label{item:dgraph.wr} $\RF : \Obj \rightarrow 2^{\T \times \T}$ is such that:
\begin{enumerate}[(a)]
\item\label{item:dgraph.wr.values} $\forall T, S \in \T.\, \forall x.\, T \xrightarrow{\RF(x)} S \implies {}
T \neq S \wedge
\exists n.\, (T \ni \WR\; x : n) \wedge (S \ni \RD \; x : n)$,
\item\label{item:dgraph.wr.initial} $\forall S \in \T.\, \forall x.\, (S \ni \RD\; x: n) \implies \exists
  T.\, T \xrightarrow{\RF(x)} S$,
\item\label{item:dgraph.wr.unique} $\forall T, T', S \in \T.\, \forall x.\, (T \xrightarrow{\RF(x)} S \wedge
  T' \xrightarrow{\RF(x)} S) \implies
  T = T'$;
\end{enumerate}  
\item\label{item:dgraph.ww} $\VO: \Obj \rightarrow 2^{\T \times \T}$ is such that for every $x \in
  \Obj$, $\VO(x)$ is a strict, total order over $\WTr_x$;
\item\label{item:dgraph.rw} $\AD : \Obj  \rightarrow 2^{\T \times \T}$ is such that $S \xrightarrow{\AD(x)} T$ 
iff $S \neq T$ and $\exists T'.\; T' \xrightarrow{\RF(x)} S \wedge T' \xrightarrow{\VO(x)} T$.
\end{enumerate}
\end{definition}

Given a dependency graph $\G = (\T, \RF, \VO, \AD)$, 
we let $\T_\G := \T$, $\RF_{\G} := \RF$, $\VO_{\G} := \VO$, $\AD_{\G} := \AD$. 
The set of all dependency graphs is denoted as $\graphs$.
Sometimes, we commit an abuse of notation and use the symbol 
$\RF$ to denote the relation $\bigcup_{x \in \Obj} \RF(x)$, 
and similarly for $\VO$ and $\AD$. The actual meaning 
of $\RF$ will always be clear from the context.

Let $\G \in \graphs$. The \emph{write-read dependency} $T \xrightarrow{\RF_{\G}(x)} S$ 
means that $S$ reads the value of object $x$ that has been written by $T$.
By Definition \ref{def:rf.exec}, for any transaction $S \in \RTr_x$ 
there exists exactly one transaction $T$ such that $T \xrightarrow{\RF_{\G}(x)} S$.
The relation $\VO_\G(x)$ establishes a total order in which updates 
over object $x$ are executed by transactions; its elements are called \emph{write-write dependencies}. 
Edges in the relation $\AD_\G(x)$ take the 
name of \emph{anti-dependencies}. $T \xrightarrow{\AD_\G(x)} S$ means that 
transaction $T$ fetches some value for object $x$, but this is later 
updated by $S$.
Given an abstract execution $\aexec$, we can extract a dependency graph $\graphof(\aexec)$ such that $\hist_{\graphof(\aexec)} = \hist_\aexec$.
\begin{definition}
\label{def:execution.dependencies}
Let $\aexec = (\T, \VIS, \AR)$ be an execution.  For $x \in \Obj$, we define
$\graphof(\aexec) = (\T, \RF_{\aexec}, \VO_{\aexec}, \AD_{\aexec})$, where:
\begin{enumerate}
\item \label{def:rf}
$
T \xrightarrow{\RF_{\aexec}(x)} S \iff
{(S \ni \RD\;x: \_)} \wedge { T = \max_{\AR}(\VIS^{-1}(S) \cap \WTr_x)};
$
\item \label{def:vo}
$
T \xrightarrow{\VO_{\aexec}(x)} S \iff 
T \xrightarrow{\AR} S \wedge T, S \in \WTr_x;
$
\item \label{def:ad}
$
T \xrightarrow{\AD_{\aexec}(x)} S \iff S \neq T \wedge (\exists T'.\, T' \xrightarrow{\RF_{\aexec}(x)} T \wedge T' \xrightarrow{\VO_{\aexec}(x)} S) ).
$
\end{enumerate}
\end{definition}

\begin{proposition}
For any valid abstract execution $\aexec$, $\graphof(\aexec)$ 
is a dependency graph.
\end{proposition}\vspace{0.2\baselineskip}

\myparagraph{Specification of Consistency Models using Dependency Graphs}
We interpret a dependency graph $\G$ 
as a labelled graph whose vertices are transactions in $\T_x$, 
and whose edges are pairs of the form $T \xrightarrow{R} S$, where $R \in \{\RF_{\G}(x), 
\VO_{\G}(x)_{\G}, \AD_{\G}(x) \mid x \in \Obj \}$. 
To specify a consistency model, we employ a two-steps approach. 
We first identify one or more conditions to be satisfied by dependency graphs. 
Such conditions require cycles of a certain form not to appear in a dependency graph. Then  
we define a consistency model by projecting the set of dependency graphs satisfying the 
imposed conditions into the underlying histories.
This style of specification is reminiscent 
of the one used in the CAT \cite{cat} language for formalising weak memory models. 
In the following we treat the relations 
$\RF_{\G}(x), \VO_{\G}(x), \AD_{\G}(x)$ both as set-theoretic relations, and 
as edges of a labelled graph.

\begin{definition}
\label{def:graph.specification}
A \emph{dependency graph based specification}, or simply g-specification, is a set $\Delta = \{\delta_1, \cdots, \delta_n\}$, 
where for each $i \in \{1, \cdots, n\}$, $\delta_i$ is a function of type
$\graphs \rightarrow 2^{(\TrSet \times \TrSet)}$ 
and satisfies $\delta_i(\G) \subseteq (\RF_{\G} \cup \VO_{\G} \cup \AD_{\G})^{\ast}$
for every $\G \in \graphs$.

Given a g-specification $\Delta$, we define $\graphs(\Delta) = \{\G \in \graphs \mid \forall \delta \in \Delta.\, \delta(\G) \cap \Id = \emptyset \}$, 
and we let $\modelof(\Delta) = \{\T \mid \exists \RF,\VO,\AD.\, (\T, \RF, \VO, \AD) \in \graphs(\Delta)\}$.
\end{definition}
The requirement imposed over the functions $\delta_1,\cdots, \delta_n$ ensures that, whenever $(T,S) \in \delta_{i}(\G)$, for 
some dependency graph $\G$, then there exists a path in $\G$, that connects $T$ to $S$. 
For $\Delta = \{\delta_i\}_{i=1}^{n}$ and $\G \in \graphs$, the requirement that $\delta_{i}(\G) \cap \Id = \emptyset$ 
means that $\G$ does not contain any cycle 
$T_0 \xrightarrow{R_0} T_1 \xrightarrow{R_1} \cdots \xrightarrow{R_{n-1}} T_n,$ 
such that $T_0 = T_n$, and $(R_0 \rcomp \cdots \rcomp R_{n-1}) \subseteq \delta_{i}(\G)$.\vspace{0.2\baselineskip}

\myparagraph{Examples of g-specifications of consistency models}
Below we give some examples of $g$-specifications for the consistency models 
presented in \S \ref{sec:abstract.executions}. 

\begin{theorem}~
\label{thm:depgraphs}
\begin{enumerate}
\item \label{thm:ser}
An execution $\aexec$ is serialisable iff $\graphof(\aexec)$ does not contain 
any cycle. That is, \linebreak $\modelof(\Sigma_{\SER}) = \modelof(\{\delta_{\SER}\})$,
where $\delta_{\SER}(\G) = (\RF_{\G} \cup \VO_{\G} \cup \AD_{\G})^{+}$. 

\item \label{thm:si}
An execution $\aexec$ is allowed by snapshot isolation iff 
$\graphof(\aexec)$ only admits cycles with at least 
two consecutive anti-dependency edge. That is,
$\modelof(\Sigma_{\SI}) = \modelof(\{\delta_{\SI}\})$,
where  $\delta_{\SI}(\G) = ((\RF_{\G} \cup \VO_{\G}) \rcomp \AD_{\G}?)^{+}$. 

\item \label{thm:psi}
An execution $\aexec$ is allowed by parallel snapshot isolation iff  
$\graphof(\aexec)$ has no cycle where all anti-dependency 
edges are over the same object. 
Let $\delta_{\PSI_0}(\G) = (\RF_{\G} \cup \VO_\G)^{+}$, 
$\delta_{\PSI(x)}(\G) = (\bigcup_{x \in \Obj} (\RF_{\G} \cup \VO_{\G})^{\ast} \rcomp \AD_\G(x))^{+}$,  
and define $\Delta_{\PSI} = \{\delta_{\PSI_0}\} \cup \{\delta_{\PSI(x)} \mid x \in \Obj\}$. 
Then, $\modelof(\Sigma_\PSI) = \modelof(\Delta_{\PSI})$.
\end{enumerate}
\end{theorem}

Theorem \ref{thm:depgraphs}\eqref{thm:ser} was proved 
in \cite{adya}. The only if condition of Theorem \ref{thm:depgraphs}\eqref{thm:si} was 
proved in \cite{fekete-tods}; we proved the if condition of Theorem \ref{thm:depgraphs}\eqref{thm:si} 
in \cite{SIanalysis}. 
Theorem \ref{thm:depgraphs}\eqref{thm:psi} improves on
the specification we gave for PSI in \cite{SIanalysis}; the latter does
not have any constraints on the objects to which anti-dependencies refer to.
We outline the proof of Theorem \ref{thm:depgraphs}\eqref{thm:psi} in \S \ref{sec:completeness}.

\section{Algebraic Laws for Weak Consistency}
\label{sec:laws}

Having two different styles for specifying consistency 
models gives rise to the following problems: 

\noindent\textbf{Weak Correspondence Problem: }
given a x-specification $\Sigma$, determine a non-trivial g-specification $\Delta$ which 
over-approximates $\Sigma$, that is such that $\modelof(\Sigma) \subseteq \modelof(\Delta)$.

\noindent\textbf{Strong Correspondence Problem: }
Given a x-specification $\Sigma$, determine an equivalent 
g-specification $\Delta$, that is such that $\modelof(\Sigma) = \modelof(\Delta)$.

We first focus on the weak correspondence problem, 
and we discuss the strong correspondence problem in \S \ref{sec:completeness}. 
This problem is not only of theoretical interest. 
Determining a g-specification $\Delta$ that 
over-approximates a x-specification $\Sigma$ corresponds to establishing one or 
more conditions 
satisfied by all cycles of dependency graphs from the set 
$\{\graphof(\aexec) \mid \aexec \in \aeset(\Sigma)\}$. 
Cycles in a 
dependency graph that 
respect such a condition 
are called \emph{$\Sigma$-critical} (or simply critical), and graphs that admit a non-$\Sigma$-critical cycle 
cannot be obtained from abstract executions in $\aeset(\Sigma)$.
One can ensure that an application running under the model $\Sigma$ 
is \emph{robust}, i.e. it only produces serialisable behaviours, by checking for the absence of 
$\Sigma$-critical cycles at static time \cite{fekete-tods,giovanni_concur16}. 
Robustness of an application can also be checked at run-time,  
by incrementally constructing the dependency graph of executions, 
and detecting the presence of $\Sigma$-critical cycles \cite{kemme}. \vspace{0.2\baselineskip}

\begin{figure} 

\scalebox{0.96}{\small
\[
\begin{array}{| l l l l | l l l l |}
\hline
\multicolumn{4}{|c|}{\txlaws\; \text{Algebraic laws for sets of transactions}} & 
\multicolumn{4}{c|}{\allexeclaws~\text{Algebraic laws for abstract Executions}}\\[5pt]
\txId & \circled{\T'} \subseteq \Id & \txComp & \circled{\T_1 \cap \T_2} = \circled{\T_1} \rcomp \circled{\T_2} & 
\WRinVIS & \RF(x) \subseteq \VIS & \WWinAR & \VO(x) \subseteq \AR\\
\txDistrR & \multicolumn{3}{l |}{(R_1 \rcomp \circled{\T'}) \cap R_2 = (R_1 \cap R_2) \rcomp \circled{\T'}} & 
\RWinAVIS & \AD(x) \subseteq \overline{\VIS^{-1}} & \VIStrans & \VIS^{+} \subseteq \VIS\\
\txDistrL & \multicolumn{3}{l |}{(\circled{\T'} \rcomp R_1) \cap R_2 = \circled{\T'} \rcomp (R \cap R_2)} &
\ARtrans & \AR^{+} \subseteq \AR & \VISinAR & \VIS \subseteq \AR\\
\cline{1-4}
\multicolumn{4}{|c|}{\deplaws~\text{Algebraic laws for (anti-)dependencies}} & 
\LWW & \multicolumn{3}{l |}{ \circled{\WTr_x} \rcomp \VIS \rcomp \AD(x) \subseteq \AR}\\[5pt]
\depWRTx & \multicolumn{3}{l |}{ \RF(x) \subseteq \circled{\WTr_x} \rcomp \RF(x) \rcomp  \circled{\RTr_x}} &
\AVISright & \multicolumn{3}{l |}{\VIS \rcomp \overline{\VIS^{-1}} \subseteq \overline{\VIS^{-1}}}\\
\depWWTx & \multicolumn{3}{l |}{ \VO(x) \subseteq \circled{\WTr_x} \rcomp \VO(x) \rcomp \circled{\WTr_x}} & 
\AVISleft & \multicolumn{3}{l |}{\overline{\VIS^{-1}} \rcomp \VIS \subseteq \overline{\VIS^{-1}}}\\
\depRWTx & \multicolumn{3}{l |}{\AD(x) \subseteq \circled{\RTr_x} \rcomp \AD(x) \rcomp \circled{\WTr_x}} & 
\AVISnotVIS & \multicolumn{3}{l |}{(\overline{\VIS^{-1}} \rcomp \VIS) \cap \Id \subseteq \emptyset}\\
\depWRIrrefl & \multicolumn{3}{l |}{\RF(x) \subseteq \RF(x) \setminus \Id} & 
\VISnotAVIS & \multicolumn{3}{l |}{(\VIS \rcomp \overline{\VIS^{-1}}) \cap \Id \subseteq \emptyset}\\
\depWWIrrefl & \multicolumn{3}{l |}{\VO(x) \subseteq \VO(x) \setminus \Id} &
\ARirrefl & \AR \cap \Id \subseteq \emptyset & & \\
\depRWIrrefl & \multicolumn{3}{l |}{\AD(x) \subseteq \AD(x) \setminus \Id} & & & & \\
\hline
\multicolumn{8}{|c|}{\cmexeclaws~\text{Algebraic laws induced by the consistency guarantee } (\rho, \pi)}\\[5pt]
\Axiom & \multicolumn{3}{l}{\rho(\VIS) \rcomp \AR \rcomp \pi(\VIS) \subseteq \VIS} & 
\CoAxiomAR & \multicolumn{3}{l |}{(\pi(\VIS) \rcomp \overline{\VIS^{-1}} \rcomp \rho(\VIS) ) \setminus \Id \subseteq \AR}\\
\CoAxiomL & \multicolumn{7}{c |}{(\AR \rcomp \pi(\VIS) \rcomp \overline{\VIS^{-1}} ) \cap \rho(\T \times \T)^{-1}
\subseteq \overline{\VIS^{-1}}}\\
\CoAxiomR & \multicolumn{7}{c |}{(\overline{\VIS^{-1}} \rcomp \rho(\VIS) \rcomp \AR) \cap \pi(\T \times \T)^{-1} 
\subseteq \overline{\VIS^{-1}}}\\ 
\hline
\end{array}
\]
}
\caption{Algebraic laws satisfied by an abstract execution $\aexec = (\T, \VIS, \AR)$. 
Here $\graphof(\aexec) = (\T, \RF, \VO, \AD)$.
The inequalities in part \textbf{(d)} are valid under the assumption that $\aexec \in \aeset(\{(\rho, \pi)\})$.
}
\label{fig:laws}
\end{figure}

\myparagraph{General Methodology} 
Let $\Sigma$ be a given x-specification. We tackle the weak correspondence problem in two steps. 

First, we identify a set of inequalities that hold for all the executions $\aexec$ satisfying 
consistency guarantees $(\rho, \pi)$ in $\Sigma$. There are two kinds of such inequalities.
The first are the inequalities in Figure \ref{fig:laws}, and the second the inequalities 
corresponding to the axioms of the Kleene 
Algebra $(2^{\TrSet \times \TrSet}, \emptyset, \Id, \cup, \rcomp, \cdot^{\ast})$ and 
the Boolean algebra $(2^{\TrSet \times \TrSet}, \emptyset, \TrSet \times \TrSet, 
\cup, \cap, \overline{\cdot})$. 
The exact meaning of the inequalities in Figure \ref{fig:laws} is discussed later in this section.  

Second, 
we exploit our inequalities to derive  
other inequalities of the form $R_\aexec \subseteq \AR_{\aexec}$ for every $\aexec \in \aeset(\Sigma)$.
Here $R_{\aexec}$ is a relation built from dependencies in $\graphof(\aexec)$, 
i.e. $R_{\aexec} \subseteq (\RF_{\aexec} \cup \VO_{\aexec} \cup \AD_{\aexec})^{\ast}$. 
Because $\AR_{\aexec}$ is acyclic (that is $\AR_{\aexec}^{+} \cap \Id \subseteq \emptyset$), we may conclude that $R_{\aexec}$ is acyclic for any $\aexec \in \aeset(\Sigma)$. In particular, we have that $\modelof(\Sigma) \subseteq \modelof(\{\delta\})$, 
where $\delta$ is a function that maps, for every abstract execution $\aexec$, 
the dependency graph $\graphof(\aexec)$ into the relation $R_{\aexec}$.

Some of the inequalities we develop, namely those in Figure \ref{fig:laws}\cmexeclaws, 
are parametric in the consistency guarantee $(\rho, \pi)$. As a consequence, 
our approach can be specialised to any consistency model that is captured by our 
framework. To show its applicability, we derive critical cycles 
for several of the consistency models that we have presented.\vspace{0.2\baselineskip}

\myparagraph{Presentation of the Laws}\hspace{4pt} 
Let $\aexec = (\T, \VIS, \AR)$, 
and $\graphof(\aexec) = (\T, \RF, \VO, \AD)$.
We now explain the inequalities in Figure \ref{fig:laws}. 
Among these, the inequalities in Figures \ref{fig:laws}\txlaws\; 
and \deplaws\; should be self-explanatory. 

Let us discuss the inequalities of Figure \ref{fig:laws}\allexeclaws. 
The inequalities \WRinVIS, \WWinAR\; and \RWinAVIS\; 
relate dependencies to either basic or derived relations of abstract executions. 
Dependencies of the form $\RF, \VO$ are included in the 
relations $\VIS, \AR$, respectively, as established by inequalities \WRinVIS\; 
and \WWinAR.
The inequality \RWinAVIS, which we prove presently, is non-standard. It 
relates anti-dependencies to a novel \emph{anti-visibility} relation 
$\overline{\VIS^{-1}}$, defined as $T \xrightarrow{\overline{\VIS^{-1}}} S$ 
iff $\neg(S \xrightarrow{\VIS} T)$. In words, $S$ is \emph{anti-visible} to $T$ 
if $T$ does not observe the effects of $S$. As we will explain later, 
anti-visibility plays a fundamental role in the development of our laws.

\noindent\myparagraph{Proof of Inequality \RWinAVIS}
\label{proof:RWinAVIS}
Suppose $T \xrightarrow{\AD(x)} S$ for some object $x \in \Obj$.
By definition, $T \neq S$, and there exists a transaction $T'$ such that $T' \xrightarrow{\RF(x)} T$ 
and $T' \xrightarrow{\VO(x)} S$. In particular, $T' \xrightarrow{\VIS} T$ and $T' \xrightarrow{\AR} S$ by 
the inequalities \WRinVIS\; and \WWinAR, respectively.
Now, if it were $S \xrightarrow{\VIS} T$, 
then we would have that $T'$ is not 
the $\AR$-supremum of the set of transactions 
visible to $T$, and writing to object $x$. 
But this contradicts the definition of $\graphof(\aexec)$, and the 
edge $T' \xrightarrow{\RF(x)} T$.
Therefore, $T \xrightarrow{\overline{\VIS^{-1}}} S$. \qed

Another non-trivial inequality is \LWW\; in Figure \ref{fig:laws}\allexeclaws.
It says that if a 
transaction $T$ reads a value for an object $x$ that is 
later updated by another transaction $S$ ($T \xrightarrow{\AD} S)$, then the 
update of $S$ is more recent (i.e. it follows 
in arbitration) than all the updates to $x$ seen by $T$. 
We prove it in \tra{\ref{app:laws}}{\nproofs}. The other inequalities 
in Figure \ref{fig:laws}\allexeclaws\; are self explanatory.

The inequalities in Figure \ref{fig:laws}\cmexeclaws\;
are specific to a consistency guarantee $(\rho, \pi)$, and
hold for an execution $\aexec$ when the execution satisfies $(\rho, \pi)$. 
The inequality \Axiom\; is just the definition of consistency
guarantee.  The next inequality \CoAxiomAR\; is where  
the novel anti-visibility relation, introduced previously, comes into play. 
While the consistency guarantee $(\rho, \pi)$ expresses
when arbitration induces transactions related by visibility, 
the inequality \CoAxiomAR\; expresses when anti-visibility induces
transactions related by arbitration. To emphasise
this correspondence, we call the inequality \CoAxiomAR\;  
\emph{co-axiom} induced by $(\rho, \pi)$. 
Later in this section, we show how by exploiting the co-axiom induced by 
several consistency guarantees, we can derive critical cycles of several 
consistency models.

\noindent\myparagraph{Proof of Inequality \CoAxiomAR}
Assume $\aexec \in \aeset(\{(\rho, \pi)\})$.
Let $T, T', S', S \in \T$ be such that $T \neq S$, $T \xrightarrow{\pi(\VIS)} T' \xrightarrow{\overline{\VIS^{-1}}} S' \xrightarrow{\rho(\VIS)} S$. 
Because $\AR$ is total, either $S \xrightarrow{\AR} T$ or $T \xrightarrow{\AR} S$. 
However, the former case is not possible. If so, we would have $S' \xrightarrow{\rho(\VIS)} S \xrightarrow{\AR} 
T \xrightarrow{\pi(\VIS)} T'$. 
because $\aexec \in \aeset(\{(\rho, \pi)\})$, by 
the inequality \Axiom, it would follow that $S' \xrightarrow{\VIS} T'$, contradicting the assumption that $T' \xrightarrow{\overline{\VIS^{-1}}} S'$. 
Therefore, it has to be $T \xrightarrow{\AR} S$. \qed

The last inequalities \CoAxiomL\; and 
\CoAxiomR\; in Figure \ref{fig:laws}\cmexeclaws\;
show that anti-visibility edges of $\aexec$ 
are also induced by the consistency guarantee $(\rho, \pi$). 
We prove them formally in \tra{\ref{app:laws}}{\nproofs}, where 
we also illustrate some of their applications.

\myparagraph{Applications}
We employ the algebraic laws of Figure \ref{fig:laws} to derive $\Sigma$-critical 
cycles for arbitrary x-specifications, using the methodology explained previously: 
given a x-specification $\Sigma$ and an abstract execution $\aexec$, we 
characterise a subset of $\AR_{\aexec}$ as a relation 
$R_{\mathsf{G}}$ built from the dependencies in $\graphof(\aexec)$ 
and relations of the form $\circled{o}$, where $o \in \Op$. 
Because $R_{\mathsf{G}} \subseteq \AR_{\aexec}$, we conclude 
that $R_{\mathsf{G}}$ is acyclic.

The inequalities \WRinVIS, \VISinAR\; and \WWinAR\; ensure that 
we can always include write-read and write-write dependencies in 
the relation $R_{\mathsf{G}}$ above. Because 
of inequalities \RWinAVIS\; and \CoAxiomAR\; (among others), 
we can include in $R_{\mathsf{G}}$ also relations that involve anti-dependencies.
The following result shows how this methodology can 
be applied to serialisability.
We use the notation $R_1 \stackrel{\mathbf{(eq)}}{\subseteq} R_2$ 
to denote that the inequality $R_1 \subseteq R_2$ follows from $\mathbf{(eq)}$.

\begin{theorem} 
\label{thm:ser.acyclic}
For all $\aexec \in \aeset(\Sigma_{\SER})$, the relation $(\RF_{\aexec} \cup \VO_{\aexec} \cup 
\AD_{\aexec})$ is acyclic.
\end{theorem}

\noindent\myparagraph{Proof} 
Recall that $\Sigma_{\SER} = \{(\rho_\Id, \rho_\Id)\}$, where $\rho_{\Id}(\_) = \Id$. We have
\begin{gather}
\AD_{\aexec} \stackrel{\depRWIrrefl}{\subseteq} 
\AD_{\aexec} \setminus \Id \stackrel{\RWinAVIS}{\subseteq} 
\overline{\VIS^{-1}_{\aexec}} \setminus \Id = (\rho_{\Id}(\VIS_{\aexec}) \rcomp 
\overline{\VIS^{-1}_{\aexec}} \rcomp \rho_{\Id}(\VIS_{\aexec})) \setminus \Id \stackrel{\CoAxiomAR}{\subseteq} \AR_{\aexec}
\label{eq:RWinAR}\\
(\RF_{\aexec} \cup \VO_{\aexec} \cup \AD_{\aexec}) \stackrel{\textbf{(c.1,c.6)}}{\subseteq} (\AR_{\aexec} \cup \VO_{\aexec} \cup \AD_{\aexec}) 
\stackrel{\WWinAR}{\subseteq} (\AR_{\aexec} \cup \AD_{\aexec}) \stackrel{\eqref{eq:RWinAR}}{\subseteq} \AR_{\aexec}
\label{eq:AllinAR} \\
\nonumber (\RF_{\aexec} \cup \VO_{\aexec} \cup \AD_{\aexec})^{+} \cap \Id \stackrel{\eqref{eq:AllinAR}}{\subseteq} \AR_{\aexec}^{+} \cap \Id
\stackrel{\ARtrans}{\subseteq} \AR_{\aexec} \cap \Id \stackrel{\ARirrefl}{\subseteq} \emptyset. 
\tag*{$\qed$} 
\end{gather}

Along the lines of the proof of Theorem \ref{thm:ser.acyclic}, we can 
characterise $\Sigma$-critical cycles for an arbitrary x-specification $\Sigma$. 
Below, we show how to apply our methodology to derive $\Sigma_{\CCSER}$-critical 
cycles. 
\begin{theorem}
\label{thm:ccser.acyclic}
Let $\aexec \in \aeset(\Sigma_{\CCSER})$. Say that a $\AD_{\aexec}$ edge in a cycle of 
$\graphof(\aexec)$ is \emph{protected} if its endpoints are connected to 
serialisable transactions via a sequence of $\RF_{\aexec}$ edges. Then 
all cycles in $\graphof(\aexec)$ have at least one unprotected $\AD_{\aexec}$ edge. 
Formally, let $\fenced{\AD_{\aexec}}$ be $(\circled{\SerTX} \rcomp (\RF_{\aexec})^{\ast} 
\rcomp \AD_{\aexec} \rcomp (\RF_{\aexec})^{\ast} \rcomp \circled{\SerTX})$. 
Then $(\RF_{\aexec} \cup \VO_{\aexec} \cup \fenced{\AD_{\aexec}})$ is acyclic.
\end{theorem}

\noindent\myparagraph{Proof}
It suffices to prove that $\fenced{\AD_{\aexec}} \subseteq \AR_{\aexec}$. The 
rest of the proof is similar to the one of Theorem \ref{thm:ser.acyclic}. 
We recall that $\Sigma_{\CCSER} = \{(\rho_S, \rho_S)\}$, where $\rho_S(\_) = \circled{\SerTX}$. 
\begin{gather}
\nonumber \RF_{\aexec}^{\ast} \rcomp \AD_{\aexec} \rcomp \RF_{\aexec}^{\ast} \stackrel{\textbf{(c.1,c.4)}}{\subseteq} \VIS_{\aexec}? \rcomp \AD_{\aexec} \rcomp \VIS_{\aexec}? 
\stackrel{\depRWIrrefl}{\subseteq} \VIS_{\aexec}? \rcomp (\AD_{\aexec} \setminus \Id) \rcomp \VIS_{\aexec}? 
\stackrel{\RWinAVIS}{\subseteq} \\
\nonumber\VIS_{\aexec}? \rcomp (\overline{\VIS_{\aexec}^{-1}} \setminus \Id) \rcomp \VIS_{\aexec}? 
\subseteq 
( (\overline{\VIS_{\aexec}^{-1}} \setminus \Id) \cup (\VIS_{\aexec} \rcomp \overline{\VIS_{\aexec}^{-1}}) ) \rcomp 
\VIS_{\aexec}? \stackrel{\VISnotAVIS}{\subseteq}\\ 
((\overline{\VIS_{\aexec}^{-1}} \setminus \Id) \cup (\VIS_{\aexec} \rcomp \overline{\VIS_{\aexec}^{-1}}) \setminus \Id) 
\rcomp \VIS_{\aexec}? \stackrel{\AVISright}{\subseteq}
 (\overline{\VIS_{\aexec}^{-1}} \setminus \Id) \rcomp \VIS_{\aexec}?
\stackrel{\textbf{(c.10,c.9)}}{\subseteq} \overline{\VIS_{\aexec}^{-1}} \setminus \Id \label{eq:VIS.RW.VIS}\\
\nonumber \circled{\SerTX} \rcomp (\overline{\VIS_{\aexec}^{-1}} \setminus \Id) \rcomp \circled{\SerTX} \stackrel{\textbf{(a.3,a.4)}}{=} 
(\circled{\SerTX} \rcomp \overline{\VIS_{\aexec}^{-1}} \rcomp \circled{\SerTX}) \setminus \Id = \\
(\rho_{S}(\VIS_{\aexec}) \rcomp \overline{\VIS_{\aexec}^{-1}} \rcomp \rho_S(\VIS_{\aexec})) \setminus \Id \stackrel{\CoAxiomAR}{\subseteq} \AR_{\aexec}
\label{eq:FencedinAR}\\
\nonumber \fenced{\AD_{\aexec}} = \circled{\SerTX} \rcomp \RF_{\aexec}^{\ast} \rcomp \AD_{\aexec} \rcomp \RF_{\aexec}^{\ast} \rcomp \circled{\SerTX} 
\stackrel{(\ref{eq:VIS.RW.VIS}, \ref{eq:FencedinAR})}{\subseteq} \AR_{\aexec}. \tag*{$\qed$}
\end{gather}

We remark that our characterisation of $\Sigma_{\CCSER}$-critical 
cycle cannot be compared to the one given in \cite{giovanni_concur16}. 
In \tr{\ref{app:laws}}{\nproofs} we show how our methodology 
can be applied to give a characterisation of $\Sigma_{\CCSER}$-critical cycles 
that is stronger than both the one presented in Theorem \ref{thm:ccser.acyclic} and the one given in \cite{giovanni_concur16}. 
We also employ our proof technique to prove both known and new derivations of critical cycles for other x-specifications.

\section{Characterisation of Simple Consistency Models}
\label{sec:completeness}

We now turn our attention to the \emph{Strong Correspondence 
Problem} presented in \S \ref{sec:laws}. Given a x-specification 
$\Sigma = \{(\rho_1, \pi_1), \cdots, (\rho_n, \pi_n)\}$ and 
a dependency graph $\G$, we want to find a sufficient and necessary 
condition for determining whether 
$\G = \graphof(\aexec)$ for some $\aexec \in \aeset(\Sigma)$. 

In this section we propose a proof technique for solving the strong correspondence problem. 
This technique applies to a particular class of x-specifications, which 
we call \emph{simple} x-specifications.
This class includes several of the consistency models we have presented.
\vspace{0.05\baselineskip}

\myparagraph{Characterisation of Simple x-specifications}
Recall that for each $x \in \Obj$,
the function $\rho_x$ of an abstract execution $\aexec$ is defined as $\rho_x(\_) = \circled{\WTr_x}$, 
and the associated axiom is $\circled{\WTr_x} \rcomp \AR_{\aexec} \rcomp \circled{\WTr_x} \subseteq \VIS_{\aexec}$.
\begin{figure}
{\small
\begin{empheq}[left=\empheqlbrace]{alignat*=6} 
\RF   \subseteq  X_V & \quad \Srf \quad \quad & 
X_V \rcomp X_V \subseteq X_V & \quad \SvisTrans \quad \quad & 
\bigcup_{\{x \mid (\rho_x, \rho_x) \in \Sigma\}} \VO(x) \subseteq X_V & \quad \Sconflict\\[5pt]
& &  &  & 
\rho(X_V) \rcomp X_A \rcomp \pi(X_V) \subseteq X_V & \quad \Saxiom\\[5.0pt]
\VO  \subseteq X_A & \quad \Svo \quad \quad &
X_V  \subseteq X_A & \quad \Svis \quad \quad &
\bigcup_{x \in \Obj} \left(\circled{\WTr_x} \rcomp X_V \rcomp \AD(x) \right) \subseteq X_A & \quad \Sext \\[5pt]
& & X_A \rcomp X_A \subseteq X_A & \quad \SarTrans \quad \quad &
\left( \pi(X_V) \rcomp X_N \rcomp \rho(X_V) \right) \setminus \Id  \subseteq X_A & \quad \Scoaxiom\\[5.0pt]
\AD \subseteq X_N & \quad \Sad \quad \quad & 
X_{V} \rcomp X_N \subseteq X_N & \quad \SavisL \quad \quad & 
X_N \rcomp X_V \subseteq X_N & \quad \SavisR
\end{empheq}
}
\vspace{-10pt}
\caption{The system of inequalities $\System_{\Sigma}(\G)$ for the simple consistency model $\Sigma$ 
and the dependency graph $\G = (\T, \RF, \VO, \AD)$.}
\label{fig:system}
\end{figure}
\begin{definition}
A x-specification $\Sigma$ is \emph{simple} 
if there exists a consistency guarantee $(\rho, \pi)$ such that 
$\Sigma \subseteq \{(\rho, \pi)\} \cup \{(\rho_{x}, \rho_{x})\}_{x \in \Obj}$. 
\end{definition}
That is, a simple x-specification $\Sigma$ 
contains at most one consistency guarantee, beside those 
of the form $(\rho_x, \rho_x)$ which express the write-conflict detection 
for some object $x \in \Obj$. 
Among the x-specifications 
that we have presented in this paper, the only non-simple one is $\Sigma_{\SI+\SER}$.

For simple x-specifications, it is possible to solve the strong correspondence problem. 
Fix a simple x-specification $\Sigma \subseteq \{(\rho, \pi)\} \cup \{(\rho_x, \rho_x) \mid x \in \Obj\}$ 
and a dependency graph $\G$. We define a system 
of inequalities $\System_{\Sigma}(\G)$ in three unknowns $X_V, X_A$ and $X_N$, and 
depicted in Figure \ref{fig:system} (the inequalities \Saxiom\; and \Scoaxiom\; are included 
in the system if and only if $(\rho, \pi) \in \Sigma$). These unknowns correspond to subsets of the visibility, arbitration 
and anti-visibility relations of the abstract execution $\aexec \in \aeset(\Sigma)$, with underlying 
dependency graph $\G$, that we wish to find. 
Note that each one of the inequalities of $\System_{\Sigma}(\G)$, with the exception of $\Sconflict$, 
follows the structure of one of the algebraic laws from Figure \ref{fig:laws}. 
We prove that, in order to ensure that the abstract execution $\aexec$ exists, it is 
sufficient to find a solution of $\System_{\Sigma}(\G)$ whose $X_A$-component is acyclic. 
In particular, this is true if and only if the $X_A$-component of the smallest solution\footnote{
A solution $(X_V = \VIS, X_A = \AR, X_N = \mathsf{AntiVIS})$ is smaller than another one 
$(X_V = \VIS', X_A = \AR', X_N = \mathsf{AntiVIS}')$ iff  
$\VIS \subseteq \VIS', \AR \subseteq \AR'$ and $\mathsf{AntiVIS} \subseteq \mathsf{AntiVIS}'$.} of $\System_{\Sigma}(\G)$ 
is acyclic.
\begin{theorem}
\label{thm:completeness}~
\begin{description}
\item[\textbf{Soundness: }] for any $\aexec \in \aeset(\Sigma)$ such that $\graphof(\aexec) = \G$, 
the triple $(X_V = \VIS_{\aexec}, X_A = \AR_{\aexec}, X_N = \overline{\VIS^{-1}_{\aexec}})$ is a 
solution of $\System_{\Sigma}(\G)$, 
\item[\textbf{Completeness: }]
Let $(X_V = \VIS_0, X_A = \AR_0, X_N = \mathsf{AntiVIS}_0)$ be the smallest solution of $\System_{\Sigma}(\G)$
. 
If $\AR_0$ is acyclic, then there exists an abstract 
execution $\aexec$ such that $\aexec \in \aeset(\Sigma)$ and $\graphof(\aexec) = \G$. \qed
\end{description}

\end{theorem}
Note that the relation $\AR_0$ need not to be total in the completeness direction of Theorem \ref{thm:completeness}.

Before discussing the proof of Theorem \ref{thm:completeness}, 
we show how it can be used to prove the equivalence of a x-specification 
and a g-specification. We give a proof of Theorem \ref{thm:depgraphs}\eqref{thm:psi}. 
Theorems \ref{thm:depgraphs}{\eqref{thm:ser}} and \ref{thm:depgraphs}{\eqref{thm:si}} 
can be proved similarly, and their proof is given in \tra{\ref{app:completeness}}{\ncompleteness}.

\noindent\myparagraph{Proof Sketch of Theorem \ref{thm:depgraphs}\eqref{thm:psi}} 
Recall that $\Delta_{\PSI} = \{\delta_{\PSI_0} \} 
\cup \{\delta_{\PSI(x)}(\G) \mid x \in \Obj\}$, where $\delta_{\PSI_0}(\G) = (\RF_{\G} \cup \VO_{\G})^{+}$, 
$\delta_{\PSI(x)}(\G) = ((\RF_{\G} \cup \VO_{\G})^{\ast} \rcomp \AD_{\G}(x))^{+}$. In \tra{\ref{app:completeness}}{\ncompleteness} 
we prove that
$\graphs(\Delta_{\PSI}) = \graphs(\{\delta_{\PSI}\})$, where 
\[
\delta_{\PSI}(\G) =  (\RF_{\G} \cup \VO_{\G})^{+} \cup \bigcup_{x \in \Obj}\left( \circled{\WTr_x} \rcomp (\RF_{\G} \cup \VO_{\G})^{\ast} \rcomp \AD_{\G}(x) \right)^{+}.
\]
Therefore, it suffices to prove that $\modelof(\Sigma_{\PSI}) = \modelof(\{\delta_{\PSI}\})$:
\begin{description}
\item[$\modelof(\Sigma_{\PSI}) \subseteq \modelof(\{\delta_{\PSI}\})$: ] given $\aexec \in \aeset(\Sigma_{\PSI})$, and let $\G := \graphof(\aexec)$,  
we need to show that $\delta_{\PSI}(\G) \cap \Id = \emptyset$. The proof follows the style of Theorems \ref{thm:ser.acyclic}\; and \ref{thm:ccser.acyclic}; 
details can be found in \tra{\ref{app:laws}}{\nproofs}, 
\item[$\modelof(\{\delta_{\PSI}\}) \subseteq \modelof(\Sigma_{\PSI})$: ] given $\G \in \graphs(\{\delta_{\PSI}\})$, let $\VIS_{\G} = (\RF \cup \VO)^{+}$; 
It is immediate to prove that the triple $(X_V = \VIS_{\G}, X_A = \delta_{\PSI}(\G), X_N = \VIS_{\G}? \rcomp \AD \rcomp \VIS_{\G}?)$ 
is a solution of $\System_{\Sigma_{\PSI}}(\G)$. Because $\delta_{\PSI}(\G)$ is acyclic, 
if we take the smallest solution $(X_V = \_, X_A = \AR_{\G}, X_N = \_ )$ of $\System_{\Sigma}(\G)$, then $\AR_\G \subseteq \delta_{\PSI}(\G)$, hence $\AR_\G$ is acyclic. 
By Theorem \ref{thm:completeness}, there exists an abstract execution $\aexec \in \aeset(\PSI)$ 
such that $\graphof(\aexec) = \G$, and in particular $\T_{\aexec} = \T_{\G}$. \qed
\end{description}

We now turn our attention to the proof of Theorem \ref{thm:completeness}. 
The proof of the soundness direction is straightforward.

\noindent\myparagraph{Proof of Theorem \ref{thm:completeness} (Soundness)} 
Let $\aexec \in \aeset(\Sigma)$, and define $\G := \graphof(\aexec)$. 
To show that the triple $(X_V = \VIS_{\aexec}, X_A = \AR_{\aexec}, X_N = 
\overline{\VIS^{-1}_{\aexec}})$ is a solution of $\System_{\Sigma}(\G)$, 
we need to show that all the inequalities from said system are satisfied, 
when the unknowns $X_A, X_V, X_N$ are replaced with $\VIS_{\aexec}, \AR_{\aexec}, \overline{\VIS^{-1}_{\aexec}}$, 
respectively. 
In practice, all the inequalities, with the exception of \Sconflict, follow from the algebraic laws of Figure \ref{fig:laws}. 
Let us prove that  $\Sconflict$ is also valid: for any $(\rho_x, \rho_x) \in \Sigma$  
we have that 
\[
\VO_{\aexec}(x) \stackrel{\depWWTx}{=} \circled{\WTr_x} \rcomp \VO_{\aexec}(x) \rcomp \circled{\WTr_x} 
\stackrel{\WWinAR}{\subseteq}
 \circled{\WTr_x} \rcomp \AR_{\aexec} \rcomp \circled{\WTr_x} \stackrel{\Axiom}{\subseteq} \VIS_{\aexec}.\qed
\]

The proof of the completeness direction of Theorem \ref{thm:completeness} 
is much less straightforward. 
Let  $(X_V = \VIS_0, X_A = \AR_0, X_N = \mathsf{AntiVIS}_0)$ 
be the smallest solution of $\System_{\Sigma}(\G)$. Assume that $\AR_0$ is acyclic.  
The challenge is that of constructing a valid abstract execution $\aexec$, i.e. whose 
arbitration order is total, 
from the dependencies in $\G$, that is included in $\aeset(\Sigma)$. We 
do this incrementally: at intermediate stages of the construction 
we get structures similar to abstract executions, but where the arbitration 
order can be partial.

\begin{definition}
\label{def:pe}
A \emph{pre-execution} $\pe = (\T_{\G}, \VIS, \AR)$ is a tuple that satisfies 
all the constraints of abstract executions, except that $\AR$ is not necessarily
total, although $\AR$ is still required to be total over the set $\WTr_x$ for 
every object $x$.
\end{definition}

The notation adopted for abstract executions naturally extends to pre-executions; 
also, for any pre-execution $\pe$, $\graphof(\pe)$ is a well-defined dependency graph. 
Given a x-specification $\Sigma$, we let $\peset(\Sigma)$ be the set of all valid pre-executions 
that satisfy all the consistency guarantees in $\Sigma$.

$\System_{\Sigma}(\G)$ is defined so that all of its solutions whose $X_A$-component 
is acyclic induce a valid pre-execution in $\peset(\Sigma)$ with underlying dependency graph $\G$.

\begin{proposition}
\label{prop:pexec}
Let $(X_V = \VIS', X_A = \AR', X_{N} = \mathsf{AntiVIS}')$ 
be a solution to $\System_{\Sigma}(\G)$. If $\AR' \cap \Id = \emptyset$, 
then $\pe = (\T_{\G}, \VIS', \AR') \in \peset(\Sigma)$; 
moreover, $\graphof(\pe) = \G$. 
\end{proposition}

\noindent\myparagraph{Proof Sketch} 
The inequalities \Svo, \Svis\; and \SarTrans\, 
together with the assumption that $\AR_0$ is acyclic, ensure that $\pe$ is a pre-execution. 
In particular, \Svo\; ensures that $\AR_0$ is a total relation over 
the set $\WTr_x$, for any $x \in \Obj$.
As we explain in \tra{\ref{app:completeness}}{\ncompleteness}, 
the inequalities \Srf, \Svo\; and \Sext\; enforce the Last Write Wins policy 
(Definition \ref{def:constraints}). 
The inequality \SvisTrans\; mandates that $\pe$ respects causality. 
Finally, the inequalities \Sconflict\; and \Saxiom\; ensure that all 
the consistency guarantees in $\Sigma$ are satisfied by $\pe$. \qed

In particular, the smallest solution $(X_V = \VIS_0, X_A = \AR_0, X_N = \mathsf{AntiVIS}_0)$ 
of \linebreak $\System_{\Sigma}(\G)$ 
induces the pre-execution $(\T_{\G}, \VIS_{0}, \AR_{0}) \in \peset(\Sigma)$. 

To construct an abstract execution $\aexec \in \aeset(\Sigma)$, with 
$\graphof(\aexec) = \G$, we define a finite chain of pre-executions $\{\pe_i,\}_{i=0}^{n}$, 
$n \geq 0$,  as follows: \textbf{(i)} let $\pe_{0} := (\hist_{\G}, \VIS_0, \AR_0)$; \textbf{(ii)} 
given $\pe_i$, $i \geq 0$, choose two different transactions 
$T_i, S_i \in \hist_{\G}$ (if any) that are not related by $\AR_{i}$, 
compute the smallest solution $(X_V = \VIS_{i+1}, X_A = \AR_{i+1}, X_N = \_)$ 
such that $\AR_{i+1} \supseteq \AR_i \cup \{(T_i, S_i)\}$, and let $\pe_{i+1} 
:= (\hist_{\G}, \VIS_{i+1}, \AR_{i+1})$; 
\textbf{(iii)} if the transactions $T_i, S_i \in \hist_{\G}$ from the previous step do not exist, 
then let $n := i$ and terminate the construction. 
Because we are assuming that $\hist_{\G}$ is finite, the construction of $\{\pe_0,\cdots, \pe_n\}$ 
always terminates.

To prove the completeness direction of Theorem \ref{thm:completeness}, we show that 
all of the pre-executions $\{\pe_0, \cdots, \pe_n\}$ in the construction outlined 
above are included in $\peset(\Sigma)$; then, because in $\pe_n = (\hist_{\G}, \VIS_n, 
\AR_{n})$ all transactions are related by $\AR_{n}$, we may conclude that $\AR_n$ is 
total, and $\pe_n \in \aeset(\Sigma)$. According to Proposition \ref{prop:pexec}, it 
suffices to show that each of the relations $\AR_i, i=0,\cdots,n$ is acyclic.
However, this is not completely trivial, because of how $\AR_{i+1}$ is defined: 
adding one edge  $(T_i, S_i)$ in $\AR_{i+1}$ may cause more edges to be 
included in $\VIS_{i+1}$, due to the inequality \Saxiom. This in turn leads 
to including more edges in $\AR_{i+1}$, thus augmenting the risk of having a 
cycle in $\AR_{i+1}$. 

In practice, the definition of $\System_{\Sigma}(\G)$ ensures that 
this scenario does not occur. 

\begin{proposition}
\label{prop:incremental}
For $i=0,\cdots, n-1$, let $\Delta\AR_i := \AR_i? \rcomp \{(T_i, S_i)\} \rcomp \AR_n?$. 
Then $\AR_{i+1} = \AR_i \cup \Delta \AR_{i}$. 
\end{proposition}
\begin{corollary}
\label{cor:incremental}
For $i=0,\cdots, n-1$, if $\AR_i \cap \Id = \emptyset$, then $\AR_{i+1} \cap \Id = \emptyset$.
\end{corollary}

\noindent\myparagraph{Proof}
Because $\AR_i \cap \Id = \emptyset$ by hypothesis, 
by Proposition \ref{prop:incremental} we only need to show that 
$\Delta \AR_i \cap \Id = \emptyset$. If $(T,T) \in \Delta \AR_i$ 
for some $T \in \T_\G$, then it must be $T \xrightarrow{\AR_i?} T_i$ 
and $S_i \xrightarrow{\AR_i?} T$. It follows that 
$S_i \xrightarrow{\AR_i?} T_i$. But this contradicts the 
hypothesis that $\AR_i$ does not relate transactions $T_i$ and $S_i$. 
Therefore, $(T, T) \not\in \Delta \AR_i$ 
for any $T \in \T_{\G}$, i.e. $\Delta \AR_i \cap \Id = \emptyset$.
\qed

We have now everything in place to prove Theorem \ref{thm:completeness}.

\noindent\myparagraph{Proof of Theorem \ref{thm:completeness} (Completeness)}
Let $\G$ be a dependency graph, and 
define the chain of pre-executions $\pe_{0} = (\T_{\G}, \VIS_0, \AR_0), 
\cdots, \pe_{n} = (\T_{\G}, \VIS_{n}, \AR_{n})$ as described above. 
We show that for any $i=0,\cdots,n$, $\pe_i \in \peset(\Sigma)$, 
and $\graphof(\pe_i) = \G$. Because $\AR_{n}$ is a total order, this implies that $\pe_n \in \aeset(\Sigma)$, 
and $\graphof(\pe_n) = \G$, as we wanted to prove. The proof is by induction on $n$.
\begin{description}
\item[Case $i = 0$: ] observe that the triple $(X_V = \VIS_0, X_A = \AR_0, X_N = \_)$ corresponds to 
the smallest solution of $\System_{\Sigma}(\G)$, hence $\AR_0$ is acyclic 
by hypothesis. It follows from Proposition \ref{prop:pexec}\; that 
$\pe_0 \in \peset(\Sigma)$, and $\graphof(\pe_0) = \G$,
\item[Case $i > 0$: ] assume that $i \leq n$; then $i - 1 < n$, and by 
induction hypothesis $\pe_{i-1} \in \peset(\Sigma)$. In particular, the relation 
$\AR_{i-1}$ is acyclic; by Corollary \ref{cor:incremental} we obtain 
that $\AR_i$ is acyclic. Finally, recall that the triple $(X_V = \VIS_i, X_A = \AR_i,  X_N = \_)$ 
is a solution of $\System_{\Sigma}(\G)$ by construction. It follows from 
Proposition \ref{prop:pexec} that $\pe_i \in \peset(\Sigma)$, and $\graphof(\pe_i) = \G$. \qed
\end{description}

\vspace{-5pt}
\section{Conclusion}
\label{sec:conclusion}

We have explored the connection between two different styles 
of specifications for weak consistency models at an algebraic 
level. We have proposed several laws which we applied to 
devise several robustness criteria for consistency models. 
To the best of our knowledge, this is the first generic proof 
technique for proving robustness criteria of weak consistency models.
We have shown that, for a particular class of consistency models, 
our algebraic approach leads to a precise characterisation of consistency models 
in terms of dependency graphs. 

\myparagraph{Related Work}
Abstract executions have been introduced by Burckhardt in \cite{ev_transactions} to 
model the behaviour of eventually consistent data-stores; 
They  have been used to capture the behaviour of 
replicated data types \cite[Gotsman et al.,][]{repldatatypes},  
geo-replicated databases \cite[Cerone et al.,][]{framework-concur} and 
non-transactional distributed storage systems \cite[Viotti et al.,][]{viotti}.

Dependency graphs have been introduced by Adya \cite{adya};
they have been used since to reason about programs running 
under weak consistency models.
Bernardi et al., used dependency graphs to derive robustness criteria 
of several consistency models \cite{giovanni_concur16}, including PSI and red-blue; in contrast with 
our work, the proofs there contained do not rely on 
a general technique.  
Brutschy et al. generalised the notion of dependency 
graphs to replicated data types, and proposed a robustness criterion 
for eventual consistency \cite{vechev_popl17}. 

Weak consistency also arises in the context of shared memory systems \cite{cat}. 
Alglave et al., proposed the CAT language for specifying weak memory models in \cite{cat}, 
which also specifies weak memory models as a set of irreflexive relations over data-dependencies 
of executions. Castellan \cite{castellan2016weak}, and Jeffrey et al. \cite{james_lics16}, 
proposed different formalisations of weak memory models via event structures. 
The problem of checking the robustness of applications has also been 
addressed for weak memory models \cite{jade_stability,TSOrobust,jade_fenceinsertion}.

The strong correspondence problem (\S \ref{sec:completeness}) is 
also highlighted by Bouajjani et al. in \cite{bouajjani_popl17}: there the authors 
emphasize the need for general techniques to identify all the \emph{bad patterns} 
that can arise in dependency-graphs like structures. We solved the 
strong correspondence problem for SI in \cite{SIanalysis}.

\bibliographystyle{abbrv}
\bibliography{bibliography2}

\iflong

\appendix

\clearpage

\makeatletter
\renewcommand{\@evenhead}{\large\sffamily\bfseries
                   \llap{\hbox to0.5\oddsidemargin{\ifx\@ArticleNo\@empty\textcolor{blue}{XX}\else\@ArticleNo\fi:\thepage\hss}}APPENDIX\hfil}
\renewcommand{\@oddhead}{\large\sffamily\bfseries APPENDIX\hfil
                  \rlap{\hbox to0.5\oddsidemargin{\hss\ifx\@ArticleNo\@empty\textcolor{blue}{XX}\else\@ArticleNo\fi:\thepage}}}              
\makeatother

\newcommand\Item[1][]{%
  \ifx\relax#1\relax  \item \else \item[#1] \fi
  \abovedisplayskip=0pt\abovedisplayshortskip=0pt~\vspace*{-\baselineskip}}

\section{Exampes of Anomalies}
\label{app:anomalies}
We give examples of several anomalies: for each 
of them we list those consistency models, among those 
considered in the paper,  that allow the anomaly, 
and those that forbid it. For the sake of clarity, 
we have removed from the pictures below a 
transaction writing the initial value $0$ to relevant objects, 
and visible to all other transactions. Also, unnecessary 
visibility and arbitration edges are omitted from figures.

\begin{description}
\item[Fractured Reads: ]
Transaction $T_2$ reads only one of the updates performed by transaction $T_1$: 
\begin{itemize}
\item \textbf{Allowed by: } No consistency model enjoying atomic visibility allows this anomaly.
\end{itemize}
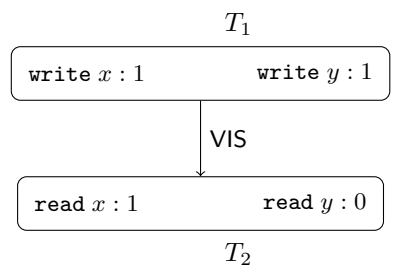
\begin{center}
\begin{tikzpicture}[every node/.style={transform shape}, font=\small]
\node (Rx0) {$\WR\; x: 1$};
\path(Rx0.center) + (3,0) node (Wy1) {$\WR\; y: 1$};
\path(Rx0.south) + (0, -1.5) node (Ry0) {$\RD\; x: 1$};
\path (Ry0.center) + (3, 0) node (Wx1) {$\RD\; y: 0$};
%
%
\begin{pgfonlayer}{background}
\node(t1) [background, fit=(Rx0) (Wy1), inner sep=0.1cm] {};
\node(t2) [background, fit=(Ry0) (Wx1), inner sep=0.1cm] {};
\path(t1.north) + (0.5,0.3) node[font=\normalsize] (T1) {$T_1$};
\path(t2.south) + (0.5,-0.3) node[font=\normalsize] (T2) {$T_2$};
\path[->] 
  (t1.south) edge node[right] {$\VIS$} (t2.north);
\end{pgfonlayer}
\end{tikzpicture}
\end{center}

\item[Violation of Causality: ]
The update of transaction $T_2$ to object $y$ depends on the value of $x$ 
written by another transaction $T_1$. For example, $T_2$ can be generated by 
the code $\mathtt{if } (x = 1) \;\mathtt{ then } \; y := 1;$. A third transaction 
$T_3$ observes the update to $y$, but not the one to $x$.
\begin{itemize}
\item \textbf{Allowed by: } None of the models discussed in the paper. 
However, some other  consistency models such as \textbf{Read Atomic} 
\cite{ramp} allow this anomaly. 
\end{itemize}
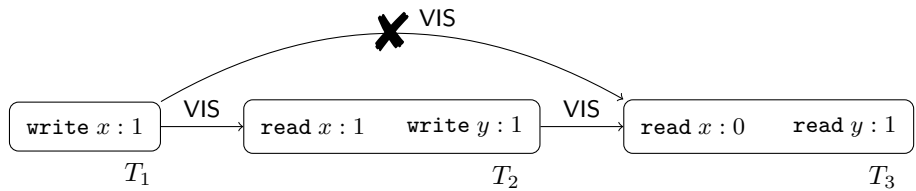
\begin{center}
\begin{tikzpicture}[every node/.style={transform shape}, font=\small]
\node (Wx1) {$\WR\; x: 1$};
\path(Wx1.center) + (3,0) node (Rx1) {$\RD\; x: 1$};
\path(Rx1.center) + (2,0) node (Wy1) {$\WR\; y: 1$};
\path (Wy1.center) + (3, 0) node (Rx0) {$\RD\; x: 0$};
\path (Rx0.center) + (2,0) node (Ry1) {$\RD\; y: 1$};
%
%
\begin{pgfonlayer}{background}
\node(t1) [background, fit=(Wx1), inner sep=0.1cm] {};
\node(t2) [background, fit=(Rx1) (Wy1), inner sep=0.1cm] {};
\node(t3) [background, fit=(Rx0) (Ry1), inner sep=0.1cm] {};
\path(t1.south) + (0.7,-0.3) node[font=\normalsize] (T1) {$T_1$};
\path(t2.south) + (1.5,-0.3) node[font=\normalsize] (T2) {$T_2$};
\path(t3.south) + (1.5,-0.3) node[font=\normalsize] (T3) {$T_3$};
\path(t2.north) + (0,0.9) node[font=\huge] (cross) {\cross};
\path[->] 
  (t1.east) edge node[above] {$\VIS$} (t2.west)
  (t2.east) edge node[above] {$\VIS$} (t3.west)
  (t1.north east) edge[bend left=30] node[above,pos=0.6] {$\VIS$} (t3.north west);
\end{pgfonlayer}
\end{tikzpicture}
\end{center}

\item[Lost Update: ] This is the abstract Execution depicted in Figure 
\ref{fig:example}, which we draw again below. Two transactions $T_1, T_2$ 
concurrently update the same object, after reading the initial value for it. 
\begin{itemize}
\item \textbf{Allowed by: } Causal Consistency, Red-blue Consistency,
\item \textbf{Forbidden by: } Parallel Snapshot Isolation, Snapshot Isolation, Serialisability.
\end{itemize}
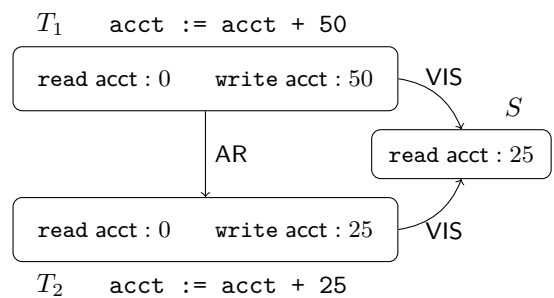
\begin{center}
\begin{tikzpicture}[every node/.style={transform shape}, font=\small]
\node(Ra1) {$\RD\; \account: 0$};
\path (Ra1.center) + (2.5, 0) node (Wa1) {$\WR\; \account: 50$};
\path (Ra1.center) + (0,-2.0) node (Ra2) {$\RD\; \account: 0$};
\path (Ra2.center) + (2.5, 0) node (Wa2) {$\WR\; \account: 25$};
\path (Wa1.center) + (2.2,-1.0) node (Rf) {$\RD\; \account: 25$};
%
%
\begin{pgfonlayer}{background}
\node(t1) [background, fit=(Ra1) (Wa1), inner sep=0.2cm] {};
\node(t2) [background, fit=(Ra2) (Wa2), inner sep=0.2cm] {};
\node(t3) [background, fit=(Rf), inner sep=0.1cm] {};
\path (t3.west) + (0.2,0.1) node (h1) {};
\path (t3.west) + (0.2,-0.1) node (h2) {};
\path(t1.south) + (-1,0) node (a12s) {};
\path(t2.north) + (-1,0) node (a12e) {};
\path(t1.south) + (1,0) node (a21e) {};
\path(t2.north) + (1,0) node (a21s) {};
\path(t2.south east) + (-0.2,0) node (wr23s){};
\path(t3.south) + (1,0) node (wr23e) {};
\path[->] 
  (t1.south) edge node[right] {$\AR$} (t2.north)
  (t1.east) edge[bend left=30] node[above] {\;\;$\VIS$} (t3.north)
  (t2.east) edge[bend right=30] node[below] {\;\;$\VIS$} (t3.south);
\path (t1.north) + (0.3,0.3) node[font=\normalsize] (code1) {\texttt{acct := acct + 50}};
\path (t2.south) + (0.3,-0.3) node[font=\normalsize] (code2) {\texttt{acct := acct + 25}};
\path(t1.north west) + (0.5,0.3) node[font=\normalsize] (T1) {$T_1$};
\path(t2.south west) + (0.5,-0.3) node[font=\normalsize] (T2) {$T_2$};
\path(t3.north east) + (-0.5,0.3) node[font=\normalsize] (T3) {$S$};
\end{pgfonlayer}
\end{tikzpicture}
\vspace{-12pt}
\end{center}

\item[Serialisable Lost Update: ] This execution is the same as the one above, but 
the two transactions $T_1, T_2$ are marked as serialisable. In the figure below, 
transactions marked as serialisable are depicted using a box with double borders. Because Causal Consistency does not distinguish 
between transactions marked as serialisable from those that are not marked as such, it allows the serialisable 
lost update. However, this anomaly is forbidden by Red-blue Consistency.
\begin{itemize}
\item \textbf{Allowed by: } Causal Consistency, 
\item \textbf{Forbidden by: } Red-blue Consistency, Parallel Snapshot Isolation, Snapshot Isolation, Serialisability.
\end{itemize}
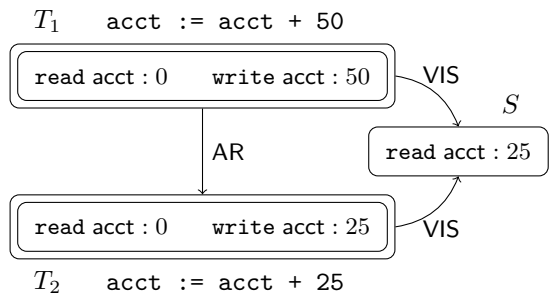
\begin{center}
\begin{tikzpicture}[every node/.style={transform shape}, font=\small]
\node(Ra1) {$\RD\; \account: 0$};
\path (Ra1.center) + (2.5, 0) node (Wa1) {$\WR\; \account: 50$};
\path (Ra1.center) + (0,-2.0) node (Ra2) {$\RD\; \account: 0$};
\path (Ra2.center) + (2.5, 0) node (Wa2) {$\WR\; \account: 25$};
\path (Wa1.center) + (2.2,-1.0) node (Rf) {$\RD\; \account: 25$};
%
%
\begin{pgfonlayer}{background}
\node(t1in) [background, fit=(Ra1) (Wa1), inner sep=0.1cm] {};
\node(t1) [background, fit=(Ra1) (Wa1), inner sep=0.2cm] {};
\node(t2in) [background, fit=(Ra2) (Wa2), inner sep=0.1cm] {};
\node(t2) [background, fit=(Ra2) (Wa2), inner sep=0.2cm] {};
\node(t3) [background, fit=(Rf), inner sep=0.1cm] {};
\path (t3.west) + (0.2,0.1) node (h1) {};
\path (t3.west) + (0.2,-0.1) node (h2) {};
\path(t1.south) + (-1,0) node (a12s) {};
\path(t2.north) + (-1,0) node (a12e) {};
\path(t1.south) + (1,0) node (a21e) {};
\path(t2.north) + (1,0) node (a21s) {};
\path(t2.south east) + (-0.2,0) node (wr23s){};
\path(t3.south) + (1,0) node (wr23e) {};
\path[->] 
  (t1.south) edge node[right] {$\AR$} (t2.north)
  (t1.east) edge[bend left=30] node[above] {\;\;$\VIS$} (t3.north)
  (t2.east) edge[bend right=30] node[below] {\;\;$\VIS$} (t3.south);
\path (t1.north) + (0.3,0.3) node[font=\normalsize] (code1) {\texttt{acct := acct + 50}};
\path (t2.south) + (0.3,-0.3) node[font=\normalsize] (code2) {\texttt{acct := acct + 25}};
\path(t1.north west) + (0.5,0.3) node[font=\normalsize] (T1) {$T_1$};
\path(t2.south west) + (0.5,-0.3) node[font=\normalsize] (T2) {$T_2$};
\path(t3.north east) + (-0.5,0.3) node[font=\normalsize] (T3) {$S$};
\end{pgfonlayer}
\end{tikzpicture}
\end{center}
\item[Long Fork: ] Two transactions $T_1$, $T_2$ 
write to different objects: two other transactions $T_3, T_4$ 
only observe the updates of $T_1, T_2$, respectively: 
\begin{itemize}
\item \textbf{Allowed by:} Causal Consistency, Red-blue Consistency, Parallel Snapshot Isolation,  
\item \textbf{Forbidden by: } Snapshot Isolation, Serialisability.
\end{itemize}
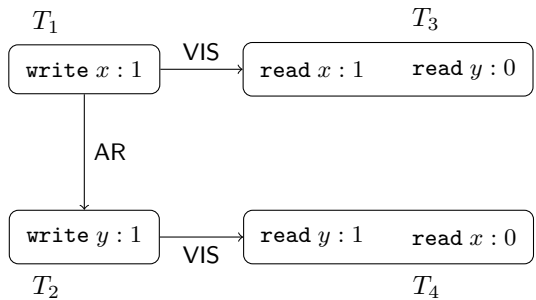
\begin{center}
\begin{tikzpicture}[every node/.style={transform shape}, font=\small]
\node(Wx1) {$\WR \; x: 1$};
\path (Wx1.center) + (3, 0) node (Rx1) {$\RD\; x: 1$};
\path(Rx1.center) + (2,0) node (Ry0) {$\RD\; y: 0$};
\path(Wx1.south) + (0, -2) node (Wy1) {$\WR\; y: 1$};
\path (Wy1.center) + (3, 0) node (Ry1) {$\RD\; y: 1$};
\path(Ry1.center) + (2,0) node (Rx0) {$\RD\; x: 0$};
%
%
\begin{pgfonlayer}{background}
\node(t1) [background, fit=(Wx1), inner sep=0.1cm] {};
\node(t2) [background, fit=(Wy1), inner sep=0.1cm] {};
\node(t3) [background, fit=(Rx1) (Ry0), inner sep=0.1cm] {};
\node(t4) [background, fit=(Ry1) (Rx0), inner sep=0.1cm]{};
\path(t1.north) + (-0.5,0.3) node[font=\normalsize] (T1) {$T_1$};
\path(t2.south) + (-0.5,-0.3) node[font=\normalsize] (T2) {$T_2$};
\path(t3.north) + (0.5,0.3) node[font=\normalsize] (T3) {$T_3$};
\path(t4.south) + (0.5,-0.3) node[font=\normalsize] (T4) {$T_4$};
\path[->] 
  (t1.east) edge node[above] {$\VIS$} (t3.west)
  (t2.east) edge node[below] {$\VIS$} (t4.west)
  (t1.south) edge node[right] {$\AR$} (t2.north);
\end{pgfonlayer}
\end{tikzpicture}
\end{center}
\item[Long Fork with Serialisable Updates: ] This is the same 
as the long fork, but the transactions $T_1, T_2$ that 
write to objects $x,y$, respectively, are marked as serialisable. 
Because Parallel Snapshot Isolation does not take serialisable 
transactions into account, it allows this anomaly.
However, Red-blue Consistency distinguishes between serialisable and 
non-serialisable transactions, hence it does not allow 
it. 
\begin{itemize}
\item \textbf{Allowed by:} Causal Consistency, Parallel Snapshot Isolation,  
\item \textbf{Forbidden by: } Red-blue Consistency, Snapshot Isolation, Serialisability.
\end{itemize}
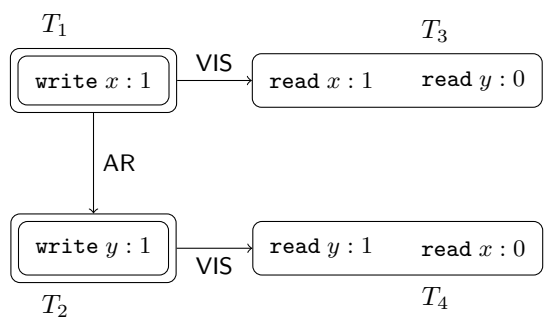
\begin{center}
\begin{tikzpicture}[every node/.style={transform shape}, font=\small]
\node(Wx1) {$\WR \; x: 1$};
\path (Wx1.center) + (3, 0) node (Rx1) {$\RD\; x: 1$};
\path(Rx1.center) + (2,0) node (Ry0) {$\RD\; y: 0$};
\path(Wx1.south) + (0, -2) node (Wy1) {$\WR\; y: 1$};
\path (Wy1.center) + (3, 0) node (Ry1) {$\RD\; y: 1$};
\path(Ry1.center) + (2,0) node (Rx0) {$\RD\; x: 0$};
%
%
\begin{pgfonlayer}{background}
\node(t11) [background, fit=(Wx1), inner sep=0.1cm] {};
\node(t22) [background, fit=(Wy1), inner sep=0.1cm] {};
\node(t3) [background, fit=(Rx1) (Ry0), inner sep=0.1cm] {};
\node(t4) [background, fit=(Ry1) (Rx0), inner sep=0.1cm]{};
\node(t1) [background, fit=(Wx1), inner sep=0.2cm] {};
\node(t2) [background, fit=(Wy1), inner sep=0.2cm] {};
\path(t1.north) + (-0.5,0.3) node[font=\normalsize] (T1) {$T_1$};
\path(t2.south) + (-0.5,-0.3) node[font=\normalsize] (T2) {$T_2$};
\path(t3.north) + (0.5,0.3) node[font=\normalsize] (T3) {$T_3$};
\path(t4.south) + (0.5,-0.3) node[font=\normalsize] (T4) {$T_4$};
\path[->] 
  (t1.east) edge node[above] {$\VIS$} (t3.west)
  (t2.east) edge node[below] {$\VIS$} (t4.west)
  (t1.south) edge node[right] {$\AR$} (t2.north);
\end{pgfonlayer}
\end{tikzpicture}
\end{center}
\textbf{Remark: } Note that Red-blue consistency forbids this anomaly, but allows the lost update anomaly 
from above. In contrast, Parallel Snapshot Isolation allows this anomaly, but forbids 
the lost-update anomaly. In other words, Red-blue Consistency and Parallel Snapshot Isolation 
are incomparable: $\aeset(\Sigma_{\CCSER}) \not\subseteq \aeset(\Sigma_{\PSI})$ 
and $\aeset(\Sigma_{\PSI}) \not\subseteq \aeset(\Sigma_{\CCSER})$.
\item[Write Skew: ]
Transactions $T_1, T_2$ read each the initial value of an object which is 
updated by the other.
\begin{itemize}
\item \textbf{Allowed by: } Causal Consistency, Red-blue Consistency, Parallel Snapshot  
Isolation, Snapshot Isolation,
\item \textbf{Forbidden by: } Serialisability.
\end{itemize}
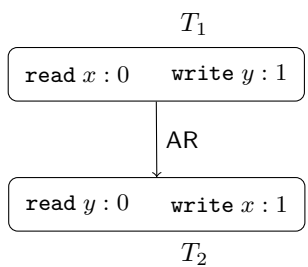
\begin{center}
\begin{tikzpicture}[every node/.style={transform shape}, font=\small]
\node (Rx0) {$\RD\; x: 0$};
\path(Rx0.center) + (2,0) node (Wy1) {$\WR\; y: 1$};
\path(Rx0.south) + (0, -1.5) node (Ry0) {$\RD\; y: 0$};
\path (Ry0.center) + (2, 0) node (Wx1) {$\WR\; x: 1$};
%
%
\begin{pgfonlayer}{background}
\node(t1) [background, fit=(Rx0) (Wy1), inner sep=0.1cm] {};
\node(t2) [background, fit=(Ry0) (Wx1), inner sep=0.1cm] {};
\path(t1.north) + (0.5,0.3) node[font=\normalsize] (T1) {$T_1$};
\path(t2.south) + (0.5,-0.3) node[font=\normalsize] (T2) {$T_2$};
\path[->] 
  (t1.south) edge node[right] {$\AR$} (t2.north);
\end{pgfonlayer}
\end{tikzpicture}
\end{center}

\end{description}

\section{Session Guarantees and Non-Causal Consistency Models}
\label{app:general}
We augment histories with sessions:  clients submit transactions 
within sessions, and the order in which they are submitted to the database is tracked 
by a \emph{session order}. We propose a variant of x-specifications that 
allows for specifying session guarantees, as well as causality guarantees 
that are weaker than causal consistency. 

\begin{definition}
Let $\T$ be a set of transactions, and let $\{\T_1, \T_2, \cdots, \T_n\}$ be 
a partition of $\T$. 
An \emph{extended history} is a pair $\mathcal{H} = (\T, \PO)$, where $\PO = \bigcup_{i =1}^{n} \PO_i$, 
and each $\PO_i$ is a strict, total order over $\T_i$.  Each of the sets $\T_i = 1,\cdots, n$ takes 
the name of \emph{session}, and we call $\PO$ the \emph{session order}.
\end{definition}
Given an extended history $\mathcal{H} = (\T, \PO)$, we let $\T_{\mathcal{H}} = \T$, and 
$\PO_{\mathcal{H}} = \PO$. 
If $(\T, \PO)$ is an extended history, and $(\T, \VIS, \AR)$ is an abstract execution, then we 
call $(\T, \PO, \VIS, \AR)$ an \emph{extended abstract execution}.
Specification functions can also be lifted to take extended abstract executions into account: 
an \emph{extended specification function} is a function $\rho: (\mathcal{H}, R) \mapsto R'$, 
such that for any extended history $\mathcal{H}$ and relation $R \subseteq \T_{\mathcal{H}} \times \T_{\mathcal{H}}$, 
$\rho(\mathcal{H}, R) = \rho(\mathcal{H}, \T_{\mathcal{H}} \times \T_{\mathcal{H}}) \cap R?$. 
An example of extended specification function is $\rho(\mathcal{H}, R) = R \setminus (\PO_{\mathcal{H}}?)$. 
An extended consistency guarantee is a pair $(\rho, \pi)$, where $\rho, \pi$ are extended specification functions.

\begin{definition}
A \emph{session guarantee} is a function $\sigma: 2^{\TrSet \times \TrSet} \rightarrow 2^{\TrSet \times \TrSet}$ 
such that, for any relation $R \subseteq \TrSet \times \TrSet$, $\sigma(R) \subseteq R?$.
A \emph{causality guarantee} is a pair $(\gamma, \beta)$, where $\gamma$ and $\beta$ are extended specification functions. 

An \emph{extended x-specification} of a consistency model is a triple $\Sigma = (\{\sigma_i\}_{i \in I}, 
\{(\gamma_j, \beta_j)\}_{j \in J},\allowbreak \{(\rho_k,~ \pi_k)\}_{k \in K})$, where $I, J, K$ are (possibly empty) 
index sets, for any $i \in I, j \in J$ and $k \in K$, $\sigma_i$ is a session guarantee, $(\gamma_j, \beta_j)$ 
is a causality guarantee, and $(\rho_k, \pi_k)$ is an extended consistency guarantee. 
\end{definition}

Note that the definition of causality and (extended) consistency guarantees are the same. However, 
they play a different role when defining the set of executions admitted by a consistency model. 

\begin{definition}
\label{def:extended.spec}
An extended abstract execution $\aexec = (\T ,\PO, \VIS, \AR)$ 
conforms to the extended specification $(\{\sigma_i\}_{i \in I}, \{(\gamma_j, \beta_j)\}_{j \in J},
\{(\rho_k, \pi_k)\}_{k \in K}$ iff 
\begin{enumerate}
\item for any $i \in I$, $\sigma_i(\PO) \subseteq \VIS$
\item for any $j \in J$, $\gamma_j(\mathcal{H}, \VIS)\rcomp \beta_j(\mathcal{H}, \VIS) \subseteq \VIS$, 
\item for any $k \in K$, $\rho_k(\mathcal{H}, \VIS) \rcomp \AR \rcomp \pi_k(\mathcal{H}, \VIS) \subseteq \VIS$.
\end{enumerate}
\end{definition}

Any x-specification can be lifted to an extended one: let 
$\gamma_{\CC}(\_, R)  = (R \setminus \Id)$\footnote{The 
difference with the identity relation 
is needed for $\gamma$ to satisfy the definition of specification function. However, we will always apply $\gamma$ to an 
irreflexive relation $R$, for which $\gamma(\_, R) = (R \setminus \Id) = R$.}.
Let also $\Sigma$ be any $x$-specification, and for any pair $(\rho, \pi) \in \Sigma$, define 
$\rho'(\_, R) = \rho(R)$, $\pi'(\_, R) = \pi(R)$. Then for any abstract $\aexec$, 
$\aexec \in \aeset(\Sigma)$ iff $\aexec$ conforms to the extended specification 
$(\emptyset, \{(\gamma_{\CC}, \gamma_{\CC})\}, \{(\rho', \pi') \mid (\rho, \pi) \in \Sigma\})$.

Dependency graphs can also be extended to take sessions into account. If 
$(\T, \PO)$ is a history, and $(\T, \RF, \VO, \AD)$ is a dependency graph, 
then $\G = (\T, \PO, \RF, \VO, \AD)$ is an \emph{extended dependency graph}. 
Given an extended abstract execution $\aexec = (\T, \PO, \VIS, \AR)$, we 
define $\graphof(\aexec) = (\T, \PO, \RF, \VO, \AD)$, where $(\T, \RF, \VO, \AD) = \graphof(\T, \VIS, \AR)$.
An extended abstract execution $\aexec = (\T, \PO, \VIS, \AR)$ with underlying extended dependency 
graph $\graphof(\aexec) = (\T, \PO, \RF, \VO, \AD)$ and conforming to the extended specification \\
$(\{\sigma_i\}_{i \in I}, \{(\gamma_j, \beta_j)\}_{j \in J}, \{(\rho_k, \pi_k)\}_{k \in K}$, 
satisfies all the Equations of Figure \ref{fig:laws}, exception made for equations, 
\AVISright\; and \AVISleft. Furthermore, sessions and causality guarantees induce 
novel inequalities, which are listed below: 
\begin{enumerate}
\item\label{eq:session} $\bigcup_{i \in I} \sigma_{i}(\PO) \subseteq \VIS$,
\item\label{eq:causal.antivisR} for any $j \in J$, $(\beta_j(\mathcal{H}, \VIS) \rcomp \overline{\VIS^{-1}}) \cap \gamma(\mathcal{H}, \T \times \T)^{-1} \subseteq \overline{\VIS^{-1}}$,
\item\label{eq:causal.antivisL} for any $j \in J$, $(\overline{\VIS^{-1}} \rcomp \gamma_j(\mathcal{H}, \VIS)) \cap \beta_j(\mathcal{H}, \T \times \T)^{-1} 
\subseteq \overline{\VIS^{-1}}$.
\end{enumerate}

Equation \eqref{eq:session} is obviously satisfied. To see why \eqref{eq:causal.antivisR} 
is satisfied by $\aexec$, let $j \in J$ and suppose that $T \xrightarrow{\beta_j(\mathcal{H}, \VIS)} V \xrightarrow{\overline{\VIS^{-1}}} S$, 
and $S \xrightarrow{\gamma_j(\mathcal{H}, \T \times \T)} T$. If it were $S \xrightarrow{\VIS} T$, then 
we would have a contradiction: because $\gamma_j$ is an extended specification function, 
$S \xrightarrow{\gamma_j(\mathcal{H}, \T \times \T)} T$ and $S \xrightarrow{\VIS} T$ imply that 
$S \xrightarrow{\gamma_j(\mathcal{H}, \VIS)} T$, and together with $T \xrightarrow{\beta_j(\mathcal{H}, \VIS)} V$ 
then we would have $S \xrightarrow{\VIS} V$, contradicting the assumption that $V \xrightarrow{\overline{\VIS^{-1}}} S$. 
Therefore it has to be $\neg (S \xrightarrow{\VIS} T)$, or equivalently $T \xrightarrow{\overline{\VIS^{-1}}} S$. 
Equation \eqref{eq:causal.antivisL} can be proved similarly.

\myparagraph{Examples of Session Guarantees} 
Below we give some examples of session guarantees, inspired by \cite{terry1994session}.

\noindent\textbf{Read Your Writes:} This guarantee states that when processing a transaction, 
a client must see previous writes in the same session. This can be easily expressed 
via the collection of consistency guarantees $\{\sigma_{\mathsf{RYW(x)}}\}_{x \in \Obj}$, 
where for  each object $x$, $\sigma_{\mathsf{RYW(x)}}(R) = \circled{\WTr_x} \rcomp R \rcomp \circled{\RTr_x}$. 
An extended abstract execution $\aexec = (\T, \PO, \VIS, \AR)$ satisfies this session guarantee 
if $\bigcup_{x \in \Obj} \circled{\WTr_x} \rcomp \PO \rcomp \circled{\RTr_x} \subseteq \VIS$, 

\noindent\textbf{Monotonic Writes:} This guarantee states that transactions writing 
at least to one object are processed in 
the same order in which the client requested them. It can be specified via the 
function $\sigma_{\mathsf{MW}}(R) = (\bigcup_{x \in \Obj} \circled{\WTr_x}) \rcomp R \rcomp 
(\bigcup_{x \in \Obj} \circled{\WTr_x})$. Any extended abstract execution $\aexec = (\T, \PO, \VIS, \AR)$ satisfies 
the monotonic writes guarantee, is such that $(\bigcup_{x \in \Obj} \circled{\WTr_x}) \rcomp \PO \rcomp
(\bigcup_{x \in \Obj} \circled{\WTr_x}) \subseteq \VIS$,

\noindent\textbf{Strong Session Guarantees:} This guarantee states that all transactions 
are processed by the database in the same order in which the client requested them. 
It can be specified via the function $\sigma_{\mathsf{SS}}(R) = R$; 
an extended abstract execution $(\T, \PO, \VIS, \AR)$ satisfies this guarantee if 
$\PO \subseteq \VIS$.

\myparagraph{Examples of Causality Guarantee: } 
We have already seen how to model causal consistency via 
the causality guarantee $(\gamma_{\CC}, \gamma_{\CC})$. 
Below we give an example of weak causality guarantee: 

\noindent\textbf{Per-object Causal Consistency:} this guarantee states that 
causality is preserved only among transactions accessing the same object. 
That is, let $\gamma_x(R) = (\circled{\WTr_x \cup \RTr_x} \rcomp R \rcomp \circled{\WTr_x \cup \RTr_x})
\setminus \Id$. The difference with the identity set is needed in order for $\gamma_x(R)$ to be a 
specification function. By definition, An extended abstract execution $\aexec = (\T, \PO, \VIS, \AR)$ that 
satisfies the per-object causal consistency guarantee, satisfies the inequality 
$\circled{\WTr_x \cup \RTr_x} \rcomp \VIS  \rcomp \circled{\WTr_x \cup \RTr_x} \rcomp \VIS 
\rcomp \circled{\WTr_x \cup \RTr_x} \subseteq \VIS$.

\section{Additional Proofs of Algebraic Laws and Robustness Criteria}
\label{app:laws}
Throughout this Section, we assume that $\aexec = (\T, \VIS, \AR)$ is 
a valid abstract execution, and  
$\graphof(\aexec) = (\T, \RF, \VO, \AD)$.

First, a result about specification functions, which was hinted at in 
the main paper: 
\begin{proposition} 
\label{prop:specfun.properties}
Let $\rho(\cdot)$ be a specification function. 
For all histories $\T$ and relations $R, R' \subseteq \T \times \T$,
\begin{enumerate}[(i)]
\item \label{prop:specfun.1} $\rho(R) \subseteq R?$; 
\item \label{prop:specfun.2} $\rho(\T \times \T) 
\cap R \subseteq \rho(R)$; 
\item \label{prop:specfun.3} $\rho(R) \cup \rho(R') = 
\rho(R \cup R')$.
\end{enumerate}
\end{proposition}

\noindent\myparagraph{Proof}
Recall that, by definition, if $\rho$ is a specification function, 
then $\rho(R) = \rho(\T \times \T) \cap R?$. It is immediate 
to observe then that \eqref{prop:specfun.1} $\rho(R) \subseteq R?$, and 
\eqref{prop:specfun.2} $\rho(\T \times \T) \cap R \subseteq \rho(\T \times \T) \cap R? = 
\rho(R)$. To prove \eqref{prop:specfun.3} note that 
\begin{gather*}
\rho(R) \cup \rho(R') = 
(\rho(\T \times \T) \cap R?) \cup (\rho(\T \times \T) \cap R'?) = 
\rho(\T \times \T) \cap (R? \cup R'?) = \\
\rho(\T \times \T) \cap (R \cup R')? = \rho(R \cup R') \tag*{\qed}
\end{gather*}

\subsection{Proof of the Algebraic Laws in Figure \ref{fig:laws}}

\begin{proposition}
All the (in)equalities of Figure \ref{fig:laws}\txlaws\; are satisfied.
\end{proposition}

\noindent\myparagraph{Proof}\hspace{4pt}
We prove each of the (in)equalities in Figure \ref{fig:laws}\txlaws\; individually. Throughout 
the proof, we let $\T', \T_1, \T_2 \subseteq \T$, and $R_1, R_2 \subseteq \T \times \T$
\begin{itemize}
\item \txId: by Definition, $\circled{\T'} = \{(T,T) \mid T \in \T'\} 
\subseteq \Id_{\T}$, 
\item \txComp: note that we can rewrite $\circled{\T_i} = \{(T,S) \mid T \in \T_1 \wedge S \in \T_1 \wedge T = S\}$, where $i=1,2$; then 
\begin{multline*}
  \circled{\T_1} \rcomp \circled{\T_2} = \{(T,S) \mid \exists V.\, (T,V) \in \circled{\T_1} \wedge (V,S) \in \circled{\T_2} \} = \\
   \{(T,S) \mid \exists V.\, T \in \T_1 \wedge V \in \T_1 \wedge T = V \wedge S \in \T_2 \wedge V \in \T_2 \wedge V = S\} = \\
   \{(T,S) \mid T \in \T_1 \wedge S \in \T_1 \wedge S = V \wedge S \in \T_2 \wedge T \in \T_2 \} =\\
   \{(T,S) \mid T \in (\T_1 \cap \T_2) \wedge S \in (\T_1 \cap \T_2) \wedge (S = T)\} = 
   \circled{\T_1 \cap \T_2}
\end{multline*}
%

\item \txDistrR: 
\begin{multline*}
(R_1 \rcomp \circled{\T'}) \cap R_2 = \{(T,S) \mid (\exists V.\, (T,V) \in R_1 \wedge V \in \T' \wedge V = S) \wedge (T,S) \in R_2\} =\\
\{(T,S) \mid (T,S) \in R_1 \cap R_2 \wedge S \in \T'\} = (R_1 \cap R_2) \rcomp \circled{\T'}
\end{multline*}

\item \txDistrL:
\begin{multline*}
(\circled{\T'} \rcomp R_1) \cap R_2 = \{(T,S) \mid = (\exists V.\, T = V \wedge T \in \T' \wedge (V,S) \in R_1) \wedge (T,S) \in R_2\} = \\
\{(T,S) \mid (T,S) \in R_1 \cap R_2 \wedge T \in \T'\} = \circled{\T'}\rcomp (R_1 \cap R_2)
\end{multline*}
\end{itemize}
\qed

\begin{proposition}
All the inequalities of Figure \ref{fig:laws}\deplaws\; are satisfied by $\aexec$.
\end{proposition}

\noindent\myparagraph{Proof} We only prove (in)equalities \depWRTx\; and \depWRIrrefl. The proof for the other 
(in)equalities is similar. 

Suppose that $T \xrightarrow{\RF(x)} S$. By Definition, $S \ni (\RD\; x:\_)$, hence 
$(S,S) \in \circled{\RTr_x}$. Also, $T \in \VIS^{-1}(S) \cap \WTr_x \subseteq \WTr_x$, 
from which $(T,T) \in \circled{\WTr_x}$ follows. Thus, $(T,S) \in \circled{\WTr_x} \rcomp 
\RF(x) \rcomp \circled{\RTr_x}$; this proves Equation \depWRTx.

To prove Equation \depWRIrrefl, first observe that because $T \xrightarrow{\RF(x)} S$, then 
$T \xrightarrow{\VIS} S$, and because $\VIS \subseteq \AR$ then also $T \xrightarrow{\AR} S$.
By definition of abstract 
execution, then $T \neq S$. Therefore, $\RF(x) \cap \Id = \emptyset$. Now we can rewrite 
\begin{gather*}
\RF(x) = (\RF(x) \cap (\Id \cup \overline{\Id})) = (\RF(x) \cap \Id) \cup \RF(x) \cap \overline{\Id} = \\
\emptyset \cup (\RF(x) \cap \overline{\Id}) = \RF(x) \cap \overline{\Id} = \RF(x) \setminus \Id. \tag*{\qed}
\end{gather*}

%

\begin{proposition}
\label{prop:deps.laws}
$\aexec$ satisfies inequalities \WRinVIS, \WWinAR\; 
and \LWW.
\end{proposition}
 
\noindent\myparagraph{Proof} 
The inequalities \WRinVIS\; and \WWinAR\; follow directly from the 
Definition of $\graphof(\aexec)$. 
It remains to prove the inequality \LWW. Let $T, S, T'$ be three transactions 
such that $T \ni (\WR\; x: \_)$, $T \xrightarrow{\VIS} S$ and $S \xrightarrow{\AD(x)} T'$; 
we need to show that $T \xrightarrow{\AR} T'$. Recall that, because $\aexec$ is an 
abstract execution, then the relation $\AR$ is total: either $T = T'$, $T' \xrightarrow{\AR} T$, 
or $T \xrightarrow{\AR} T'$. It is not possible that $T = T'$, because otherwise we would have 
$S \xrightarrow{\AD(x)} T$ and $T \xrightarrow{\VIS} S$ 
\setlength{\intextsep}{0pt}
\begin{mywrapfigure}[9]{r}{0.55\textwidth}
\begin{tikzpicture}[thick, scale=1.0, every node/.style={transform shape}, font=\normalsize]
\node(T) {$T \ni \WR\; x: \_ $};
\path(T.east) + (2,0) node (S) {$S \ni \RD\;x: \_$};
\path(S.north) + (0,1.3) node (T1) {$T' \ni \WR\;x: \_ $}; 
\path(S.south) + (0,-1.3) node (S1) {$S' \ni \WR\;x: \_$};
\path(S.north) + (-0.5,0) node (adStoTs) {};
\path(T.north) + (0.5,0) node (adStoTe) {};

\path[thick,->] 
   (T.east) edge[thick] node[above]{$\VIS$}  (S.west)
   (S.north) edge[thick] node[right] {$\AD(x)$} (T1.south)
   (S1.north) edge[thick] node[right] {$\RF(x)$} (S.south)
   (S1.north east) edge[thick, bend right=30] node[right] {$\VO(x)$} (T1.south east);
\path[thick,dashed, ->]
   (T1.west) edge[thick, bend right=30] node[left] {$\AR, \VO(x)$\;\;} (T.north)
   (S1.west) edge[thick, bend left=30] node[left] {$\VO(x)$\;\;} (T.south)
   (adStoTs.center) edge[thick, bend right=30] node[above] {$\AD(x)$} (adStoTe.center);
\end{tikzpicture}
\end{mywrapfigure}
\setlength{\intextsep}{10pt plus 6pt minus 3pt}
(equivalently, $\neg(S \xrightarrow{\overline{\VIS^{-1}}} T)$), 
contradicting the inequality \RWinAVIS.
It cannot be that 
$T' \xrightarrow{\AR} T$ either: in the picture 
to the right, we have given a 
graphical 
representation of this scenario, where dashed edges 
represent the consequences of having $T' \xrightarrow{\AR} T$.
In this case, $T \in \WTr_x$ by hypothesis; because $S \xrightarrow{\AD(x)} T'$, 
we also have that $T' \in \WTr_x$; because $T, T' \in \WTr_x$, and 
$T' \xrightarrow{\AR} T$, the definition of $\graphof(\aexec)$ implies that 
it has to be $T' \xrightarrow{\VO(x)} T$. 
Since $S \xrightarrow{\AD(x)} T'$, 
then $S' \xrightarrow{\RF(x)} S$, and $S' \xrightarrow{\VO(x)} T'$ for some $S'$; 
because $\VO(x)$ is transitive, then 
$S' \xrightarrow{\VO(x)} T$. We have proved that $S' \xrightarrow{\RF(x)} S$, 
and $S' \xrightarrow{\VO(x)} T$. By definition, it follows that $S \xrightarrow{\AD(x)} T$: 
together with the hypothesis $T \xrightarrow{\VIS} S$, we get a contradiction 
because the inequality \RWinAVIS\; is violated. We have proved that it 
cannot be $T = T'$, nor $T' \xrightarrow{\AR} T$. Therefore $T \xrightarrow{\AR} T'$, 
as we wanted to prove.
\qed

\begin{proposition}
\label{prop:avis.comp}
$\aexec$ satisfies inequalities \AVISright\; and \AVISleft.
\end{proposition}

\noindent\myparagraph{Proof}
We only prove the inequality \AVISright, as 
the inequality \AVISleft\; can be proved in a similar manner.
\setlength{\intextsep}{0pt}
\begin{mywrapfigure}[5]{r}{0.4\textwidth}
\begin{tikzpicture}[thick, scale=1.0, every node/.style={transform shape}, font=\normalsize]
\node(T) {$T$};
\path(T.east) + (2,0) node (V) {$V$};
\path(V.east) + (2,0) node (S) {$S$}; 
\path(V.east) + (1,0) node[font=\huge] (cross) {\cross};

\path[thick,->] 
   (T.east) edge[thick] node[above]{$\VIS$}  (V.west)
   (S.west) edge[thick] node[above, pos=0.75] {$\VIS$} (V.east);
\path[thick,dashed, ->]
   (S.north) edge[thick, bend right=15] node[above] {$\VIS$} (T.north)
   (S.south) edge[thick, bend left = 15] node[below] {$\VIS$} (V.south);
\end{tikzpicture}
\end{mywrapfigure}
\setlength{\intextsep}{10pt plus 6pt minus 3pt}
Suppose that $T \xrightarrow{\VIS} V \xrightarrow{\overline{\VIS^{-1}}} S$.
We prove that $\neg( S \xrightarrow{\VIS} T)$, or 
equivalently $(T \xrightarrow{\overline{\VIS^{-1}}} S)$, by contradiction. 
Let then $S \xrightarrow{\VIS} T$. Because $\aexec$ respects causality, 
$S \xrightarrow{\VIS} T \xrightarrow{\VIS} V$ implies that $S \xrightarrow{\VIS} V$. 
But $V \xrightarrow{\overline{\VIS^{-1}}} S$ by hypothesis, 
which causes the contradiction. A graphical representation of the 
proof is given to the right; here dashed edges are implied by the assumption that 
$S \xrightarrow{\VIS} T$. \qed

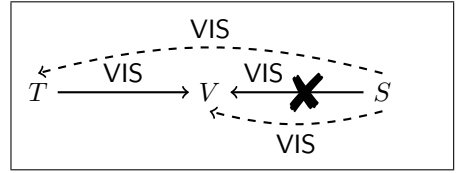
\begin{proposition}
$\aexec$ satisfies all the inequalities of Figure \ref{fig:laws}\allexeclaws.
\end{proposition}

\noindent\myparagraph{Proof} We have proved that $\aexec$ satisfies the inequalities \WRinVIS, \WWinAR\; 
and \LWW\; in Proposition \ref{prop:deps.laws}. The Proof of the inequality \RWinAVIS\; was given 
at Page \pageref{proof:RWinAVIS}. The inequalities \ARtrans, \VISinAR, 
and \ARirrefl\; are trivial consequences of the definition of abstract execution. The inequalities 
\VIStrans\; is satisfied because we are assuming that $\aexec$ respects causality. 
The inequality \VISnotAVIS\; is a trivial 
consequence of the fact that, for any relation $R \subseteq \T \times \T$, 
$\overline{R^{-1}} = \{(T,S) \mid (S,T) \notin R\} $; then 
\begin{multline*}
(R \rcomp \overline{R^{-1}}) \cap \Id = \{(T,T) \mid \exists S.\, (T,S) \in R \wedge 
(S,T) \in \overline{R^{-1}}\} = \\ 
\{(T,T) \mid \exists S.\, (T,S) \in R \wedge (T, S) \notin R \} = \emptyset
\end{multline*}
The inequality \AVISnotVIS\; can be proved similarly.
Finally, the inequalities \AVISright\; and \AVISleft\; are satisfied,  
as we have proved in Proposition \ref{prop:avis.comp}. \qed
%

\begin{proposition}
\label{prop:coaxiom.avis}
If $\aexec$ satisfies the consistency guarantee $(\rho, \pi)$, 
then it also satisfies the inequalities \CoAxiomL\; and \CoAxiomR.
\end{proposition}

\noindent\myparagraph{Proof}
We only prove the inequality \CoAxiomL. The proof for the inequaiton 
\CoAxiomR\; is similar.
Let $T, T', S', S \in \T$ be such that $T \xrightarrow{\AR} T'$, $T' \xrightarrow{\pi(\VIS)} S'$, 
$S' \xrightarrow{\overline{\VIS^{-1}}} S$, and $S \xrightarrow{\rho(\T \times \T)} T$. 
\setlength{\intextsep}{0pt}
\begin{mywrapfigure}[7]{r}{0.44\textwidth}
\begin{tikzpicture}[thick, scale=0.9, every node/.style={transform shape}, font=\normalsize]
\node(T) {$T$};
\path(T.east) + (1.5,0) node (T1) {$T'$};
\path(T1.east) + (2,0) node (S1) {$S'$}; 
\path(S1.east) + (2,0) node (S) {$S$};
\path(S1.east) + (1.2,0) node[font=\huge] (cross) {\cross};

\path[thick,->] 
   (T.east) edge[thick] node[above]{$\AR$}  (T1.west)
   (T1.east) edge[thick] node[above] {$\pi(\VIS)$} (S1.west)
   (S.west) edge[thick] node[above, pos=0.7] {$\VIS$} (S1.east)
   (S.south) edge[thick, bend left=15] node[below] {$\rho(\T \times \T)$} (T.south);
\path[thick,dashed, ->]
   (S.north) edge[thick, bend right=45] node[above] {$\rho(\T \times \T) \cap \VIS \subseteq \rho(\VIS)$} (T.north)
   (S.north west) edge[thick, bend right = 30] node[above, pos=0.8] {$\VIS$} (S1.north);
\end{tikzpicture}
\end{mywrapfigure}
\setlength{\intextsep}{10pt plus 6pt minus 3pt}
We need to prove that $T \xrightarrow{\overline{\VIS^{-1}}} S$, or equivalently that $\neg(S \xrightarrow{\VIS} T)$. 
The proof goes by contradiction: suppose that $S \xrightarrow{\VIS} T$. 
Then we 
have that $S \xrightarrow{\rho(\T \times \T) \cap \VIS} T$, and by 
Proposition \ref{prop:specfun.properties} it follows that $S \xrightarrow{\rho(\VIS)} T$. 
We have $S \xrightarrow{\rho(\VIS)} T \xrightarrow{\AR} T' \xrightarrow{\pi(\VIS)} S'$. 
Because $\aexec \in \aeset(\{\rho, \pi\})$, then $S \xrightarrow{\VIS} S'$ 
by Inequality \Axiom. But $S' \xrightarrow{\overline{\VIS^{-1}}} S$ by hypothesis, hence 
the contradiction. A graphical representation of the proof is given to the right: here dashed 
edges are implied by the assumption that $S \xrightarrow{\VIS} T$. 
\qed

\begin{proposition}
If $\aexec$ satisfies the consistency guarantee $(\rho, \pi)$, then it satisfies 
all the inequalities of Figure \ref{fig:laws}\cmexeclaws, relatively to said 
consistency guarantee. 
\end{proposition}

\noindent\myparagraph{Proof}
Because $\aexec$ satisfies the consistency guarantee $(\rho, \pi)$ by hypothesis, 
then it satisfies the inequality \Axiom. It also satisfies the inequality \CoAxiomAR,  
as we showed in \S \ref{sec:laws}. Finally, it satisfies inequalities \CoAxiomL\; and 
\CoAxiomR\; by Proposition \ref{prop:coaxiom.avis}.
\qed

\subsection{Additional Algebraic Laws}
Here we prove some additional algebraic laws that can 
be proved from the laws of Figure \ref{fig:laws}, 
and from the axioms of the Kleene Algebra and boolean algebra of set relations. 
In the following, we assume that $\aexec = (\T, \VIS, \AR)$ is an abstract 
execution, and $\graphof(\aexec) = (\T, \RF, \VO, \AD)$. 
Given two relations $R_1, R_2 \subseteq \T \times \T$, we recall 
that we use the notation $R_1 \stackrel{\textbf{(eq)}}{\subseteq} R_2$ ($R_1 \stackrel{\textbf{(eq)}}{=} R_2$) to denote 
the fact that $R_1 \subseteq R_2$ ($R_1 = R_2$) follows from the (in)equality \textbf{(eq)}. 
Sometimes we omit the complete sequence of steps needed to derive an inequality, when these 
can be easily inferred. For example, we write $\RF \stackrel{\WRinVIS}{\subseteq} \VIS$, instead 
of the whole sequence of inclusions needed to prove such an inequality, namely 
\[
\RF = \bigcup_{x \in \Obj} \RF(x) \stackrel{\WRinVIS}{\subseteq} \bigcup_{x \in \Obj} \VIS = \VIS.
\]

\begin{proposition}
\label{prop:irrefl.change}
For all relations $R_1, R_2 \subseteq \T \times \T$, 
\begin{equation}
\label{eq:acyclic.comp} (R_1 \rcomp R_2) \cap \Id \subseteq \emptyset 
\implies (R_2 \rcomp R_1) \cap \Id \subseteq \emptyset 
\end{equation}
\end{proposition}

\noindent\myparagraph{Proof}
Suppose $(R_1 \rcomp R_2) \cap \Id \subseteq \emptyset$. 
For any $T \in \T$, there exists no $S \in \T$ such that 
$(T,S) \in R_1$ and $(S,T) \in R_2$. In particular, 
there exists no $S \in T$ such that $(S,T) \in R_2, (T,S) \in R_1$, 
for all $T \in \T$: equivalently, $(S,S) \notin (R_2 \rcomp R_1)$ for all $S \in \T$. 
That is, $(R_2 \rcomp R_1) \cap \Id \subseteq \emptyset$.
\qed


\begin{proposition}
For any set $\T' \subseteq \T$, 
\begin{equation}
\label{eq:tx.dupl}
\circled{\T'} = \circled{\T'} \rcomp \circled{\T'}.
\end{equation}
\end{proposition} 

\myparagraph{Proof}
$
\circled{\T'} = \circled{\T' \cap \T'} \stackrel{\txComp}{=} \circled{\T'} \rcomp \circled{\T'}.
$
\qed
%

\begin{proposition}
For any relation $R \subseteq \T \times \T$, 
\begin{equation}
(R \cap \Id = \emptyset) \iff (R \subseteq R \setminus \Id). \label{eq:acycl.irrefl}
\end{equation}
\end{proposition}

\myparagraph{Proof}
Suppose $R \cap \Id = \emptyset$. Then 
\begin{gather*}
R = R \cap (\Id \cup \overline{\Id}) = (R \cap \Id) \cup (R \cap \overline{\Id}) = 
\emptyset \cup (R \setminus \Id) = (R \setminus \Id).
\end{gather*}

Now, suppose that $R \subseteq R \setminus \Id$. Then 
\begin{gather*}
(R \cap \Id) \subseteq (R \setminus \Id) \cap \Id = (R \cap \overline{\Id}) \cap \Id = 
R \cap (\overline{\Id} \cap \Id) = R \cap \emptyset = \emptyset \tag*{\qed}
\end{gather*}

Most of the time we will omit applications of the implications given by equation \eqref{eq:acycl.irrefl}. 
For example, we write $\AR \stackrel{\ARirrefl}{\subseteq} \AR \setminus \Id$ 
instead of 
\[
\AR \cap \Id \stackrel{\ARirrefl}{\subseteq} \emptyset \stackrel{\eqref{eq:acycl.irrefl}}{\implies} 
\left(\AR \subseteq (\AR \setminus \Id) \right).
\]
Other examples of inequalities that we can prove using equation \eqref{eq:acycl.irrefl} 
are given below:
\begin{gather}
\VIS \stackrel{\textbf{(c.6,c.12)}}{\subseteq} \VIS \setminus \Id \label{eq:VIS.irrefl}\\
\nonumber(\VIS \rcomp \overline{\VIS^{-1}}) \stackrel{\VISnotAVIS}{\subseteq} (\VIS \rcomp \overline{\VIS^{-1}}) \setminus \Id \\
\nonumber(\overline{\VIS^{-1}} \rcomp \VIS) \stackrel{\AVISnotVIS}{\subseteq} (\overline{\VIS^{-1}} \rcomp \VIS) \setminus \Id.
\end{gather}


%

\begin{proposition}
Let $\Sigma$ be a x-specification such that $(\rho_x, \rho_x) \in \Sigma$, 
for some object $x \in \Obj$. If $\aexec \in \aeset(\Sigma)$, then 
\begin{equation}
\label{eq:WWconf.one}
\VO(x) \subseteq \VIS.
\end{equation}
\end{proposition}

\myparagraph{Proof}
Recall that $\rho_x(\_) = \circled{\WTr_x}$. 
Because $(\rho_x, \rho_x) \in \Sigma$, then
\begin{gather*}
\VO(x) \stackrel{\depWWTx}{\subseteq} \circled{\WTr_x} \rcomp \VO(x) \rcomp \circled{\WTr_x} 
\stackrel{\WWinAR}{\subseteq} \circled{\WTr_x} \rcomp \AR \rcomp \circled{\WTr_x} 
=\\ 
\rho_x(\VIS) \rcomp \AR \rcomp \rho_x(\VIS) \stackrel{\Axiom}{\subseteq} \VIS.
\end{gather*}
\qed

\begin{corollary}
Let $\Sigma$ be a consistency model such that $(\rho_x, \rho_x) \in \Sigma$ 
for all $x \in \Obj$. If $\aexec \in \aeset(\Sigma)$, then 
\begin{equation}
\label{eq:WWconf}
\VO \subseteq \VIS.
\end{equation}
\end{corollary}

\myparagraph{Proof}
If $\aexec \in \aeset(\Sigma)$, then 
\begin{equation*}
\VO = \bigcup_{x \in \Obj} \VO(x) \stackrel{\eqref{eq:WWconf.one}}{\subseteq} \bigcup_{x \in \Obj} \VIS \subseteq \VIS \tag*{\qed}
\end{equation*}

\begin{corollary}
\label{cor:DEPSinVIS}
Let $\Sigma$ be a consistency model such that $(\rho_{x}, \rho_{x}) \in \Sigma$ 
for any $x \in \Obj$. If $\aexec \in \aeset(\Sigma)$, then 
\begin{equation}
\label{eq:WWconf.VIS}
(\RF \cup \VO)^{+} \subseteq \VIS.
\end{equation}
\end{corollary}

\myparagraph{Proof}
If $\aexec \in \aeset(\Sigma)$, then 
\begin{equation*}
(\RF \cup \VO)^{+} \stackrel{\WRinVIS, \eqref{eq:WWconf}}{\subseteq} \VIS^{+} 
\stackrel{\VIStrans}{\subseteq} \VIS \tag*{\qed}
\end{equation*}

Some proofs of the robustness criteria we present require the following theorem from 
Kleene Algebra: 
\begin{theorem}[\cite{kleenecomplete}]
\label{thm:kleene}
For any relations $R_1, R_2 \subseteq \T \times \T$, 
\begin{gather}
(R_1 \rcomp R_2)^{+} = R_1 \rcomp (R_2 \rcomp R_1)^{\ast} \rcomp R_2. \label{eq:kleene.unfoldplus}
\end{gather}
\end{theorem}
\myparagraph{Proof}
Recall that $(R_1 \rcomp R_2)^{+} = \bigcup_{n > 0} (R_1 \rcomp R_2)^{n}$, and 
$(R_2 \rcomp R_1)^{\ast} = \bigcup_{n \geq 0} (R_2 \rcomp R_1)^{n}$. We 
prove, by induction on $n$, that for all $n > 0$, $(R_1 \rcomp R_2)^{n} = (R_1 \rcomp (R_2 \rcomp R_1)^{n-1} 
\rcomp R_2)$. Then we have 
\begin{gather*}
(R_1 \rcomp R_2)^{+} = \bigcup_{n > 0} (R_1 \rcomp R_2)^{n} = \bigcup_{n > 0} \left( R_1 \rcomp (R_2 \rcomp R_1)^{n - 1} \rcomp R_2 \right) = \\
\bigcup_{n \geq 0} \left( R_1 \rcomp (R_2 \rcomp R_1)^{n} \rcomp R_2 \right) = 
\left( R_1 \rcomp \left( \bigcup_{n \geq 0} (R_2 \rcomp R_1)^{n} \right) \rcomp R_2 \right) = (R_1 \rcomp (R_2 \rcomp R_1)^{\ast} \rcomp R_2).
\end{gather*}
\begin{description}
\item[Case $n = 1$:] 
\begin{gather*}
(R_1 \rcomp R_2) = R_1 \rcomp \Id \rcomp R_2 = R_1 \rcomp (R_2 \rcomp R_1)^{0} \rcomp R_2
\end{gather*}
\item[Case $n > 1$:] suppose that 
\begin{gather}
(R_1 \rcomp R_2)^{n-1} = R_1 \rcomp (R_2 \rcomp R_1)^{n-2} \rcomp R_2 \tag{IH} \label{eq:kleene.ih}.
\end{gather}
Then
\begin{gather*}
(R_1 \rcomp R_2)^{n} = (R_1 \rcomp R_2) \rcomp (R_1 \rcomp R_2)^{n-1} \stackrel{\eqref{eq:kleene.ih}}{=} 
(R_1 \rcomp R_2) \rcomp (R_1 \rcomp (R_2 \rcomp R_1)^{n-2} \rcomp R_2) = \\
(R_1 \rcomp (R_2 \rcomp R_1) \rcomp (R_2 \rcomp R_1)^{n-2} \rcomp R_2) = 
(R_1 \rcomp (R_2 \rcomp R_1)^{n-1} \rcomp R_2). \tag*{\qed}
\end{gather*}
\end{description}

\subsection{Robustness Criteria of x-Specifications}
In this Section we show several applications of the algebraic laws for 
inferring robustness criteria for several x-specification. 
We start by giving alternative proofs of previously known results 
(theorems \ref{thm:SI.acyclic} and \ref{thm:PSI.acyclic}). Then 
we present and prove novel robustness criteria for other x-specifications 
(theorems \ref{thm:rb.acyclic} and \ref{thm:cpser.acyclic}).

\begin{theorem}[\cite{fekete-tods}]
\label{thm:SI.acyclic} 
For all $\aexec \in \aeset(\Sigma_{\SI})$, every cycle in $\graphof(\aexec)$ has two consecutive 
$\AD_{\aexec}$ edges. That is, 
$((\RF_{\aexec} \cup \VO_{\aexec}) \rcomp \AD_{\aexec}?)$ is acyclic.
\end{theorem}

\noindent\myparagraph{Proof}
Recall that $\Sigma_{\SI} = \{(\rho_{\Id}, \rho_{\SI})\} \cup \{ (\rho_x, \rho_x)\}_{x \in \Obj}$, 
where $\rho_{\Id}(\_) = \Id$ and $\rho_{\SI}(R) = R \setminus \Id$.
If $\aexec \in \aeset(\Sigma_{\SI})$, then
\begin{gather}
\hspace{-10pt}{\textcolor[rgb]{0.6,0.6,0.61}{\rule{0.67em}{0.33em}}\;} 
{\color{red} (\VIS_{\aexec} \rcomp \AD_{\aexec}) \subseteq \AR_{\aexec}: } \label{eq:SI.VISnotRW}\\
\nonumber (\VIS_{\aexec} \rcomp \AD_{\aexec}) \stackrel{\RWinAVIS}{\subseteq} (\VIS_{\aexec} \rcomp \overline{\VIS^{-1}_{\aexec}}) 
\stackrel{\VISnotAVIS}{\subseteq} (\VIS_{\aexec} \rcomp \overline{\VIS^{-1}_{\aexec}}) \setminus \Id 
\stackrel{\eqref{eq:VIS.irrefl}}{\subseteq} ((\VIS_{\aexec} \setminus \Id) \rcomp \overline{\VIS^{-1}_{\aexec}}) \setminus \Id = \\
\nonumber (\rho_{\SI}(\VIS_{\aexec}) \rcomp \overline{\VIS^{-1}_{\aexec}} \rcomp \rho_{\Id}(\VIS_{\aexec}) ) \setminus \Id \stackrel{\CoAxiomAR}{\subseteq} 
\AR_{\aexec}\\  
\nonumber\\
\hspace{-10pt}{\textcolor[rgb]{0.6,0.6,0.61}{\rule{0.67em}{0.33em}}\;}
{\color{red} ((\RF_{\aexec} \cup \VO_{\aexec}) \rcomp \AD_{\aexec}?) \subseteq \AR_{\aexec}: } \label{eq:SI.maybeRW}\\
\nonumber ((\RF_{\aexec} \cup \VO_{\aexec}) \rcomp \AD_{\aexec}?) \stackrel{\eqref{eq:WWconf.VIS}}{\subseteq} (\VIS_{\aexec} \rcomp \AD_{\aexec}?)^{+} = 
(\VIS_{\aexec} \cup (\VIS_{\aexec} \rcomp \AD_{\aexec})) \stackrel{\VISinAR,\eqref{eq:SI.VISnotRW}}{\subseteq} \AR_{\aexec}\\
\nonumber\\
\nonumber 
\hspace{-10pt}{\textcolor[rgb]{0.6,0.6,0.61}{\rule{0.67em}{0.33em}}\;}
{\color{red} ((\RF_{\aexec} \cup \VO_{\aexec}) \rcomp \AD_{\aexec}?)^{+} \cap \Id \subseteq \emptyset: } \\
\nonumber ((\RF_{\aexec} \cup \VO_{\aexec}) \rcomp \AD_{\aexec}?)^{+} \cap \Id \stackrel{\eqref{eq:SI.maybeRW}}{\subseteq} \AR_{\aexec}^{+} \cap \Id \stackrel{\ARtrans}{\subseteq} 
\AR_{\aexec} \cap \Id \stackrel{\ARirrefl}{\subseteq} \emptyset \tag*{\qed}
\end{gather}

\begin{theorem}[\cite{giovanni_concur16}]
\label{thm:PSI.acyclic}
For all $\aexec \in \aeset(\Sigma_{\PSI})$, 
it is not possible that all anti-dependencies in a cycle of $\graphof(\aexec)$ are over the same object\footnote{This 
implies that all cycles have at least two anti-dependencies.}: 
$(\RF_{\aexec} \cup \VO_{\aexec})^{\ast} \rcomp \AD(x)$ is acyclic for all $x \in \Obj$.
\end{theorem}

\myparagraph{Proof}
Recall that $\Sigma_{\PSI} = \{(\rho_x, \rho_x)\}_{x \in \Obj}$, where $\rho_x(\_) = \circled{\WTr_x}$. 
Then 
\begin{gather}
\hspace{-10pt}{\textcolor[rgb]{0.6,0.6,0.61}{\rule{0.67em}{0.33em}}\;}
{\color{red} (\circled{\WTr_x} \rcomp \overline{\VIS^{-1}_{\aexec}} \rcomp \circled{\WTr_x}) \setminus \Id \subseteq \AR_{\aexec}:}  
\label{eq:PSI.coaxiom}\\
\nonumber (\circled{\WTr_x} \rcomp \overline{\VIS^{-1}_{\aexec}} \rcomp \circled{\WTr_x}) \setminus \Id = 
(\rho_x(\VIS_{\aexec}) \rcomp \overline{\VIS^{-1}_{\aexec}} \rcomp \rho_x(\VIS_{\aexec})) \setminus \Id \stackrel{\CoAxiomAR}{\subseteq} \AR_{\aexec}\\
\nonumber\\ 
\hspace{-10pt}{\textcolor[rgb]{0.6,0.6,0.61}{\rule{0.67em}{0.33em}}\;}
{\color{red} \circled{\WTr_x} \rcomp (\RF_{\aexec} \cup \VO_{\aexec})^{\ast} \rcomp \AD_{\aexec}(x) \subseteq \AR_{\aexec}:} 
\label{eq:PSI.RWx2AR}\\
\nonumber\circled{\WTr_x} \rcomp (\RF_{\aexec} \cup \VO_{\aexec})^{\ast} \rcomp \AD_{\aexec}(x) \stackrel{\eqref{eq:WWconf.VIS}}{\subseteq}
\circled{\WTr_x} \rcomp \VIS_{\aexec}? \rcomp \AD_{\aexec}(x) =\\
\nonumber  (\circled{\WTr_x} \rcomp \AD_{\aexec}(x)) \cup (\circled{\WTr_x} \rcomp \VIS_{\aexec} \rcomp \AD_{\aexec}(x)) 
\stackrel{\depRWTx}{\subseteq}\\ 
\nonumber
(\circled{\WTr_x} \rcomp \AD_{\aexec}(x) \rcomp \circled{\WTr_x}) \cup (\circled{\WTr_x} \rcomp \VIS_{\aexec} \rcomp \AD_{\aexec}(x) \rcomp \circled{\WTr_x}) \stackrel{\depRWIrrefl}{\subseteq} \\
\nonumber
(\circled{\WTr_x} \rcomp (\AD_{\aexec}(x) \setminus \Id) \rcomp \circled{\WTr_x}) \cup (\circled{\WTr_x} \rcomp \VIS_{\aexec} \rcomp \AD_{\aexec}(x) \rcomp \circled{\WTr_x}) 
\stackrel{\RWinAVIS}{\subseteq}\\
\nonumber
(\circled{\WTr_x} \rcomp (\overline{\VIS^{-1}_{\aexec}} \setminus \Id) \rcomp \circled{\WTr_x}) \cup 
(\circled{\WTr_x} \rcomp \VIS_{\aexec} \rcomp \overline{\VIS^{-1}_{\aexec}} \rcomp \circled{\WTr_x}) \stackrel{\VISnotAVIS}{\subseteq} \\
\nonumber
(\circled{\WTr_x} \rcomp (\overline{\VIS^{-1}_{\aexec}} \setminus \Id) \rcomp \circled{\WTr_x}) \cup 
(\circled{\WTr_x} \rcomp (\VIS_{\aexec} \rcomp \overline{\VIS^{-1}_{\aexec}}) \setminus \Id \rcomp \circled{\WTr_x}) \stackrel{\AVISright}{\subseteq} \\
\nonumber (\circled{\WTr_x} \rcomp (\overline{\VIS^{-1}_{\aexec}} \setminus \Id) \rcomp \circled{\WTr_x}) 
\stackrel{\txDistrR,\txDistrL}{=}
(\circled{\WTr_x} \rcomp \overline{\VIS^{-1}_{\aexec}} \rcomp \circled{\WTr_x}) \setminus \Id \stackrel{\eqref{eq:PSI.coaxiom}}{\subseteq} \AR_{\aexec}\\
\nonumber\\
\hspace{-10pt}{\textcolor[rgb]{0.6,0.6,0.61}{\rule{0.67em}{0.33em}}\;}
{\color{red} \circled{\WTr_x} \rcomp \left((\RF_{\aexec} \cup \VO_{\aexec})^{\ast} \rcomp \AD_{\aexec}(x) \right)^{+} \subseteq \AR_{\aexec}:} 
\label{eq:PSI.AR}\\
\nonumber\circled{\WTr_x} \rcomp \left((\RF_{\aexec} \cup \VO_{\aexec})^{\ast} \rcomp \AD_{\aexec}(x) \right)^{+} \stackrel{\depRWTx}{=}\\
\nonumber\circled{\WTr_x} \rcomp \left((\RF_{\aexec} \cup \VO_{\aexec})^{\ast} \rcomp \AD_{\aexec}(x) \rcomp \circled{\WTr_x} \right)^{+} 
\stackrel{\eqref{eq:kleene.unfoldplus}}{=}\\
\nonumber \circled{\WTr_x} \rcomp (\RF_{\aexec} \cup \VO_{\aexec})^{\ast} \rcomp \AD_{\aexec}(x) \rcomp (\circled{\WTr_x} \rcomp 
(\RF_{\aexec} \cup \VO_{\aexec})^{\ast} \rcomp \AD_{\aexec}(x))^{\ast} \rcomp \circled{\WTr_x} =\\
\nonumber
(\circled{\WTr_x} \rcomp (\RF_{\aexec} \cup \VO_{\aexec})^{\ast} \rcomp \AD_{\aexec}(x))^{+} \rcomp \circled{\WTr_x} \stackrel{\eqref{eq:PSI.RWx2AR}}{\subseteq}\\
\nonumber
(\AR_{\aexec}^{+} \rcomp \circled{\WTr_x}) \stackrel{\txId}{\subseteq} \AR_{\aexec}^{+} \stackrel{\ARtrans}{\subseteq} \AR_{\aexec}\\
\nonumber\\
\hspace{-10pt}{\textcolor[rgb]{0.6,0.6,0.61}{\rule{0.67em}{0.33em}}\;}
{\color{red} ( \circled{\WTr_x} \rcomp \left((\RF_{\aexec} \cup \VO_{\aexec})^{\ast} \rcomp \AD_{\aexec}(x) \right)^{+}) \cap \Id \subseteq \emptyset:} 
\label{eq:PSI.acyclic.1}\\
\nonumber ( \circled{\WTr_x} \rcomp \left((\RF_{\aexec} \cup \VO_{\aexec})^{\ast} \rcomp \AD_{\aexec}(x) \right)^{+}) \cap \Id \stackrel{\eqref{eq:PSI.AR}}{\subseteq} 
\AR_{\aexec} \cap \Id \stackrel{\ARirrefl}{\subseteq} \emptyset\\
\nonumber\\
\hspace{-10pt}{\textcolor[rgb]{0.6,0.6,0.61}{\rule{0.67em}{0.33em}}\;}
{\color{red} ((\RF_{\aexec} \cup \VO_{\aexec})^{\ast} \rcomp \AD_{\aexec}(x))^{+} \rcomp \circled{\WTr_x} = 
((\RF_{\aexec} \cup \VO_{\aexec})^{\ast} \rcomp \AD_{\aexec}(x))^{+}:} \label{eq:PSI.noFinalWrite}\\
\nonumber ((\RF_{\aexec} \cup \VO_{\aexec})^{\ast} \rcomp \AD_{\aexec}(x))^{+} \rcomp \circled{\WTr_x} = \\
\nonumber ((\RF_{\aexec} \cup \VO_{\aexec})^{\ast} \rcomp \AD_{\aexec}(x))^{\ast} \rcomp ((\RF_{\aexec} \cup \VO_{\aexec})^{\ast} \rcomp \AD_{\aexec}(x) \rcomp \circled{\WTr_x}) 
\stackrel{\depRWTx}{=}\\
\nonumber (\RF_{\aexec} \cup \VO_{\aexec})^{\ast} \rcomp \AD_{\aexec}(x))^{\ast} \rcomp ((\RF_{\aexec} \cup \VO_{\aexec})^{\ast} \rcomp \AD_{\aexec}(x))= \\
\nonumber ((\RF_{\aexec} \cup \VO_{\aexec})^{\ast} \rcomp \AD_{\aexec}(x))^{+}\\
\nonumber\\
\nonumber
\hspace{-10pt}{\textcolor[rgb]{0.6,0.6,0.61}{\rule{0.67em}{0.33em}}\;}
{\color{red} \left( (\RF_{\aexec} \cup \VO_{\aexec})^{\ast} \rcomp \AD_{\aexec}(x) \right)^{+} \cap \Id \subseteq \emptyset: }\\
\nonumber \left(\circled{\WTr_x} \rcomp ((\RF_{\aexec} \cup \VO_{\aexec})^{\ast} \rcomp \AD_{\aexec}(x))^{+}\right) \cap \Id \stackrel{\eqref{eq:PSI.acyclic.1}}{\subseteq} 
\emptyset \stackrel{\eqref{eq:acyclic.comp}}{\implies} \\
\nonumber \left( (\RF_{\aexec} \cup \VO_{\aexec}^{\ast}) \rcomp \AD_{\aexec}(x))^{+} \rcomp \circled{\WTr_x}\right) 
\cap \Id \subseteq \emptyset \stackrel{\eqref{eq:PSI.noFinalWrite}}{\implies} \\
\nonumber \left( (\RF_{\aexec} \cup \VO_{\aexec}^{\ast} \rcomp \AD_{\aexec}(x) \right)^{+} \cap \Id \subseteq \emptyset. \tag*{\qed}
\end{gather}
\begin{definition}
\label{def:rb.protected}
Let $\aexec \in \Sigma_{\CCSER}$, and suppose that 
$\graphof(\aexec)$ contains a cycle $T_0 \xrightarrow{R_0} \cdots \xrightarrow{R_{n-1}} T_n$, 
where $T_0 = T_n$ and $R_i \in \{\RF_{\aexec}, \VO_{\aexec}, \AD_{\aexec}\}$ for any $i=0,\cdots,n-1$. 
We recall the following definition of protected anti-dependency edge in the cycle, and also introduce the notion of 
protected $\VO$-dependencies. 
\begin{itemize} 
\item an anti-dependency edge $R_i = \AD_{\aexec}$ is \emph{protected} if there exist two integers $j,k = 0,\cdots, n-1$ 
such that $(T_{(i - j)\,\mathsf{mod}\,{n}}) \ni \SerTX, (T_{((i+1) + k)\,\mathsf{mod}\,n}) \ni \SerTX$, and for all $h = (i-j), \cdots, (i+k+1)$, 
$R_{h\,\mathsf{mod}\,n} = \RF_{\aexec}$; in other words, in the cycle the endpoints of the $R_i$ anti-dependency edge 
are connected to serialisable transactions by a sequence of $\RF$-dependencies, 
\item a $\VO$-dependency edge $R_i = \VO_{\aexec}$ is \emph{protected} if tere exist two integers 
$j,k = 0,\cdots, n-1$ such that $(T_{(i - j)\,\mathsf{mod}\,n}) \ni \SerTX, (T_{((i+1) + k)\,\mathsf{mod}\, n}) \ni \SerTX$, and for all $h = (i-j), \cdots, (i+k+1)$, 
$R_{h\,\mathsf{mod}\,n} \in \{\RF_{\aexec}, \VO_{\aexec}\}$; in other words, in the cycle the endpoints of the $R_i$ dependency edge 
are connected to serialisable transactions by a sequence of both $\RF$-dependencies and $\VO$-dependencies.
\end{itemize}
\end{definition}

\begin{theorem}
\label{thm:rb.acyclic}
Let $\aexec \in \aeset(\Sigma_{\CCSER})$. Then any cycle in $\graphof(\aexec)$ contains 
at least one unprotected anti-dependency edge, and another edge that is either an unprotected 
anti-dependency, or an unprotected $\VO$-dependency. 
Formally, given a relation $R \subseteq \T_{\aexec} \times \T_{\aexec}$, 
let $\fenced{R} = \circled{\SerTX} \rcomp \RF_{\aexec}^{\ast} \rcomp R \rcomp \RF_{\aexec}^{\ast} 
\rcomp \circled{\SerTX}$. then 
\[
\left( (\RF_{\aexec} \cup \fenced{(\RF_{\aexec} \cup \VO_{\aexec})^{+}} \cup \fenced{\AD_{\aexec}})^{+} \rcomp \AD_{\aexec}\right) \cap \Id \subseteq \emptyset.
\]
\end{theorem}

\noindent\myparagraph{Proof}
Recall that $\Sigma_{\CCSER} = \{(\rho_S, \rho_S)\}$, where $\rho_S(\_) = \circled{\SerTX}$. 
In the proof of Theorem \ref{thm:ccser.acyclic} we proved the following fact: 
\begin{gather}
\fenced{\AD_{\aexec}} \subseteq \AR_{\aexec} \label{eq:RB.protectedRW},
\end{gather}
which we will need to prove Theorem \ref{thm:rb.acyclic}. 
We have 
\begin{gather}
\hspace{-10pt}{\textcolor[rgb]{0.6,0.6,0.61}{\rule{0.67em}{0.33em}}\;}
{\color{red} \fenced{\AD_{\aexec}} = \circled{\SerTX} \rcomp \fenced{\AD_{\aexec}} \rcomp \circled{\SerTX}:} \label{eq:addSer}\\
\nonumber \fenced{\AD_{\aexec}} = \circled{\SerTX} \rcomp \RF_{\aexec}^{\ast} \rcomp \AD_{\aexec} \rcomp \RF_{\aexec}^{\ast} \rcomp \circled{\SerTX} \stackrel{\eqref{eq:tx.dupl}}{=}\\ 
\nonumber
\circled{\SerTX} \rcomp \circled{\SerTX} \rcomp \RF_{\aexec}^{\ast} \rcomp \AD_{\aexec} \rcomp \RF_{\aexec}^{\ast} \rcomp \circled{\SerTX} \rcomp \circled{\SerTX} = 
\circled{\SerTX} \rcomp \fenced{\AD_{\aexec}} \rcomp \circled{\SerTX}\\
\nonumber\\
\hspace{-10pt}{\textcolor[rgb]{0.6,0.6,0.61}{\rule{0.67em}{0.33em}}\;}
{\color{red} \circled{\SerTX} \rcomp \AR_{\aexec} \rcomp \circled{\SerTX} \subseteq \VIS_{\aexec}:} \label{eq:RB.axiom}\\
\nonumber\circled{\SerTX} \rcomp \AR_{\aexec} \rcomp \circled{\SerTX} = \rho_S(\VIS_{\aexec}) \rcomp \AR_{\aexec} \rcomp \rho_S(\VIS_{\aexec}) 
\stackrel{\Axiom}{\subseteq} \VIS_{\aexec}\\
\nonumber\\
\hspace{-10pt}{\textcolor[rgb]{0.6,0.6,0.61}{\rule{0.67em}{0.33em}}\;}
{\color{red}\fenced{\AD_{\aexec}} \subseteq \VIS_{\aexec}: } \label{eq:RB.protectedRWinVIS}\\
\nonumber \fenced{\AD_{\aexec}} \stackrel{\eqref{eq:addSer}}{=} \circled{\SerTX} \rcomp \fenced{\AD_{\aexec}} \rcomp \circled{\SerTX} 
\stackrel{\eqref{eq:RB.protectedRW}}{\subseteq} \circled{\SerTX} \rcomp \AR_{\aexec} \rcomp \circled{\SerTX} \stackrel{\eqref{eq:RB.axiom}}{\subseteq} \VIS_{\aexec}\\
\nonumber\\
\hspace{-10pt}{\textcolor[rgb]{0.6,0.6,0.61}{\rule{0.67em}{0.33em}}\;}
{\color{red} \fenced{(\RF_{\aexec} \cup \VO_{\aexec})^{+}} \subseteq \VIS_{\aexec}:} \label{eq:RB.protectedWW}\\
\nonumber \fenced{( \RF_{\aexec} \cup \VO_{\aexec})^{+}} = \circled{\SerTX} \rcomp \RF_{\aexec}^{\ast} \rcomp (\RF_{\aexec} \cup \VO_{\aexec})^{+} \rcomp 
\RF_{\aexec}^{\ast} \rcomp \circled{\SerTX} = \\
\nonumber \circled{\SerTX} \rcomp (\RF_{\aexec} \cup \VO_{\aexec})^{+} \rcomp \circled{\SerTX} \stackrel{\WRinVIS,\VISinAR}{\subseteq} 
\circled{\SerTX} \rcomp (\AR_{\aexec} \cup \VO_{\aexec})^{+} \rcomp \circled{\SerTX} \stackrel{\WWinAR}{\subseteq}\\ 
\nonumber
\circled{\SerTX} \rcomp \AR_{\aexec}^{+} \rcomp \circled{\SerTX} \stackrel{\ARtrans}{\subseteq} 
\circled{\SerTX} \rcomp \AR_{\aexec} \rcomp \circled{\SerTX} \stackrel{\eqref{eq:RB.axiom}}{\subseteq} \VIS_{\aexec}\\
\nonumber
\end{gather}
\begin{gather}
\hspace{-10pt}{\textcolor[rgb]{0.6,0.6,0.61}{\rule{0.67em}{0.33em}}\;}
{\color{red} (\RF_{\aexec} \cup \fenced{(\RF_{\aexec} \cup \VO_{\aexec})^{+}} \cup \fenced{\AD_{\aexec}})^{+} \subseteq \VIS_{\aexec}:} 
\label{eq:RB.VIS}\\
\nonumber (\RF_{\aexec} \cup \fenced{(\RF_{\aexec} \cup \VO_{\aexec})^{+}} \cup \fenced{\AD_{\aexec}})^{+} \stackrel{\WRinVIS}{\subseteq}\\
\nonumber (\VIS_{\aexec} \cup \fenced{(\RF_{\aexec} \cup \VO_{\aexec})}^{+} \cup \fenced{\AD_{\aexec}})^{+} \stackrel{\eqref{eq:RB.protectedWW}}{\subseteq}\\
\nonumber
(\VIS_{\aexec} \cup \fenced{\AD_{\aexec}})^{+} \stackrel{\eqref{eq:RB.protectedRW}}{\subseteq} \VIS_{\aexec}^{+} \stackrel{\VIStrans}{\subseteq} \VIS_{\aexec}\\
\nonumber\\
\nonumber \hspace{-10pt}{\textcolor[rgb]{0.6,0.6,0.61}{\rule{0.67em}{0.33em}}\;}
{\color{red} ((\RF_{\aexec} \cup \fenced{(\RF_{\aexec} \cup \VO_{\aexec})^{+}} \cup \fenced{\AD_{\aexec}})^{+} \rcomp \AD_{\aexec}) \cap \Id \subseteq \emptyset:}\\
\nonumber ((\RF_{\aexec} \cup \fenced{(\RF_{\aexec} \cup \VO_{\aexec})^{+}} \cup \fenced{\AD_{\aexec}})^{+} \rcomp \AD_{\aexec}) \cap \Id 
\stackrel{\eqref{eq:RB.VIS}}{\subseteq} (\VIS_{\aexec} \rcomp \AD_{\aexec}) \cap \Id \stackrel{\RWinAVIS}{\subseteq}\\
(\VIS_{\aexec} \rcomp \overline{\VIS^{-1}_{\aexec}}) \cap \Id \stackrel{\VISnotAVIS}{\subseteq} \emptyset. \tag*{\qed}
\end{gather}
%
So far, none of the robustness criteria that we have derived has 
exploited the inequalities \CoAxiomL\; and \CoAxiomR\; from 
Figure \ref{fig:laws}. Here we give another example of x-specification, 
for which we can derive a robustness criterion which makes use of the 
inequalities \CoAxiomL\; and \CoAxiomR.
Such a x-specification is given by $\Sigma_{\CP}= \{(\rho_{\Id}, \rho_{\SI}), 
(\rho_{S}, \rho_{S})\}$. The set of executions $\aeset(\Sigma_{\CP})$ coincides with the 
definition of the \emph{Consistent Prefix} consistency model given in \cite{giovanni_concur16}. 
The x-specification $\Sigma_{\CP}$ can be thought as a weakening of 
$\Sigma_{\SI+\SER}$ which does not have any write conflict detection. 


\begin{theorem}
\label{thm:cpser.acyclic}
Let $\aexec = (\T, \VIS, \AR) \in \aeset(\Sigma_{\CP})$.
We say that a path $T_0 \xrightarrow{R_0} \cdots \xrightarrow{R_{n-1}} T_{n}$ of $\graphof(\aexec)$, 
is \emph{critical} if $T_0 \neq T_n$, both $T_0, T_n \ni \SerTX$, only one of the edges 
$R_i, 0 \leq i < n$ is an anti-dependency, and none of the edges 
$R_j, 0 \leq j < i$ is a $\VO$-edge (note that if $j > i$, we allow $R_j = \VO_{\aexec}$). 
Then all cycles of $\graphof(\aexec)$ have at least one anti-dependency 
edge 
that is not contained within a critical sub-path of the cycle.

Formally, let $\mathsf{CSub}_{\aexec} = (\circled{\SerTX} \rcomp \RF^{\ast} \rcomp \AD \rcomp (\VO \cup \RF)^{\ast} \rcomp \circled{\SerTX}) \setminus \Id$, 
where $\graphof(\aexec) = (\T, \RF, \VO, \AD)$. 
Then $(\RF \cup \VO \cup \mathsf{CSub}_{\aexec})$ 
is acyclic. 
\end{theorem}

\noindent\myparagraph{Proof}
By Definition, $\Sigma_{\CP} = \{(\rho_S, \rho_S), (\rho_{\Id}, \rho{\SI})\}$, 
where $\rho_{S}(\_) = \circled{\SerTX}$, $\rho_{\Id}(\_) = \Id$ and 
$\rho_{\SI}(R) = R \setminus \Id$. This implies that $\rho_{\SI}(\T \times \T)^{-1} = 
((\T \times \T) \setminus \Id)^{-1} = (\T \times \T) \setminus \Id$, and 
for any relation $R \subseteq \T \times \T$, 
\begin{gather}
R \cap \rho_{\SI}(\T \times \T)^{-1} = R \setminus \Id \label{eq:CP.reverserho}.
\end{gather}
For $\aexec \in \aeset(\Sigma_{\CP})$, we have:
\begin{gather}
\hspace{-10pt}{\textcolor[rgb]{0.6,0.6,0.61}{\rule{0.67em}{0.33em}}\;}
{\color{red} \circled{\SerTX} \rcomp \overline{\VIS_{\aexec}^{-1}} \rcomp \circled{\SerTX}) \setminus \Id \subseteq \AR_{\aexec}:} \label{eq:cpser.coaxiom2}\\
\nonumber
(\circled{\SerTX} \rcomp \overline{\VIS_{\aexec}^{-1}} \rcomp \circled{\SerTX}) \setminus \Id = 
((\rho_{S}(\VIS_{\aexec}) \rcomp \overline{\VIS_{\aexec}^{-1}} \rcomp \rho_S(\VIS_{\aexec})) \setminus \Id 
\stackrel{\CoAxiomAR}{\subseteq} \AR_{\aexec}\\
\nonumber\\
\hspace{-10pt}{\textcolor[rgb]{0.6,0.6,0.61}{\rule{0.67em}{0.33em}}\;}
{\color{red} (\overline{\VIS_{\aexec}^{-1}} \rcomp \AR_{\aexec}) \setminus \Id \subseteq \overline{\VIS^{-1}_{\aexec}}:} \label{eq:cpser.moreavis}\\
\nonumber (\overline{\VIS_{\aexec}^{-1}} \rcomp \AR_{\aexec}) \setminus \Id \stackrel{\eqref{eq:CP.reverserho}}{=} 
(\overline{\VIS_{\aexec}^{-1}} \rcomp \AR_{\aexec}) \cap \rho_{\SI}(\T \times \T)^{-1} = \\
\nonumber (\overline{\VIS_{\aexec}^{-1}}\rcomp \rho_{\Id}(\VIS_{\aexec}) \rcomp \AR_{\aexec}) \cap \rho_{\SI}(\T \times \T)^{-1})
\stackrel{\CoAxiomR}{\subseteq} \overline{\VIS^{-1}_{\aexec}}
\end{gather}
%
\begin{gather} 
\hspace{-10pt}{\textcolor[rgb]{0.6,0.6,0.61}{\rule{0.67em}{0.33em}}\;}
{\color{red} \mathsf{CSub}_{\aexec} \subseteq \AR_{\aexec}:} \label{eq:cpser.csub}\\
\nonumber
\mathsf{CSub}_{\aexec} = (\circled{\SerTX} \rcomp \RF_{\aexec}^{\ast} \rcomp \AD_{\aexec} \rcomp (\VO_{\aexec} \cup \RF_{\aexec})^{\ast} \rcomp 
\circled{\SerTX}) \setminus \Id \stackrel{\WRinVIS}{\subseteq} \\
\nonumber
(\circled{\SerTX} \rcomp \VIS_{\aexec}^{\ast} \rcomp \AD_{\aexec} \rcomp (\VO_{\aexec} \cup \VIS_{\aexec})^{\ast} \rcomp \circled{\SerTX}) \setminus \Id \stackrel{\WWinAR}{\subseteq} \\
\nonumber
(\circled{\SerTX} \rcomp \VIS^{\ast} \rcomp \AD_{\aexec} \rcomp (\AR_{\aexec} \cup \VIS_{\aexec})^{\ast} \rcomp \circled{\SerTX}) \setminus \Id \stackrel{\VISinAR}{\subseteq} \\
\nonumber
(\circled{\SerTX} \rcomp \VIS_{\aexec}^{\ast} \rcomp \AD_{\aexec} \rcomp \AR_{\aexec}^{\ast} \rcomp \circled{\SerTX}) \setminus \Id \stackrel{\VIStrans,\ARtrans}{\subseteq}\\
\nonumber
(\circled{\SerTX} \rcomp \VIS_{\aexec}? \rcomp \AD_{\aexec} \rcomp \AR_{\aexec}? \rcomp \circled{\SerTX}) \setminus \Id \stackrel{\RWinAVIS}{\subseteq} \\
\nonumber
(\circled{\SerTX} \rcomp \VIS_{\aexec}? \rcomp \overline{\VIS_{\aexec}^{-1}} \rcomp \AR_{\aexec}? \rcomp \circled{\SerTX)} \setminus \Id \stackrel{\AVISright}{\subseteq}\\
\nonumber
(\circled{\SerTX} \rcomp \overline{\VIS_{\aexec}^{-1}} \rcomp \AR_{\aexec}? \rcomp \circled{\SerTX}) \setminus \Id \subseteq \\
\nonumber
\left( (\circled{\SerTX} \rcomp \overline{\VIS_{\aexec}^{-1}} \rcomp \AR_{\aexec}? \rcomp \circled{\SerTX}) \setminus \Id \right) \setminus \Id \stackrel{\txDistrL, \txDistrR}{\subseteq} \\
\nonumber
\left( \circled{\SerTX} \rcomp (\overline{\VIS_{\aexec}^{-1}} \rcomp \AR_{\aexec}?) \setminus \Id \rcomp \circled{\SerTX} \right) \setminus \Id \stackrel{\eqref{eq:cpser.moreavis}}{\subseteq} \\
\nonumber
( \circled{\SerTX} \rcomp \overline{\VIS_{\aexec}^{-1}} \rcomp \circled{\SerTX} ) \setminus \Id \stackrel{\eqref{eq:cpser.coaxiom2}}{\subseteq} \AR_{\aexec}\\
\nonumber\\
\nonumber \hspace{-10pt}{\textcolor[rgb]{0.6,0.6,0.61}{\rule{0.67em}{0.33em}}\;}
{\color{red} (\RF_{\aexec} \cup \VO_{\aexec} \cup \mathsf{CSub}_{\aexec})^{+} \cap \Id \subseteq \emptyset:}\\
\nonumber
(\RF_{\aexec} \cup \VO_{\aexec} \cup \mathsf{CSub}_{\aexec})^{+} \cap \Id \stackrel{\WRinVIS,\VISinAR}{\subseteq}\\
\nonumber 
(\AR_{\aexec} \cup \VO_{\aexec} \cup \mathsf{CSub}_{\aexec})^{+} \cap \Id \stackrel{\WWinAR}{\subseteq} 
(\AR_{\aexec} \cup \mathsf{CSub}_{\aexec})^{+} \cap \Id \stackrel{\eqref{eq:cpser.csub}}{\subseteq}\\
\nonumber 
\AR_{\aexec}^{+} \cap \Id \stackrel{\ARtrans}{\subseteq} \AR_{\aexec} \cap \Id \stackrel{\ARirrefl}{\subseteq} \emptyset. \tag*{\qed}
\end{gather}

\section{Proofs of Results for Simple x-Specifications}
\label{app:completeness}

Let $X \subseteq \Obj$ and suppose that $(\rho, \pi)$ is a consistency guarantee.
Throughout this section we will work with the (simple) x-specification  
$\Sigma = \{(\rho_{x}, \rho_{x})\}_{x \in X} \cup \{(\rho, \pi)\}$, 
although all the results apply to the x-specification $\Sigma' = \{(\rho_{x}, \rho_{x})\}_{x \in X}$ 
which does not contain any consistency guarantee, aside from those enforcing the write conflict detection 
property over the objects included in $X$. 
%

\subsection{Proof of Proposition \ref{prop:pexec}}

Let $\G = (\T, \RF, \VO, \AD)$ be a dependency graph. 
%

Recall the following definition of valid pre-execution:
\begin{definition}
\label{app.def:pexec}
a pre-execution is a quadruple $\pe = (\T, \VIS, \AR)$ such that 
\begin{enumerate}
\item \label{pe:VISinAR} $\VIS \subseteq \AR$, 
\item \label{pe:VISARorders} $\VIS$ and $\AR$ are strict partial orders, 
\item \label{pe:totalWrites}for any object $x \in \Obj$, $\AR$ is total over the set $\WTr_x$,
\item \label{pe:LWW}$\pe$ satisfies the Last Write Wins property: for any $T \in \T$, if
$T \ni (\RD\; x: n)$ then  $S := \max_{\AR}(\VIS^{-1}(T) \cap \WTr_x)$ 
is well defined, and $S \ni \WR\;x: n$.
\end{enumerate}
\end{definition}



The proof of Proposition \ref{prop:pexec} relies 
on the following auxiliary result: 

\begin{proposition}
\label{prop:graph.ext}
Let $(X_V = \VIS, X_A = \AR, X_N = \mathsf{AntiVIS})$ be a solution of $\System_{\Sigma}(\G)$. 
If $\AR \cap \Id$ is acyclic, then $\pe = (\T, \VIS, \AR)$ is a valid pre-execution.
\end{proposition}

\noindent\myparagraph{Proof}
Because $(X_V = \VIS, X_A = \AR, X_{N} = \mathsf{AntiVIS})$ is a solution of $\System_{\Sigma}(\G)$, 
all the inequalities in the latter are satisfied when substituting the relations $\VIS, \AR, \mathsf{AntiVIS}$ 
for the unknowns 
$X_V, X_A, X_N$, respectively. 
We prove that all the properties \eqref{pe:VISinAR}-\eqref{pe:LWW} from Definition \ref{app.def:pexec} is satisfied 
by $\pe = (\T, \VIS, \AR)$. 
\begin{enumerate}
\item \label{proof:pe.VISinAR}$\VIS \subseteq \AR$: this follows directly from the inequality \Svis,
\item \label{proof:pe.VISARorders} $\VIS, \AR$ are strict partial orders (i.e. they are irreflexive and transitive): the relation $\AR$ is irreflexive by hypothesis, 
and transitive because of the inequality \SarTrans. The relation $\VIS$ is irreflexive 
because of the inequality \Svis\; and the assumption that $\AR \cap \Id \subseteq \emptyset$; 
$\VIS$ is also transitive because of the inequality \SvisTrans,
\item \label{proof:pe.totalWrites} $\AR$ is a strict total order order 
over the set $\WTr_x$, for any $x \in \Obj$: we prove that 
$\AR \cap (\WTr_x \times \WTr_x) = \VO(x)$; then the claim 
follows because $\VO(x)$ is a strict total order over $\WTr_x$ 
by definition. Let then $x \in \Obj$. For all $T,T' \in \T$ such 
that $T \xrightarrow{\VO(x)} T'$, we have that $T \in \WTr_x, T' \in \WTr_x$, 
and $T \xrightarrow{\AR} T'$ because of the inequality \Svo. This proves 
that $\VO(x) \subseteq \AR \cap (\WTr_x \times \WTr_x)$.
%
%
%
For the opposite implication, let $T, T' \in \T$ be transactions such that 
$T \in \WTr_x, T' \in \WTr_x$ and $T \xrightarrow{\AR} T'$. 
Because $\VO(x)$ is a strict total order over $\WTr_x$ by hypothesis, 
then either $T = T', T' \xrightarrow{\VO(x)} T$, or $T \xrightarrow{\VO(x)} T'$. 
Because $T \xrightarrow{\AR} T'$ and because $\AR \cap \Id \subseteq \emptyset$ by 
hypothesis, then $ T \neq T'$. Also, it cannot be $T' \xrightarrow{\VO(x)} T$. By the inequality 
\Svo\; this would imply that $T' \xrightarrow{\AR} T$, and because of the assumption $T \xrightarrow{\AR} T'$ 
and the inequality \SarTrans, this would mean that $T' \xrightarrow{\AR} T'$, contradicting the assumption 
that $\AR \cap \Id \subseteq \emptyset$, 
%

\item\label{proof:pe.LWW} $\pe$ satisfies the Last Write Wins property:
let $T \in \T$ be a transaction such that $T \ni (\RD\; x: n)$. By Definition \ref{def:rf.exec} there exists 
a transaction $S$ such that $S \ni \WR\; x: n$ and $S \xrightarrow{\RF(x)} T$. By Equation \Srf, 
we have that $\RF \subseteq \VIS$, hence
$S \xrightarrow{\VIS} T$. Because $S \xrightarrow{\VIS} T$ and $S \ni  (\WR\;x: n)$, we have that 
$S \in (\VIS^{-1}(T) \cap \WTr_x)$, and in particular 
$(\VIS^{-1}(T) \cap \WTr_x) \neq \emptyset$. 
Because $(\VIS^{-1}(T) \cap \WTr_x) \neq \emptyset$, and 
because by \eqref{proof:pe.totalWrites} above we have that $\AR \cap (\WTr_x \times \WTr_x) = \VO(x)$, 
then the entity $S' = \max_{\AR}(\VIS^{-1}(T) \cap \WTr_x)$ is well-defined. 
It remains to prove that $S' \ni (\WR\; x: n)$. To this end, we show that  
that $S = S'$ (recall that $S$ is the unique transaction such that $S \xrightarrow{\RF(x)} T$), 
and observe that $S \ni (\WR\;x: n)$, from which the claim follows.
Because $S,S' \in \WTr_x$ and $\VO(x)$ coincides with the restriction of 
$\AR$ to the set $\WTr_x$, we obtain that either $S' \xrightarrow{\AR} S$, $S \xrightarrow{\AR} S'$ or $S = S'$. 
The first case is not possible, because $S \in \VIS^{-1}(T) \cap \WTr_x$, and $S' = \max_{\AR}(\VIS^{-1}(T) \cap \WTr_x)$. 
The second case is also not possible: 
if $S \xrightarrow{\AR} S'$ then $S \xrightarrow{\VO(x)} S'$; together with $S \xrightarrow{\RF(x)} T$ this implies 
that there is an anti-dependency edge $T \xrightarrow{\AD(x)} S'$; now we have that $S' \in \WTr_x$, 
and $S' \xrightarrow{\VIS} T \xrightarrow{\AD(x)} S'$: that is, $(S', S') \in \circled{\WTr_x} \rcomp \VIS \rcomp \RF(x)$.  
By the inequation \Sext, this implies that $S' \xrightarrow{\AR} S'$, 
contradicting the assumption that $\AR \cap \Id \subseteq \emptyset$. We are left with the only possibility
$S = S'$, which is exactly what we wanted to prove.
\qed
\end{enumerate}
~\\[5pt]
\noindent\myparagraph{Proof of Proposition \ref{prop:pexec}}
%
Let $\pe := (\T, \VIS, \AR)$. By Proposition \ref{prop:graph.ext} we know 
that $\pe$ is a valid pre-execution. We need to show that 
$\pe \in \peset(\Sigma)$, and $\graphof(\pe)$ is well-defined 
and equal to $\G = (\T, \RF, \VO, \AD)$. To show that $\pe \in \peset(\Sigma)$, 
we need to show the following: 

\begin{enumerate}
\item $\pe$ satisfies the consistency guarantee $(\rho_x, \rho_x)$ for any object $x \in X$: 
that is, given $x \in X$, then $\circled{\WTr_x} \rcomp \AR \rcomp \circled{\WTr_x} 
\subseteq \VIS$
Let then $x \in X$, and consider two transactions $T,S$ 
be such that $T \xrightarrow{\AR} S$, and $T,S \in \WTr_x$: we show that 
$T \xrightarrow{\VIS} S$. Because $\AR \cap \Id \subseteq \emptyset$, then 
$T \neq S$. Also, it cannot be $S \xrightarrow{\VO(x)} T$: by inequation \Svo\; 
this would imply that $S \xrightarrow{\AR} T$; by inequation \SarTrans\; 
and the assumption that $T \xrightarrow{\AR} S$, this 
would lead to $S \xrightarrow{\AR} S$, contradicting the assumption that $\AR \cap \Id 
= \emptyset$. We have proved that $T,S \in \WTr_x$, $T \neq S$ and $\neg(S \xrightarrow{\VO(x)} T)$: 
since $\VO(x)$ is a total order over the set $\WTr_x$, it must be $T \xrightarrow{\VO(x)} S$. It 
follows from the inequation \Sconflict\; that $T \xrightarrow{\VIS} S$,

\item $\rho(\VIS) \rcomp \AR \rcomp \pi(\VIS) \subseteq \VIS$; this 
inequality is directly enforced by the inequation \Saxiom.
\end{enumerate}

Therefore, $\pe$ is a valid pre-execution that 
satisfies all the consistency guarantees of the x-specification 
$\Sigma = \{(\rho_{\WTr_x}, \rho_{\WTr_x})\}_{x \in X} \cup \{(\rho, \pi)\}$. 
By definition, $\pe \in \peset(\Sigma)$.
%
%
%

Next, we show that $\graphof(\pe)$ is well-defined and equal to $\G$. 
To this end, let $\G' := \graphof(\pe)$. The proof that $\G'$ is a well-defined 
dependency graph is analogous to the one given for abstract 
executions in \citep[][extended version, Proposition 23]{SIanalysis}.

It remains to prove that $\G' = \G$; to this end, it suffices to show that 
for any $x \in \Obj$, $\RF_{\G}(x) = \RF_{\G'}(x)$, and $\VO_{\G}(x) = \VO_{\G'}(x)$.

Let $T, S$ be two entities such that $T \xrightarrow{\RF_{\G}(x)} S$. 
By definition, $S \ni (\RD\;x: n)$, and $T \ni (\WR\; x: n)$ for some 
$n$. Also, let $T' \ni (\WR\; x: n)$ be the entity such 
that $T' \xrightarrow{\RF_{\G'}(x)} S$, which exists because $S \ni (\RD\; x: n)$ 
and $\G'$ is a well-defined dependency graph. 
By definition, $T' = \max_{\AR}(\VIS^{-1}(S) \cap \WTr_x)$, and 
in particular $T' \xrightarrow{\VIS} S$.

Since $T, T' \ni (\WR\; x: n)$, we have that either $T = T'$, $T \xrightarrow{\VO_{\G}(x)} 
T'$, or $T' \xrightarrow{\VO_{\G}(x)} T$. We prove that the first case is the only possible one:  
\begin{itemize}
\item if $T \xrightarrow{\VO_{\G}(x)} T'$, 
then by definition, the edges 
$T \xrightarrow{\RF_{\G}(x)} S$ and $T \xrightarrow{\VO_{\G}(x)} T'$ induce 
the anti-dependency $S \xrightarrow{\AD_{\G}(x)} T'$. However, 
now we have that $T' \ni (\WR\;x: \_)$, $T' \xrightarrow{\VIS} S$ 
and $S \xrightarrow{\AD_{\G}(x)} T'$: by the inequation \Sext, 
it follows that $T' \xrightarrow{\AR} T'$, contradicting the assumption 
that $\AR \cap \Id \subseteq \emptyset$, 
\item if $T' \xrightarrow{\VO_{\G}(x)} T$, then note that by the inequation \Svo\;  
it has to be $T' \xrightarrow{\AR} T$; also, because of the dependency 
$T \xrightarrow{\RF_{\G}(x)} S$ and the inequality \Srf, it has to be 
$T \xrightarrow{\VIS} S$; but this contradicts the assumption that 
$T' = \max_{\AR}(\VIS^{-1}(S) \cap \WTr_x)$. 
\end{itemize}
We are left with the case $T = T'$, from which $T \xrightarrow{\RF_{\G'}(x)} S$ follows. 

Next, suppose that $T' \xrightarrow{\RF_{\G'}(x)} S$. Then $S \ni \RD\; x:n$ for some $n$, 
and because $\G$ is a dependency graph, there exists an entity $T$ such 
that $T \xrightarrow{\RF_{\G}(x)} S$. We can proceed as in the previous case 
to show that $T = T'$, hence $T' \xrightarrow{\RF_{\G}(x)} T$.

Finally, we need to show that $\VO_{\G'}(x) = \VO_{\G}(x)$. First, note 
that if $T \xrightarrow{\VO_{\G}(x)} S$, then $T,S \in \WTr_x$. 
By the inequation \Svo\; we obtain that $T \xrightarrow{\AR} S$, 
so that $T \xrightarrow{\VO_{\G'}(x)} S$ by definition of $\graphof(\pe)$. 

If $T \xrightarrow{\VO_{\G'}(x)} S$, then it has to be the case that $T \xrightarrow{\AR} S$, 
$T,S \in \WTr_x$. Since $\VO_{\G}(x)$ is total over $\WTr_x$, then either 
$T = S, S \xrightarrow{\VO_{\G}(x)} T$ or $T \xrightarrow{\VO_{\G}(x)} S$. However, 
the first case is not possible because it would imply $T \xrightarrow{\AR} T$, contradicting 
the assumption that $\AR \cap \Id \subseteq \emptyset$.  
The second case is not possible either, because by the inequality \Svo\; 
we would get that $S \xrightarrow{\AR} T \xrightarrow{\AR} S$, and by 
the inequality \SarTrans\; $S \xrightarrow{\AR} S$, again contradicting 
the assumption that $\AR \cap \Id \subseteq \emptyset$. We are left with 
$T \xrightarrow{\VO_{\G}(x)} S$, 
as we wanted to prove. 

The fact that $\AD_{\G} =\AD_{\G'}$ follows from the observation that, 
for any object $x \in \Obj$, $\AD_{\G}(x) = \RF_{\G}^{-1}(x) \rcomp \VO_{\G}(x) = 
\RF_{\G'}^{-1}(x) \rcomp \VO_{\G'}(x) = \AD_{\G'}(x)$.
\qed

\subsection{Proof of Proposition \ref{prop:incremental}}
In the following, we let $\G = (\T, \RF, \VO, \AD)$, and 
we assume that $(X_V = \VIS, X_A = \AR, X_N = \mathsf{AntiVIS})$ 
is a solution of $\System_{\Sigma(\G)}$ such that $\AR \cap \Id = \emptyset$. 
Also, we assume that there exist two transactions $T,S$ such that 
$T \neq S$, $\neg(T \xrightarrow{\AR} S)$, and $\neg(S \xrightarrow{\AR} T)$.
The proof of Proposition \ref{prop:incremental} is a direct consequence 
of the following result, which we will prove in this section: 

\begin{proposition}
\label{prop:incremental.all}
Define the following relations: 
\begin{itemize} 
\item $\partial A = \{(T,S)\}$,
\item $\Delta A = \AR? \rcomp \partial A \rcomp \AR?$, 
\item $\AR_{\nu} = \AR \cup \Delta \AR$,
\item $\partial V = \rho(\VIS) \rcomp \Delta A \rcomp \pi(\VIS)$, 
\item $\Delta V = \VIS? \rcomp \partial V \rcomp \VIS?$, 
\item $\VIS_{\nu} = \VIS \cup \Delta V$, 
\item $\mathsf{AntiVIS}_{\nu} = \VIS_{\nu}? \rcomp \AD \rcomp \VIS_{\nu}?$.
\end{itemize}
Then $(X_V = \VIS_{\nu}, X_A = \AR_{\nu}, X_N = \mathsf{AntiVIS}_{\nu})$ 
is a solution to $\System_{\Sigma}(\G)$. Furthermore, it is the smallest solution 
for which the relation corresponding to the unknown $X_A$  
contains the relation $(\AR \cup \partial A)$.
\end{proposition}

Before proving Proposition \ref{prop:incremental.all}, we need to 
prove several technical lemmas.

\begin{lemma}[$\partial$-Cut]
\label{lem:delta.cut}
For any relations 
$R, P, Q \subseteq \T \times \T$ we have that 
$(R \rcomp \partial A \rcomp Q \rcomp \partial A \rcomp P) \subseteq  
(R \rcomp \partial A \rcomp P)$, and 
$(R \rcomp \partial V \rcomp Q \rcomp \partial V \rcomp P) \subseteq 
(R \rcomp \partial V \rcomp P)$. 
\end{lemma}

\myparagraph{Proof}
Recall that $\partial A = \{(T,S)\}$, where $T,S$ are not related by $\AR$. 
That is, whenever $T'' \xrightarrow{\partial A} S''$, for some $T'', S'' \in \T$, 
then $T'' = T, S'' = S$. 
It follows that $(T', S') \in (R \rcomp \partial A \rcomp Q 
\rcomp \partial A \rcomp P)$ if and only if 
$T' \xrightarrow{R} T \xrightarrow{\partial A} S \xrightarrow{Q} 
T \xrightarrow{\partial A} S \xrightarrow{P} S'$. As a consequence,  
$T' \xrightarrow{R} T \xrightarrow{\partial A} S \xrightarrow{P} S'$, 
as we wanted to prove. 

Next, recall that $\partial V = \rho(\VIS) \rcomp \Delta A \rcomp \pi(\VIS)$, 
where $\Delta A = \AR? \rcomp \partial A \rcomp \AR?$. That is, 
$\partial V = \rho(\VIS) \rcomp \AR? \rcomp \partial A \rcomp \AR? \rcomp \pi(\VIS)$.
If we apply the statement above to the relations $R' := (R \rcomp \rho(\VIS) \rcomp \AR?)$, 
$Q' := (\AR? \rcomp \pi(\VIS) \rcomp Q \rcomp \rho(\VIS) \rcomp \AR?)$, 
$P' := (\AR? \rcomp \pi(\VIS) \rcomp P)$, we obtain that 
\begin{gather*}
R \rcomp \partial V \rcomp Q \rcomp \partial V \rcomp P \hfill =\\
(R \rcomp \rho(\VIS) \rcomp \AR?) \rcomp \partial A \rcomp (\AR? \rcomp \pi(\VIS) \rcomp Q \rcomp
 \rho(\VIS) \rcomp \AR?) \rcomp \partial A \rcomp (\AR? \rcomp \pi(\VIS) \rcomp P) \hfill = \\
 R' \rcomp \partial A \rcomp Q' \rcomp \partial A \rcomp P' \hfill\subseteq\\
R' \rcomp \partial A \rcomp P' = \\
R \rcomp \rho(\VIS) \rcomp \AR? \rcomp \partial A \rcomp \AR? \rcomp \pi(\VIS) \rcomp P =\\
R \rcomp \partial V \rcomp P \tag*{\qed}
\end{gather*}

\begin{corollary}
\label{cor:trans}
The relations $\AR_{\nu}$ and $\VIS_{\nu}$ are transitive.
\end{corollary}

\myparagraph{Proof} 
We only show the result for $\AR_{\nu}$. 
The statement relative to $\VIS_{\nu}$ can be proved 
analogously.

It suffices to show that $\AR_{\nu} \rcomp \AR_{\nu} = (\AR \cup \Delta A) 
\rcomp (\AR \cup \Delta A) \subseteq 
(\AR  \cup \Delta \AR)$. By distributivity of 
$\rcomp$ with respect to $\cup$, this reduces to prove the following 
four inclusions: 
\begin{itemize}
\item $(\AR \rcomp \AR) \subseteq (\AR \cup \Delta \AR)$. Recall 
that $(X_V = \VIS, X_A = \AR, X_N = \mathsf{AntiVIS})$ is a solution of 
$\System_{\Sigma}(\G)$, hence by the inequation \SarTrans\; $\AR \rcomp \AR \subseteq \AR$. 
It follows immediately that $\AR \rcomp \AR \subseteq \AR \cup \Delta \AR$.

\item $(\AR \rcomp \Delta A) \subseteq (\AR \cup \Delta A)$:  
recall that $\Delta A = \AR? \rcomp \partial A \rcomp \AR?$. 
Because of the inequation \SarTrans, we have that $\AR \rcomp \AR? \subseteq \AR?$, 
Therefore 
\begin{gather*}
\AR \rcomp \Delta A = \\ 
\AR \rcomp (\AR? \rcomp \partial A \rcomp \AR?) = \\
\AR? \rcomp \partial A \rcomp \AR? = \\
 \Delta A \subseteq \AR \cup \Delta A
\end{gather*}

\item $\Delta A \rcomp \AR \subseteq (\AR \cup \Delta A)$: 
This case is symmetric to the previous one.

\item $(\Delta A \rcomp \Delta A) \subseteq (\AR \cup \Delta A)$: 
\begin{gather*}
\Delta A \rcomp \Delta A = \\ 
(\AR? \rcomp \partial A \rcomp \AR?) \rcomp (\AR? \rcomp \partial A \rcomp \AR?) = \\
\AR? \rcomp \partial A \rcomp (\AR? \rcomp \AR?) \rcomp \partial A \rcomp \AR? \stackrel{\mathsf{Lem.} \eqref{lem:delta.cut}}{\subseteq}\\
\AR? \rcomp \partial A \rcomp \AR? = \\ 
\Delta A \subseteq \AR \cup \Delta \AR
\end{gather*}
where the inequation above has been obtained by applying a $\partial$-cut (Lemma \ref{lem:delta.cut}). \qed
\end{itemize}

\begin{lemma}[$\Delta$-extraction ($\rho$ case)]
\label{lem:rho.shift}
\[
\begin{array}{lcl}
\rho(\VIS_{\nu}) &\subseteq& 
\rho(\VIS) \cup \left( \VIS? \rcomp \rho(\VIS) \rcomp \Delta A \right)\\
\rho(\VIS_\nu) &\subseteq&
\rho(\VIS) \cup \left(\Delta A \rcomp \pi(\VIS) \rcomp \VIS? \right).
\end{array}
\]
We refer to the first inequality as \emph{right $\Delta$-extraction}, 
and to the second inequality as \emph{left $\Delta$-extraction}.
\end{lemma}

\begin{lemma}[$\Delta$-extraction ($\pi$ case)]
\label{lem:pi.shift}
\[
\begin{array}{lcl}
\pi(\VIS_{\nu}) & \subseteq& 
\pi(\VIS) \cup \left(\VIS? \rcomp \rho(\VIS) \rcomp \Delta A \right)\\
\pi(\VIS_{\nu}) &\subseteq& 
\pi(\VIS) \cup \left(\Delta A \rcomp \pi(\VIS) \rcomp \VIS? \right).
\end{array}
\]
\end{lemma}

\myparagraph{Proof}
We only show how to prove the first inequation of Lemma \ref{lem:rho.shift}. 
The proof of the second inequation of Lemma \ref{lem:rho.shift}, 
and the proof of Lemma \ref{lem:pi.shift}, are similar. 

Recall that $\VIS_{\nu} = \VIS \cup \Delta V$. 
By Proposition \ref{prop:specfun.properties}\eqref{prop:specfun.3}, we have that 
\[
\rho(\VIS_{\nu}) = \rho(\VIS) \cup \rho(\Delta V),
\]
by unfolding the definition of specification function to the RHS, and by applying the distributivity of $\cap$ over $\cup$, we get
\[
\rho(\VIS_{\nu}) = (\rho(\T \times \T) \cap \VIS?) \cup (\rho(\T \times \T) \cap \Delta V?) = \rho(\T \times \T) \cap (\VIS? \cup \Delta V?) 
\]
Note that for any relation $R_1, R_2$, $R_1? \cup R_2? = R_1? \cup R_2$, hence we can elide 
the reflexive closure in the term $(\Delta V)?$ of the equality above
\[
\rho(\VIS_{\nu}) = \rho(\T \times \T) \cap (\VIS? \cup \Delta V)
\]
By applying the distributivity of $\cap$ over $\cup$, and then by applying the definition of specification 
function, we get
\begin{multline*}
\rho(\VIS_{\nu}) = (\rho(\T \times \T) \cap \VIS?) \cup (\rho(\T \times \T) \cap \Delta V) =\\
\rho(\VIS) \cup (\rho(\T \times \T) \cap \Delta V) \subseteq \rho(\VIS) \cup (\Delta V)
\end{multline*}

Because $(X_V = \VIS, X_A = \AR, X_N = \mathsf{AntiVIS})$ is a solution of $\System_{\Sigma}(\G)$, by Equation \Svis\; 
we obtain that $\VIS? \subseteq \AR?$. Also, by Proposition \ref{prop:specfun.properties}\eqref{prop:specfun.1} 
we have that $\pi(\VIS) \subseteq \VIS? \subseteq \AR?$. Finally, the inequation \SarTrans 
states that $\AR \rcomp \AR \subseteq \AR$, from which $\AR? \rcomp AR? \subseteq \AR?$ follows. 
By putting all these together, we get
\begin{gather*}
\rho(\VIS_{\nu}) \subseteq \rho(\VIS) \cup \Delta V = \\
\rho(\VIS) \cup (\VIS? \rcomp \rho(\VIS) \rcomp \Delta A \rcomp \pi(\VIS) \rcomp \VIS?) \subseteq \\
\rho(\VIS) \cup (\VIS? \rcomp \rho(\VIS) \rcomp (\AR? \rcomp \partial A \rcomp \AR?) \rcomp \AR? \rcomp \AR?)\\
\rho(\VIS) \cup (\VIS? \rcomp \rho(\VIS) \rcomp (\AR? \rcomp \partial A \rcomp \AR?) ) = 
\rho(\VIS) \cup (\VIS? \rcomp \rho(\VIS) \rcomp \Delta A).
\end{gather*}
as we wanted to prove.
\qed

\begin{lemma}
\label{lem:delta.approx}
\[
\begin{array}{lcl}
\partial \VIS &\subseteq& \Delta A \rcomp \pi(\VIS)\\
\partial \VIS & \subseteq& \rho(\VIS) \rcomp \Delta A
\end{array}
\]
\end{lemma}

\myparagraph{Proof}
Recall that $\partial V = \rho(\VIS) \rcomp \Delta A \rcomp \pi(\VIS)$.
We prove the first inequality as follows:
\[
\begin{array}{lcl}
\Delta V &=& \rho(\VIS) \rcomp \Delta A \rcomp \pi(\VIS)\\
&=& \rho(\VIS) \rcomp \AR? \rcomp \partial A \rcomp \AR? \rcomp \pi(\VIS)\\
&\subseteq& \AR? \rcomp \AR? \rcomp \partial A \rcomp \AR? \rcomp \pi(\VIS)\\
&\subseteq& \AR? \rcomp \partial A \rcomp \AR? \rcomp \pi(\VIS)\\
&=&  \Delta A \rcomp \pi(\VIS)
\end{array}
\]
where we have used the fact that $\rho(\VIS) = \rho(\T \times \T) \cap \VIS? 
\subseteq \VIS? \subseteq \AR?$, because of the definition of specification function 
and because of Inequation \Svis.
\qed

The next step needed to prove Proposition \ref{prop:incremental.all} 
is that of verifying that by substituting $\AR_{\nu}$ for $X_A$, 
$\VIS_{\nu}$ for $X_V$, and $\mathsf{AntiVIS}_{\nu}$ for $X_N$, 
each of the inequations in $\System_{\Sigma}(\G)$ is satisfied. 
The next propositions show that this is 
indeed the case.

\begin{proposition}
\label{prop:incremental.vis}
\[
\VIS_{\nu} \subseteq 
\AR_{\nu}
\]
%
\end{proposition}

\myparagraph{Proof}
Recall that $\VIS_{\nu} = \VIS \cup \Delta V$, 
$\AR_{\nu} = \AR \cup \Delta A$. To prove that 
$\VIS_{\nu} \subseteq \AR_{\nu}$, it suffices to show 
that $\VIS \subseteq (\AR \cup \Delta A)$, and 
$\Delta V \subseteq (\AR \cup \Delta A)$. 

The inequation $\VIS \subseteq \AR \cup \Delta A$ follows immediately the fact that 
$(X_V = \VIS, X_A = \AR, X_N = \mathsf{AntiVIS})$ is 
a solution of $\System_{\Sigma}(\G)$, and from the inequation 
\Svis\; - $\VIS \subseteq \AR$.

It remains to prove that $\Delta V \subseteq \AR \cup \Delta A$. 
In fact, we prove a stronger result, namely $\Delta V \subseteq \Delta A$. This is done as follows: 
\begin{gather*}
\Delta V = \VIS? \rcomp \partial V \rcomp \VIS? = 
\VIS? \rcomp \rho(\VIS) \rcomp \Delta A \rcomp \pi(\VIS) \rcomp \VIS? = \\
\VIS? \rcomp \rho(\VIS) \rcomp \AR? \rcomp \partial A \rcomp \AR? \rcomp \pi(\VIS) \rcomp \VIS? \subseteq\\
\VIS? \rcomp \VIS? \rcomp \AR? \rcomp \partial A \rcomp \AR? \rcomp \VIS? \rcomp \VIS? \stackrel{\SvisTrans}{\subseteq}\\
\VIS? \rcomp \AR? \rcomp \partial A \rcomp \AR? \rcomp \VIS? \stackrel{\Svis}{\subseteq}\\
\AR? \rcomp \AR? \rcomp \partial A \rcomp \AR? \rcomp \AR? \stackrel{\SarTrans}{\subseteq} \\
\AR? \rcomp \partial A \rcomp \AR? = \Delta A. \tag*{\qed}
\end{gather*}
%
%

\begin{proposition}
\label{prop:incremental.axiom}
\[
\rho(\VIS_{\nu}) \rcomp \AR_{\nu} \rcomp \pi(\VIS_{\nu}) \subseteq \VIS_{\nu}.
\]
\end{proposition}

\myparagraph{Proof}
First, we perform a right $\Delta$-extraction (Lemma \ref{lem:rho.shift}) of 
$\rho(\VIS_{\nu})$,
and a left $\Delta$-extraction (Lemma \ref{lem:pi.shift}) of 
$\pi(\VIS_{\nu})$. This gives us the following inequation: 
\begin{multline*} 
\rho(\VIS_{\nu}) \rcomp \AR_{\nu} \rcomp \pi(\VIS_{\nu}) \subseteq \\
(\rho(\VIS) \cup (\VIS? \rcomp \rho(\VIS) \rcomp \Delta A) ) \rcomp \AR_{\nu} 
\rcomp ( \pi(\VIS) \cup (\Delta A \rcomp \pi(\VIS) \rcomp \VIS?)
\end{multline*}
and we rewrite the RHS of the above by applying the distributivity of $\cup$ over ${}\rcomp{}$. 
\[
\begin{array}{lclc}
\rho(\VIS_{\nu}) \rcomp \AR_{\nu} \rcomp \pi(\VIS_{\nu}) &\subseteq& 
\rho(\VIS) \rcomp \AR_{\nu} \rcomp \pi(\VIS) &\cup \\
&& \rho(\VIS) \rcomp \AR_{\nu} \rcomp (\Delta A \rcomp \pi(\VIS) \rcomp \VIS?) & \cup\\
&& \VIS? \rcomp \rho(\VIS) \rcomp \Delta A \rcomp \AR_{\nu} \rcomp \pi(\VIS) & \cup \\
&& \VIS? \rcomp \rho(\VIS) \rcomp \Delta A \rcomp \AR_{\nu} \rcomp \Delta A \rcomp \pi(\VIS) \rcomp \VIS?
\end{array}
\]
We show that each of the components of the union of the RHS of the inequation above is 
included in $\VIS_{\nu}$, from which we get the desired result $\rho(\VIS) \rcomp 
\AR_{\nu} \rcomp \pi(\VIS) \subseteq \VIS_{\nu}$. 

\begin{itemize}
\item $\rho(\VIS) \rcomp \AR_{\nu} \rcomp \pi(\VIS) \subseteq \VIS_{\nu}$.
Recall that $\AR_{\nu} = \AR \cup \Delta A$, 
from which we get that 
\[ 
\rho(\VIS) \rcomp \AR_{\nu} \rcomp \pi(\VIS) = 
(\rho(\VIS) \rcomp \AR \rcomp \pi(\VIS)) \cup 
\rho(\VIS) \rcomp \Delta A \rcomp \pi(\VIS).
\]
We prove that each of the components of the union in the RHS above are included in $\VIS_{\nu}$. 
First, observe that 
\[ 
\rho(\VIS) \rcomp \AR \rcomp \pi(\VIS) \subseteq \VIS \subseteq (\VIS \cup \Delta V) = \VIS_{\nu}
\]
 because of Inequation \Saxiom. Also, we have that  
\[
\rho(\VIS) \rcomp \Delta A \rcomp \pi(\VIS) 
= \partial V \subseteq \VIS? \rcomp \partial V \rcomp \VIS? = \Delta V \subseteq \VIS \cup \Delta V = \VIS_{\nu}
\]
and in this case there is nothing left to prove.
\item $\rho(\VIS) \rcomp \AR_{\nu} \rcomp (\Delta A \rcomp \pi(\VIS) \rcomp \VIS?) \subseteq \VIS_{\nu}$. 
Again, by unfolding the definition of $\AR_{\nu}$ and by applying the distributivity of $\cup$ over 
${}\rcomp{}$, we obtain that 
\[
\begin{array}{lcll}
\rho(\VIS) \rcomp \AR_{\nu} \rcomp \Delta A \rcomp \pi(\VIS) \rcomp \VIS? &=&
\rho(\VIS) \rcomp \AR \rcomp \Delta A \rcomp \pi(\VIS) \rcomp \VIS? &\cup \\
&& \rho(\VIS) \rcomp \Delta A \rcomp \Delta A \rcomp \pi(\VIS) \rcomp \VIS?&
\end{array}
\] 
We prove that each of the components of the union in the RHS above is included in $\VIS_{\nu}$. 
\begin{gather*} 
\rho(\VIS) \rcomp \AR \rcomp \Delta A \rcomp \pi(\VIS) \rcomp \VIS? = \\
\rho(\VIS) \rcomp \AR \rcomp (\AR? \rcomp \partial A \rcomp \AR?) \rcomp \pi(\VIS) \rcomp \VIS? \stackrel{\SarTrans}{\subseteq} \\
\rho(\VIS) \rcomp \AR? \rcomp \partial A \rcomp \AR? \rcomp \pi(\VIS) \rcomp \VIS? = \\ 
\rho(\VIS) \rcomp \Delta A \rcomp \pi(\VIS) \rcomp \VIS? = \\
\partial V \rcomp \VIS? \subseteq \VIS? \rcomp \partial V \rcomp \VIS? =\\ 
\Delta V \subseteq \VIS \cup \Delta V = \VIS_{\nu}\\
\\
\rho(\VIS) \rcomp \Delta A \rcomp \Delta A \rcomp \pi(\VIS) \rcomp \VIS? = \\
\rho(\VIS) \rcomp \AR? \rcomp \partial A \rcomp \AR? \rcomp \AR? \rcomp \partial A \rcomp \AR? \rcomp \pi(\VIS) \rcomp \VIS? 
\stackrel{\mathsf{Lem.} \ref{lem:delta.cut}}{\subseteq}\\
\rho(\VIS) \rcomp \AR? \rcomp \partial A \rcomp \AR? \rcomp \pi(\VIS) \rcomp \VIS? =\\
\rho(\VIS) \rcomp \Delta A \rcomp \pi(\VIS) \rcomp \VIS? = \\ 
\partial V \rcomp \VIS? \subseteq \VIS? \rcomp \partial V \rcomp \VIS? = \Delta V \subseteq \VIS \cup \Delta V = \VIS_{\nu}.
\end{gather*}

\item $\VIS? \rcomp \rho(\VIS) \rcomp \Delta A \rcomp \AR_{\nu} \rcomp \pi(\VIS) \subseteq \VIS_{\nu}$. 
As for the two cases above, we unfold $\AR_{\nu}$ and distribute the resulting union over ${}\rcomp{}$: 
this leads to 
\[
\begin{array}{lcll}
\VIS? \rcomp \rho(\VIS) \rcomp \Delta A \rcomp \AR_{\nu} \rcomp \pi(\VIS) &=& 
\VIS? \rcomp \rho(\VIS) \rcomp \Delta A \rcomp \AR \rcomp \pi(\VIS) &\cup \\
&& \VIS? \rcomp \rho(\VIS) \rcomp \Delta A \rcomp \Delta A \rcomp \pi(\VIS). &
\end{array}
\]
Then we prove that each of the two terms in the union on the RHS above is included 
in $\VIS_{\nu}$: 
\begin{gather*}
\VIS? \rcomp \rho(\VIS) \rcomp \Delta A \rcomp \AR \rcomp \pi(\VIS) = \\
\VIS? \rcomp \rho(\VIS) \rcomp \AR? \rcomp \partial A \rcomp \AR? \rcomp \AR \rcomp \pi(\VIS) \stackrel{\SarTrans}{\subseteq}\\
\VIS? \rcomp \rho(\VIS) \rcomp \AR? \rcomp \partial A \rcomp \AR? \rcomp \pi(\VIS) = \\
\VIS? \rcomp \rho(\VIS) \rcomp \Delta A \rcomp \pi(\VIS) = \\
\VIS? \rcomp \partial V \subseteq\\ 
\VIS? \rcomp \partial V \rcomp \VIS? = \Delta V \subseteq \VIS \cup \Delta V = \VIS_{\nu}\\
~\\
\VIS? \rcomp \rho(\VIS) \rcomp \Delta A \rcomp \Delta A \rcomp \pi(\VIS) = \\
\VIS? \rcomp \rho(\VIS) \rcomp \AR? \rcomp \partial A \rcomp \AR? \rcomp AR? \rcomp \partial A \rcomp \AR? \rcomp \pi(\VIS) 
\stackrel{\mathsf{Lem.} \ref{lem:delta.cut}}{\subseteq}\\
\VIS? \rcomp \rho(\VIS) \rcomp \AR? \rcomp \partial A \rcomp  \AR? \rcomp \pi(\VIS) = \\
\VIS? \rcomp \rho(\VIS) \rcomp \Delta A \rcomp \rho(\VIS) = \\
\VIS? \rcomp \partial V \subseteq \\
 \VIS? \rcomp \partial V \rcomp \VIS? = \Delta V \subseteq \VIS \cup \Delta V = \VIS_{\nu}
\end{gather*}

\item $\VIS? \rcomp \rho(\VIS) \rcomp \Delta A \rcomp \AR_{\nu} \rcomp \Delta A \rcomp \pi(\VIS) \rcomp \VIS?$
in this case we have the following: 
\begin{gather*} 
\VIS? \rcomp \rho(\VIS) \rcomp \Delta A \rcomp \AR_{\nu} \rcomp \Delta A \rcomp \pi(\VIS) \rcomp \VIS? = \\
\VIS? \rcomp \rho(\VIS) \rcomp \AR? \rcomp \partial A \rcomp \AR? \rcomp \AR_{\nu} \rcomp \AR? \rcomp \partial A \rcomp \AR? \rcomp \pi(\VIS) \rcomp \VIS?
\stackrel{\mathsf{Lem.} \ref{lem:delta.cut}}{\subseteq}\\
\VIS? \rcomp \rho(\VIS) \rcomp \AR? \rcomp \partial A \rcomp \AR? \rcomp \pi(\VIS) \rcomp \VIS? = \\
\VIS? \rcomp \partial V \rcomp \VIS? = \Delta V \subseteq \VIS \cup \Delta V = \VIS_{\nu}. \tag*{\qed}
\end{gather*}
\end{itemize}

\begin{proposition}
\label{prop:incremental.coaxiom}
\[
\Big( \pi(\VIS_{\nu}) \rcomp \mathsf{AntiVIS}_{\nu} \rcomp \rho(\VIS_{\nu}) \Big) \setminus \Id \subseteq \AR_{\nu}.
\]
\end{proposition}

\myparagraph{Proof} 
Recall that $\mathsf{AntiVIS}_{\nu} = \VIS_{\nu}? \rcomp \AD \rcomp \VIS_{\nu?}$. 
Thus, we need to prove that 
\[
\Big(\pi(\VIS_{\nu}) \rcomp \VIS_{\nu}? \rcomp \AD \rcomp \VIS_{\nu?} \rcomp \rho(\VIS_{\nu}) \Big) \setminus \Id \subseteq \AR_{\nu}.
\]
We start by performing a $\Delta$-extraction both for the specification functions $\pi$ and $\rho$: 
\begin{gather*}
\Big(\pi(\VIS_{\nu}) \rcomp \VIS_{\nu}? \rcomp \AD \rcomp \VIS_{\nu?} \rcomp \rho(\VIS_{\nu}) \Big) \setminus \Id \subseteq \\
\left( \pi(\VIS) \cup (\Delta A \rcomp \pi(\VIS) \rcomp \VIS?)) \rcomp \VIS_{\nu}? \rcomp \AD \rcomp \VIS_{\nu}? 
\rcomp ((\VIS? \rcomp \rho(\VIS) \rcomp \Delta A) \cup \rho(\VIS)) \right) \setminus \Id =\\
\\
(\pi(\VIS) \rcomp \VIS_{\nu}? \rcomp \AD \rcomp \VIS_{\nu}? \rcomp \rho(\VIS)) \setminus \Id \cup\\
(\Delta A \rcomp \pi(\VIS) \rcomp \VIS? \rcomp \VIS_{\nu}? \rcomp \AD \rcomp \VIS_{\nu}? \rcomp \rho(\VIS)) \setminus \Id \cup\\
(\pi(\VIS) \rcomp \VIS_{\nu}? \rcomp \AD \rcomp \VIS_{\nu}? \rcomp \VIS? \rcomp \rho(\VIS) \rcomp \Delta A) \setminus \Id \cup\\
(\Delta A \rcomp \pi(\VIS) \rcomp \VIS? \rcomp \VIS_{\nu?} \rcomp \AD \rcomp \VIS_{\nu}? \rcomp \VIS? \rcomp \rho(\VIS) \rcomp \Delta A) 
\setminus \Id
\end{gather*}
We prove that each of the four terms of the union above is included in $\AR_{\nu}$. To this end, 
it suffices to prove the following:
\begin{equation}
\label{eq:coaxiom.middle}
(\pi(\VIS) \rcomp \VIS_{\nu}? \rcomp \AD \rcomp \VIS_{\nu}? \rcomp \rho(\VIS)) \setminus \Id \subseteq \AR_{\nu}?
\end{equation}
In fact, if the inequation \eqref{eq:coaxiom.middle} is satisfied, we obtain that 
\begin{itemize}
\item $(\pi(\VIS) \rcomp \VIS_{\nu} \rcomp \AD \rcomp \VIS_{\nu}? \rcomp \rho(\VIS)) \setminus \Id \subseteq \AR_{\nu}$: 
\begin{gather*} 
(\pi(\VIS) \rcomp \VIS_{\nu}? \rcomp \AD \rcomp \VIS_{\nu}? \rcomp \rho(\VIS) ) \setminus \Id \stackrel{\eqref{eq:coaxiom.middle}}{\subseteq}\\ 
\AR_{\nu}? \setminus \Id = (\AR_{\nu} \cup \Id) \setminus \Id = \AR_{\nu} \setminus \Id \subseteq \AR_{\nu}, 
\end{gather*}
\item $(\Delta A \rcomp \pi(\VIS) \rcomp \VIS? \rcomp \VIS_{\nu}? \rcomp \AD \rcomp \VIS_{\nu}? \rcomp \rho(\VIS)) \setminus \Id \subseteq \AR_{\nu}$: 
\begin{gather*}
(\Delta A \rcomp \pi(\VIS) \rcomp \VIS? \rcomp \VIS_{\nu}? \rcomp \AD \rcomp \VIS_{\nu}? \rcomp \rho(\VIS)) \setminus \Id \subseteq\\
(\Delta A \rcomp \pi(\VIS) \rcomp \VIS_{\nu}? \rcomp \VIS_{\nu}? \rcomp \AD \rcomp \VIS_{\nu}? \rcomp \rho(\VIS)) \setminus \Id 
\stackrel{\mathsf{Cor.} \eqref{cor:trans}}{\subseteq}\\ 
(\Delta A \rcomp \pi(\VIS) \rcomp \VIS_{\nu}? \rcomp \AD \rcomp \VIS_{\nu}? \rcomp \rho(\VIS)) \setminus \Id
\stackrel{\eqref{eq:coaxiom.middle}}{\subseteq}\\
(\Delta A \rcomp \AR_{\nu}?) \setminus \Id = \\
(\Delta A \rcomp (\AR \cup \Delta A)?) \setminus \Id = \\
(\Delta A \rcomp (\AR? \cup \Delta A)) \setminus \Id = \\
(\Delta A \rcomp \AR?) \setminus \Id \cup (\Delta A \rcomp \Delta A) \setminus \Id = \\
(\AR? \rcomp \partial A \rcomp \AR? \rcomp \AR?) \setminus \Id \cup (\AR? \rcomp \partial A \rcomp \AR? \rcomp \AR? \rcomp \partial A \rcomp \AR?) \setminus \Id 
\stackrel{\mathsf{Lem.} \ref{lem:delta.cut}}{\subseteq}\\ 
(\AR? \rcomp \partial A \rcomp \AR? \rcomp \AR?) \setminus \Id \cup (\AR? \rcomp \partial A \rcomp \AR?) \setminus \Id \stackrel{\SarTrans}{\subseteq}\\
(\AR? \rcomp \partial A \rcomp \AR?) \setminus \Id = (\Delta A) \setminus \Id \subseteq \AR \cup \Delta A = \AR_{\nu}
\end{gather*}

\item $(\pi(\VIS) \rcomp \VIS_{\nu}? \rcomp \AD \rcomp \VIS_{\nu}? \rcomp \VIS? \rcomp \rho(\VIS) \rcomp \Delta A) \setminus \Id \subseteq \AR_{\nu}$: 
\begin{gather*}
(\pi(\VIS) \rcomp \VIS_{\nu}? \rcomp \AD \rcomp \VIS_{\nu}? \rcomp \VIS? \rcomp \rho(\VIS) \rcomp \Delta A) \setminus \Id \subseteq \\
(\pi(\VIS) \rcomp \VIS_{\nu}? \rcomp \AD \rcomp \VIS_{\nu}? \rcomp \VIS_{\nu}? \rcomp \rho(\VIS) \rcomp \Delta A) \setminus \Id 
\stackrel{\mathsf{Cor.} \eqref{cor:trans}}{\subseteq} \\ 
(\pi(\VIS) \rcomp \VIS_{\nu}? \rcomp \AD \rcomp \VIS_{\nu}? \rcomp \rho(\VIS) \rcomp \Delta A) \setminus \Id \stackrel{\eqref{eq:coaxiom.middle}}{\subseteq} 
(\AR_{\nu}? \rcomp \Delta A) \setminus \Id = \\
((\AR \cup \Delta A)? \rcomp \Delta A) \setminus \Id = \\
((\AR? \cup \Delta A) \rcomp \Delta A) \setminus \Id = \\
(\AR? \rcomp \Delta A) \setminus \Id \cup (\Delta A \rcomp \Delta A) \setminus \Id \subseteq\\ 
(\Delta A) \setminus \Id \subseteq \AR \cup \Delta A = \AR_{\nu}
\end{gather*}
\item $(\Delta A \rcomp \pi(\VIS) \rcomp \VIS? \rcomp \VIS_{\nu}? \rcomp \AD \rcomp \VIS_{\nu}? \rcomp \VIS? \rcomp \rho(\VIS) \rcomp \Delta A) \setminus \Id 
\subseteq \AR_{\nu}$: here it suffices to apply a $\partial$-cut (Lemma \ref{lem:delta.cut}) to obtain the result: 
\begin{gather*}
(\Delta A \rcomp \pi(\VIS) \rcomp \VIS? \rcomp \VIS_{\nu}? \rcomp \AD \rcomp \VIS_{\nu}? \rcomp \VIS? \rcomp \rho(\VIS) \rcomp \Delta A) \setminus \Id  
\subseteq \\
\Delta A \rcomp \pi(\VIS) \rcomp \VIS? \rcomp \VIS_{\nu}? \rcomp \AD \rcomp \VIS_{\nu}? \rcomp \VIS? \rcomp \rho(\VIS) \rcomp \Delta A = \\
\AR? \rcomp \partial A \rcomp \AR? \rcomp \pi(\VIS) \rcomp \VIS? \rcomp \VIS_{\nu}? \rcomp \AD \rcomp \VIS_{\nu}? \rcomp \VIS? \rcomp \rho(\VIS) \rcomp
\AR? \rcomp \partial A \rcomp \AR? \stackrel{\mathsf{Lem.} \eqref{lem:delta.cut}}{\subseteq}\\
\AR? \rcomp \partial A \rcomp \AR? = \Delta A \subseteq \AR \cup \Delta A = \AR_{\nu}
\end{gather*}
\end{itemize}

Let then prove the inequality \eqref{eq:coaxiom.middle}: we have that 
\begin{gather*} 
\pi(\VIS) \rcomp \VIS_{\nu}? \rcomp \AD \rcomp \VIS_{\nu}? \rcomp \rho(\VIS) = \\
\pi(\VIS) \rcomp (\VIS \cup \Delta V)? \rcomp \AD \rcomp (\VIS \cup \Delta V)? \rcomp \rho(\VIS) = \\
~\\
\pi(\VIS) \rcomp \VIS? \rcomp \AD \rcomp \VIS? \rcomp \rho(\VIS) \cup \\
\pi(\VIS) \rcomp \Delta V \rcomp \AD \rcomp \VIS? \rcomp \rho(\VIS) \cup \\
\pi(\VIS) \rcomp \VIS? \rcomp \AD \rcomp \Delta V \rcomp \rho(\VIS) \cup \\
\pi(\VIS) \rcomp \Delta V \rcomp \AD \rcomp \rho(\VIS) \rcomp \rho(\VIS)
\end{gather*}
We prove that each of the terms in the union above is included in $\AR_{\nu}?$. 
\begin{itemize}
\Item
\begin{equation}
\label{eq:coaxiom.basic}
\hspace{-9pt}\pi(\VIS) \rcomp \VIS? \rcomp \AD \rcomp \VIS? \rcomp \rho(\VIS) \subseteq \AR_{\nu}?:
\end{equation}
\begin{gather*}
\pi(\VIS) \rcomp \VIS? \rcomp \AD \rcomp \VIS? \rcomp \rho(\VIS)\subseteq \\
((\pi(\VIS) \rcomp \VIS? \rcomp \AD \rcomp \VIS? \rcomp \rho(\VIS)) \setminus \Id) \cup \Id \stackrel{\Sad,\SavisL,\SavisR}{\subseteq}\\
((\pi(\VIS) \rcomp \mathsf{AntiVIS} \rcomp \rho(\VIS)) \setminus \Id) \cup \Id \stackrel{\Scoaxiom}{\subseteq}\\
\AR \cup \Id \subseteq \AR \cup \Delta A \cup \Id = \AR_{\nu} \cup \Id = \AR_{\nu}?
\end{gather*}
\item $\pi(\VIS) \rcomp \Delta V \rcomp \AD \rcomp \VIS? \rcomp \rho(\VIS) \subseteq \AR_{\nu}?$:
\begin{gather*}
\pi(\VIS) \rcomp \Delta V \rcomp \AD \rcomp \VIS? \rcomp \rho(\VIS) = \\
\pi(\VIS) \rcomp \VIS? \rcomp \partial V \rcomp \VIS? \rcomp \AD \rcomp \VIS? \rcomp \rho(\VIS) = \\
\pi(\VIS) \rcomp \VIS? \rcomp \rho(\VIS) \rcomp \Delta A \rcomp \pi(\VIS) \rcomp \VIS? \rcomp \AD \rcomp \VIS? \rcomp \rho(\VIS) \subseteq \\
\VIS? \rcomp \VIS? \rcomp \VIS? \rcomp \Delta A \rcomp \pi(\VIS) \rcomp \VIS? \rcomp \AD \rcomp \VIS? \rcomp \rho(\VIS) \stackrel{\Svis,\SarTrans}{\subseteq}\\
\AR? \rcomp \Delta A \rcomp \pi(\VIS) \rcomp \VIS? \rcomp \AD \rcomp \VIS? \rcomp \rho(\VIS) \stackrel{\eqref{eq:coaxiom.basic}}{\subseteq} \\
\AR? \rcomp \Delta A \rcomp \AR_{\nu}? = \\
\AR? \rcomp \Delta A \rcomp (\AR? \cup \Delta A) = \\
(\AR? \rcomp \Delta A \rcomp \AR?) \cup (\AR? \rcomp \Delta A \rcomp \Delta A) = \\
(\AR? \rcomp \AR? \rcomp \partial A \rcomp \AR?) \cup (\AR? \rcomp \AR? \rcomp \partial A \rcomp \AR? \rcomp \AR? \rcomp \partial A \rcomp \AR?) 
\stackrel{\SarTrans}{\subseteq}\\
(\AR? \rcomp \partial A \rcomp \AR?) \cup (\AR? \rcomp \partial A \rcomp \AR? \rcomp \partial A \rcomp \AR?) \stackrel{\mathsf{Lem.} \eqref{lem:delta.cut}}{\subseteq}\\
(\AR? \rcomp \partial A \rcomp \AR?) = \Delta A \subseteq \AR \cup \Delta A = \AR_{\nu} \subseteq \AR_{\nu}?
\end{gather*}
\item 
$\pi(\VIS) \rcomp \VIS? \rcomp \AD \rcomp \Delta V \rcomp \rho(\VIS) \subseteq \AR_{\nu}?$: 
\begin{gather*}
\pi(\VIS) \rcomp \VIS? \rcomp \AD \rcomp \Delta V \rcomp \rho(\VIS) = \\
\pi(\VIS) \rcomp \VIS? \rcomp \AD \rcomp \VIS? \rcomp \partial V \rcomp \VIS? \rcomp \rho(\VIS) = \\
\pi(\VIS) \rcomp \VIS? \rcomp \AD \rcomp \VIS? \rcomp \rho(\VIS) \rcomp \Delta A \rcomp \rho(\VIS) \rcomp \VIS? \rcomp \rho(\VIS) \subseteq\\
\pi(\VIS) \rcomp \VIS? \rcomp \AD \rcomp \VIS? \rcomp \rho(\VIS) \rcomp \Delta A \rcomp \VIS? \rcomp \VIS? \rcomp \VIS? \stackrel{\Svis,\SarTrans}{\subseteq}\\
\pi(\VIS) \rcomp \VIS? \rcomp \AD \rcomp \VIS? \rcomp \rho(\VIS) \rcomp \Delta A \rcomp \AR? \stackrel{\eqref{eq:coaxiom.basic}}{\subseteq}\\
\AR_{\nu}? \rcomp \Delta A \rcomp \AR? = \\
(\AR? \rcomp \Delta A) \rcomp \Delta A \rcomp \AR? = \\
(\AR? \rcomp \Delta A \rcomp \AR?) \cup (\Delta A \rcomp \Delta A \rcomp \AR?) \stackrel{\mathsf{Lem.} \eqref{lem:delta.cut}}{\subseteq}\\
(\AR? \rcomp \Delta A \rcomp \AR?) \cup (\Delta A \rcomp \AR?) = \\
(\AR? \rcomp \Delta A \rcomp \AR?) = \Delta A \subseteq \AR \cup \Delta A = \AR_{\nu} \subseteq \AR_{\nu}?
\end{gather*}
\item 
$\pi(\VIS) \rcomp \Delta V \rcomp \AD \rcomp \Delta V \rcomp \rho(\VIS) \subseteq \AR_{\nu}?$:
\begin{gather*}
\pi(\VIS) \rcomp \Delta V \rcomp \AD \rcomp \Delta V \rcomp \rho(\VIS) \subseteq \\
\VIS? \rcomp \Delta V \rcomp \AD \rcomp \Delta V \rcomp \VIS? = \\
\VIS? \rcomp \VIS? \rcomp \partial V \rcomp \VIS? \rcomp \AD \rcomp \VIS? \rcomp \partial V \rcomp \VIS? \rcomp \VIS? 
\stackrel{\mathsf{Lem.} \eqref{lem:delta.cut}}{\subseteq}\\
\VIS? \rcomp \VIS? \rcomp \partial V \rcomp \VIS? \rcomp \VIS? \stackrel{\SvisTrans}{\subseteq}\\
\VIS? \rcomp \partial V \rcomp \VIS? \stackrel{\Svis}{\subseteq}\\
\VIS? \rcomp \VIS? \rcomp \Delta A \rcomp \VIS? \rcomp \VIS? \stackrel{\SvisTrans, \Svis}{\subseteq}\\
\AR? \rcomp \Delta A \rcomp \AR? = \Delta A \subseteq \AR \cup \Delta A = \AR_{\nu} \subseteq \AR_{\nu}?. \tag*{\qed}
\end{gather*}
\end{itemize}

\begin{proposition}
\label{prop:incremental.ext}
\[
\bigcup_{x \in \Obj} \circled{\WTr_x} \rcomp \VIS_{\nu} \rcomp \AD(x) \subseteq \AR_{\nu}.
\]
\end{proposition}

\myparagraph{Proof}
Let $T',U,S'$ be such that $T' \in \WTr_x$, $T' \xrightarrow{\VIS_{\nu}} U \xrightarrow{\AD(x)} S'$ 
for some object $x \in \Obj$. We need to show that $T' \xrightarrow{\AR_{\nu}} S'$. 
By definition, $\VIS_{\nu} = \VIS \cup \Delta V$. Thus,  
$T' \xrightarrow{\VIS} U$ or $T' \xrightarrow{\Delta V} U$.
If $T' \xrightarrow{\VIS} U$, then 
$T' \xrightarrow{\VIS} U \xrightarrow{\AD(x)} S'$ and $T' \in \WTr_x$. By the inequation \Sext\;  
we have that $T' \xrightarrow{\AR} S'$, which implies the desired
$T' \xrightarrow{\AR_{\nu}} S'$.

Suppose then that $T' \xrightarrow{\Delta V} U$.
By unfolding the definition of $\Delta V$, we have that 
\[
T' \xrightarrow{\VIS? \rcomp \rho(\VIS)} T'' \xrightarrow{\AR?} T 
\xrightarrow{\partial A} S \xrightarrow{\AR?} U' \xrightarrow{\pi(\VIS) \rcomp \VIS?} U \xrightarrow{\AD(x)} S'.
\]
Recall that by definition of $\partial A$, the transactions $T$ and $S$ are not related by $\AR$.
Note that, since $U \xrightarrow{\AD(x)} S'$, then $U \in \RTr_x, S' \in \WTr_x$. 
Recall that $\VO(x)$ is a total order over $\WTr_x$. Therefore, we have 
three possible cases: $T' \xrightarrow{\VO(x)} S'$, $T' = S'$ or  $T' \xrightarrow{\VO(x)} S'$. These 
cases are analysed separately.

\begin{itemize}
\item $T' \xrightarrow{\VO(x)} S'$: by the inequality \Svo we have that $T' \xrightarrow{\AR} S'$. Thus, $T' \xrightarrow{\AR_{\nu}} S'$.

\item $T' = S'$: this case is not possible. We first prove
that $U' \neq T''$. Suppose $U' = T''$. Then $S \xrightarrow{\AR?} 
U' = T'' \xrightarrow{\AR?} T$, that is $S \xrightarrow{\AR?} T$. 
But by hypothesis, $T$ and $S$ are not related by $\AR$, 
hence we get a contradiction.

Let then $U' \neq T''$. Since we have 
\[
U' \xrightarrow{\pi(\VIS) \rcomp \VIS?} U \xrightarrow{\AD(x)} S' = T' \xrightarrow{\VIS? \rcomp \rho(\VIS)} T''
\]
we have that $U' \xrightarrow{\AR} T''$ by the inequality \Scoaxiom.
Thus, $S \xrightarrow{\AR?} U' \xrightarrow{\AR} T'' \xrightarrow{\AR?} T$, 
or equivalently $S \xrightarrow{\AR} T$. Again, this contradict the 
assumption that $S$ and $T$ are not related by $\AR$.

\item $S' \xrightarrow{\VO(x)} T'$: this case is also not possible. 
Recall that $U \xrightarrow{\AD(x)} S'$; 
that is, there exists an entity $U''$ such that $U'' \xrightarrow{\RF(x)} U$, 
$U'' \xrightarrow{\VO(x)} S'$. By the transitivity of $\VO(x)$, we have 
that $U'' \xrightarrow{\VO(x)} T'$. Thus, $U \xrightarrow{\AD(x)} T'$. 
We can proceed as in the case above to show that this implies 
$S \xrightarrow{\AR} T$, contradicting the assumption that 
$T$ and $S$ are not related by $\AR$. \qed
\end{itemize}

Finally, we prove the following:
\begin{proposition}
\label{prop:underapproximation}
The triple $(X_V = \VIS_{\nu}, X_A = \AR_{\nu}, X_N = \mathsf{AntiVIS}_{\nu})$ 
is included in the least 
solution to $\System_{\Sigma}(\G)$ for which the relation corresponding to the unknown $X_A$ includes 
the relation $\AR \cup \partial A$. 
\end{proposition}

\noindent\myparagraph{\textbf{Proof.}} 
Let $(X_V = \VIS', X_A = \AR', X_N = \mathsf{AntiVIS}')$ be a solution 
to $\System_{\Sigma}(\G)$ such that $(\AR \cup \partial A) \subseteq \AR'$. 
We need to show that $\AR_{\nu} \subseteq \AR'$, $\VIS_{\nu} \subseteq \VIS'$, 
and $\mathsf{AntiVis}_{\nu} \subseteq \mathsf{AntiVIS}'$.
\begin{itemize}
\item $\AR_{\nu} \subseteq \AR'$: 
note that we have that 
\[
\Delta A = \AR? \rcomp \partial A \rcomp \AR? \subseteq
\AR' \rcomp \AR' \rcomp \AR' \stackrel{\SarTrans}{\subseteq} \AR'
\]
from which it follows that $\AR_{\nu} = \AR \cup \Delta \AR \subseteq ( \AR' \cup \AR') = \AR'$. 

\item $\VIS_{\nu} \subseteq \VIS'$: 
Observe that for any solution $(X_V = \VIS'', X_A = \AR'', X_N = \mathsf{AntiVIS}'')$ of 
$\System_{\Sigma}(\G)$, the relation $\VIS'$ is determined uniquely by $\AR''$: 
specifically, $\VIS'' = \mu V. \mathcal{F}(V, \AR'')$, where 
\[
\mathcal{F}(V, \AR'') = \left( \RF \cup \left( \bigcup_{\{ x \mid (\rho_x, \rho_x) \in \Sigma\}} \VO(x) \right)
\cup (\rho(V) \rcomp \AR'' \rcomp \pi(V)) \right)^{+}
\]
the functional $\mathcal{F}$ is monotone in its second argument, which means 
that the inequation $\AR_{\nu} \subseteq \AR'$ also implies that 
$\VIS_{\nu} \subseteq \VIS'$. 

\item $\mathsf{AntiVIS}_{\nu} \subseteq \VIS'$. 
Observe that, for any solution $(X_V = \VIS'', X_A = \AR'', X_N = \mathsf{AntiVIS}'')$, 
the relation $\mathsf{AntiVIS}''$ is determined uniquely by $\VIS''$. Specifically, 
we have that $\mathsf{AntiVIS}'' = \mathcal{F}(\VIS'')$, where 
$\mathcal{F}(\VIS'') = \VIS''? \rcomp \AD \rcomp \VIS''?$. 
The functional $\mathcal{F}$ is monotone, from which it follows that the inequation 
$\VIS_{\nu} \subseteq \VIS'$, proved above,  implies that $\mathsf{AntiVis}_{\nu} 
\subseteq \mathsf{AntiVIS}'$.
\end{itemize}
%
\qed

\noindent\myparagraph{Proof of Proposition \ref{prop:incremental.all}}
We need to show that $(X_V = \VIS_{\nu}, X_A = \AR_{\nu}, X_N = \mathsf{AntiVIS}_{\nu})$ 
is a solution of $\System_{\G}(\Sigma)$. By Proposition \ref{prop:underapproximation}, 
it follows that it is the smallest solution for which the relation corresponding to the unknown 
$X_A$ includes $\AR \cup \partial A$.
 
Obviously we have that $\RF \subseteq \VIS \subseteq \VIS_{\nu}$, 
and 
$\bigcup \{\VO(x) \mid (\rho_x, \rho_x) \in \Sigma \} \subseteq \VIS \subseteq \VIS_{\nu}$: the inequations \Srf 
and \Sconflict are satisfied.
The validity of inequation \SvisTrans\; follows from Corollary \ref{cor:trans}. 
The inequation 
\Saxiom\; is also satisfied, as we have proved in Proposition \ref{prop:incremental.axiom}. 
 
The inequality \Svo is satisfied because $\VO \subseteq \AR \subseteq \AR_{\nu}$, 
and the inequation \Svis\; has been proved in Proposition \ref{prop:incremental.vis}. 
The validity of the inequation \SarTrans\; also follows from Corollary \ref{cor:trans}. 
The inequation \Scoaxiom\; and \Sext\; are satisfied, 
as we have proved in propositions \ref{prop:incremental.coaxiom} and \ref{prop:incremental.ext}. 

Finally, the inequation \Sad\; is satisfied because $\AD \subseteq \VIS_{\nu}? \rcomp \AD \rcomp \VIS_{\nu}? = 
\mathsf{AntiVIS}_{\nu}$; the inequation \SavisL\; is satisfied because 
$\VIS_{\nu} \rcomp \mathsf{AntiVIS}_{\nu} = \VIS_{\nu} \rcomp \VIS_{\nu}? \rcomp \AD \rcomp \VIS_{\nu}? 
\subseteq \VIS_{\nu}? \rcomp \AD \rcomp \VIS_{\nu}? = \mathsf{AntiVIS}_{\nu}$ (recall that $\VIS_{\nu}$ is 
transitive by Corollary \ref{cor:trans}), and similarly we can prove that the inequation \SavisR\; is also satisfied.
\qed

\subsection{Proof of Theorem \ref{thm:depgraphs}}
Throughout this section we let $\G = (\T, \RF, \VO, \AD)$. 

\subsubsection{Proof of Theorem \ref{thm:depgraphs}\eqref{thm:ser}} 
Recall that $\Sigma_{\SER} = \{(\rho_S, \rho_S)\}$, where $\rho_S(R) = \Id$. 
The instantiation of inequations $\Saxiom$\; and $\Scoaxiom$, in $\System_{\Sigma_{\SER}}(\G)$ 
gives rise to the inequations $X_A \subseteq X_V$ and $X_N \setminus \Id \subseteq X_A$.

Let $\VIS = \AR = \mathsf{AntiVIS} = (\RF \cup \VO \cup \AD)^{+}$.
We prove that $(X_V = \VIS, X_A = \AR, X_{N} = \mathsf{AntiVIS})$ 
is a solution to $\System_{\Sigma_{\SER}}(\G)$: to this end, 
we show that by substituting each of the unknowns for the relation $(\RF \cup \VO \cup \AD)^{+}$ in $\System_{\Sigma_{\SER}}(\G)$, 
then each of the inequations of such a system is satisfied.
Clearly $\RF \subseteq \VIS$, hence equation \Srf\; is satisfied. 
Because there is no consistency guarantee of the form $(\rho_x, \rho_x) \in \Sigma_{\SER}$, 
the inequation \Sconflict\; is trivially satisfied.
Inequation \SvisTrans\; is also satisfied. 
$\VIS \rcomp \VIS = (\RF \cup \VO \cup \AD)^{+} \rcomp (\RF \cup \VO \cup \AD)^{+} = 
(\RF \cup \VO \cup \AD)^{+} = \VIS$. Inequation \Saxiom\; requires that 
$\AR \subseteq \VIS$: this is also satisfied, as $\AR = (\RF \cup \VO \cup \AD)^{+} = \VIS$. 

Inequation \Svo\; is trivially satisfied: $\VO \subseteq (\RF \cup \VO \cup \AD)^{+} = \AR$. 
Inequation \Svis\; is also satisfied: $\VIS = (\PO \cup \RF \cup \AD)^{+} = \AR$, hence 
$\VIS \subseteq \AR$.  
Inequation \Scoaxiom\; is satisfied as well: $\mathsf{AntiVIS} \setminus \Id = (\RF \cup \VO \cup \AD)^{+} \setminus \Id 
\subseteq (\RF \cup \VO \cup \AD)^{+} = \AR$. 
Inequation \Sext\; is also satisfied: $\bigcup_{x \in \Obj} \circled{\WTr_x} \rcomp \VIS \rcomp \AD(x) \subseteq 
\VIS \rcomp \AD = (\RF \cup \VO \cup \AD)^{+} \rcomp \AD \subseteq (\RF \cup \VO \cup \AD)^{+} = \AR$.

Inequation \Sad\; is obviously satisfied, as $\AD \subseteq (\RF \cup \VO \cup \AD)^{+} = \mathsf{AntiVIS}$. 
For inequation \SavisL\;, note that $\VIS \rcomp \mathsf{AntiVIS} = (\RF \cup \VO \cup \AD)^{+} \rcomp 
(\RF \cup \VO \cup \AD)^{+} \subseteq (\RF \cup \VO \cup \AD)^{+} = \mathsf{AntiVIS}$, and it can be 
shown that Inequation \SavisR\; is satisfied in a similar way.

The proof that the solution $(X_V = \VIS, X_A = \AR, X_N = \mathsf{AntiVIS})$ is the 
smallest solution of $\System_{\Sigma_{\SER}}(\G)$ can be obtained as in the proof 
of Theorem \ref{thm:ser.acyclic}. 
\qed

\subsubsection{Proof of Theorem \ref{thm:depgraphs}\eqref{thm:si}.}
Recall that $\Sigma_{\SI} = \{(\rho_x, \rho_x)\}_{x \in \Obj} \cup \{(\rho_{\Id}, \rho_{\SI})\}$, 
where $\rho_x(R) = \circled{\WTr_x}$, $\rho_{\SI}(R) = R \setminus \Id$. 
By instantiating inequation \Sconflict\; to $\Sigma_{\SI}$ we obtain $\VO \subseteq X_V$, 
while by instantiating inequations \Saxiom\; and \Scoaxiom\; to the consistency guarantee $(\rho_{\Id}, \rho_{\SI})$, 
we obtain $X_A \rcomp (X_V \setminus \Id) \subseteq X_V$, and 
$((X_V \setminus \Id) \rcomp X_N ) \setminus \Id \subseteq X_A$. 

Let $\AR = ((\RF \cup \VO) \rcomp \AD?)^{+}$, $\VIS = \AR? \rcomp (\RF \cup \VO)$, 
$\mathsf{AntiVIS} = \VIS? \rcomp \AD \rcomp \VIS?$. Then $(X_V = \VIS, X_A = \AR, 
X_N = \mathsf{AntiVIS})$ is a solution of $\System_{\Sigma_{\SI}}(\G)$. 
We can prove that it is the smallest such solution in the same way as in Theorem 
\ref{thm:SI.acyclic}. 

We need to show that, by substituting $\VIS, \AR, \mathsf{AntiVIS}$ for $X_V, X_A, X_N$  
respectively, in \newline $\System_{\Sigma_{\SI}}(\G)$, all the inequations are satisfied. Here 
we give the details only for the most important of them. A full proof of this statement 
can be found in \cite{SIanalysis}.

\begin{itemize}
\item $\AR \rcomp (\VIS \setminus \Id) \subseteq \VIS$: 
\begin{gather*}
\AR \rcomp (\VIS \setminus \Id) \subseteq \AR \rcomp \VIS = \\
((\RF \cup \VO) \rcomp \AD?)^{+} \rcomp ((\RF \cup \VO) \rcomp \AD?)^{\ast} \rcomp (\RF \cup \VO) \subseteq\\ 
((\RF \cup \VO) \rcomp \AD?)^{\ast} \rcomp (\RF \cup \VO) = \AR? \rcomp (\RF \cup \VO) = \VIS
\end{gather*}
\item $((\VIS \setminus \Id) \rcomp \mathsf{AntiVIS}) \setminus \Id \subseteq \AR$: 
\begin{gather*} 
((\VIS \setminus \Id) \rcomp \mathsf{AntiVIS}) \setminus \Id \subseteq \\
\VIS \rcomp \mathsf{AntiVIS} = \VIS \rcomp \VIS? \rcomp \AD \rcomp \VIS? = \\
\VIS \rcomp \AD \rcomp \VIS? = \\
(((\RF \cup \VO) \rcomp \AD?)^{\ast} \rcomp (\RF \cup \VO)) \rcomp \AD \rcomp \VIS? \subseteq \\
((\RF \cup \VO) \rcomp \AD?)^{+} \rcomp \VIS? = \\
\AR \cup \VIS? \subseteq \AR
\end{gather*}
where we have used the fact that $\AR \rcomp \VIS \subseteq \VIS$, which we have proved previously.\qed
\end{itemize}

\subsubsection{Proof of Theorem \ref{thm:depgraphs}\eqref{thm:psi}.}

%
\begin{proposition}
\label{prop:psi.correct}
Let $\VIS = (\RF \cup \VO)^{+}$, $\AR = \VIS \cup \bigcup_{x \in \Obj} \left( \circled{\WTr_x} \rcomp \VIS? \rcomp \AD(x) \right)^{+}$, 
$\mathsf{AntiVIS} = \VIS? \rcomp \AD \rcomp \VIS?$. If $\AR$ is irreflexive, then 
$(X_V = \VIS, X_A = \AR, X_N = \mathsf{AntiVIS})$ is a solution of $\System_{\Sigma_{\PSI}}(\G)$.
Furthermore, it is the smallest such solution.
\end{proposition}
%

\myparagraph{Proof}
Recall that $\Sigma_{\PSI} = \{(\rho_x, \rho_x)\}_{x \in \Obj}$. 
Therefore, the system of inequations \newline $\System_{\Sigma_{\PSI}}(\G)$ 
does not contain inequations \Saxiom\; and \Scoaxiom, and inequation \Sconflict\; 
is instantiated to $\VO \subseteq \VIS$.
We prove that, under the assumption that $\AR$ is irreflexive, 
the triple $(X_V = \VIS, X_A = \AR, X_N = \mathsf{AntiVIS})$ is 
a solution of $\System_{\Sigma_{\PSI}}(\G)$ by 
showing that, by substituting $\VIS, \AR$ and $\mathsf{AntiVIS}$ for 
$X_V, X_A$ and $X_N$ in $\System_{\Sigma_\PSI}(\G)$, respectively, 
all the inequations are satisfied.
The fact that the triple $(X_V = \VIS, X_A = \AR, X_N = \mathsf{AntiVIS})$ 
is the smallest solution of $\System_{\Sigma_{\PSI}}(\G)$
can be proved in the same way as in the proof of Theorem \ref{thm:PSI.acyclic}. 

First, we observe that if $\AR$ is irreflexive, then 
for any $x \in \Obj$, $\circled{\WTr_x} \rcomp \VIS? \rcomp \AD(x) \subseteq \VO(x)$. 
To see why this is true, recall that $\VO(x)$ is a strict, total order over $\WTr_x$. 
Suppose that $T \ni \WR\; x: \_$, $T \xrightarrow{\VIS?} S' \xrightarrow{\AD(x)} S$. 
Note that, since $\circled{\WTr_x} \rcomp \VIS? \rcomp \AD(x) \subseteq \AR$, 
and we are assuming that the latter is irreflexive, it cannot be $T = S$. 
By definition of $\AD(x)$, $S \ni \WR\; x: \_$. Therefore, either $T \xrightarrow{\VO(x)} S$, 
or $S \xrightarrow{\VO(x)} T$. However, if it were $S \xrightarrow{\VO(x)} T$, we would 
have $S \ni \; \WTr_x$, $S \xrightarrow{\VO(x)} T \xrightarrow{\VIS?} S' \xrightarrow{\AD(x)} S$: 
because $\VIS = (\RF \cup \VO)^{+}$, $\VO(x) \rcomp \VIS? \subseteq \VIS?$, hence 
$S \xrightarrow{\VIS?} S' \xrightarrow{\AD(x)} S$, and because $S \ni \WR\;x:\_$, it would 
follow that $S \xrightarrow{\AR} S$, contradicting the hypothesis that $\AR$ is irreflexive. 
Therefore, it must be $T \xrightarrow{\VO(x)} S$. 

We have proved that, if $\AR$ is irreflexive, then for any $x \in \Obj$, $\circled{\WTr_x} 
\rcomp \VIS? \rcomp \AD(x) \subseteq \VO$. An immediate consequence of this fact 
is the following: 
\begin{equation}
\label{eq:ar.psi}
\bigcup_{x \in \Obj} \left(\circled{\WTr_x} \rcomp \VIS? \rcomp \AD(x) \right)^{+} \subseteq \VO
\end{equation}

Next, we prove that each of the inequations in $\System_{\Sigma_{\PSI}}$ are 
satisfied when $\VIS, \AR, \mathsf{AntiVIS}$ are substituted for $X_V, X_A, X_N$, 
respectively.

\begin{description}
\item[Inequation \Srf:] $\RF \subseteq \VIS$. This is true, because $\RF \subseteq (\RF \cup \VO)^{+} = \VIS$, 
\item[Inequation \SvisTrans:] $\VIS \rcomp \VIS \subseteq \VIS$. This is trivially satisfied: $\VIS \rcomp \VIS = (\RF \cup \VO)^{+} 
\rcomp (\RF \cup \VO)^{+} \subseteq (\RF \cup \VO)^{+} = \VIS$, 
\item[Inequation \Sconflict:] $\VO \subseteq \VIS$. This can be proved as above: $\VO \subseteq (\RF \cup \VO)^{+} \subseteq \VIS$, 
\item[Inequation \Svo:] $\VO \subseteq \AR$. We have already proved that $\VO \subseteq \VIS$, hence it suffices to show 
that $\VIS \subseteq \AR$; this is done below, 
\item[Inequation \Svis:] $\VIS \subseteq \AR$. We have that 
\[
\VIS \subseteq \VIS \cup \bigcup_{x \in \Obj} \left( \circled{\WTr_x} \rcomp \VIS? \rcomp \AD(x)\right)^{+}  = \AR,
\]
\item[Inequation \Sext:] $\bigcup_{x \in \Obj} \circled{\WTr_x} \rcomp \VIS \rcomp \AD(x) \subseteq \AR$. This inequation is 
trivially satisfied by the definition of $\AR$: 
\begin{gather*}
\bigcup_{x \in \Obj} \circled{\WTr_{x}} \rcomp \VIS \rcomp \AD(x) \subseteq\\
\bigcup_{x \in \Obj} \circled{\WTr_x} \rcomp \VIS? \rcomp \AD(x) \subseteq \AR
\end{gather*}
\item[Inequation \SarTrans:] $\AR \rcomp \AR \subseteq \AR$. We have that 
\begin{gather*}
\AR \rcomp \AR = \\ \left(\VIS \cup \bigcup_{x \in \Obj} \left(\circled{\WTr_x} \rcomp \VIS? \rcomp \AD(x)\right)^{+}\right) \rcomp 
\left(\VIS \cup \bigcup_{x \in \Obj} \left(\circled{\WTr_x} \rcomp \VIS? \rcomp \AD(x)\right)^{+}\right) \stackrel{\eqref{eq:ar.psi}}{\subseteq}\\
(\VIS \cup \VO) \rcomp (\VIS \cup \VO) \stackrel{\Svo}{=} \VIS \rcomp \VIS \stackrel{\Svis}{\subseteq} \AR
\end{gather*}
\item[Inequation \Sad:] $\AD \subseteq \mathsf{AntiVIS}$. We have that 
$\AD \subseteq \VIS? \rcomp \AD \rcomp \VIS? = \mathsf{AntiVIS}$, 
\item[Inequation \SavisL:] $\VIS? \rcomp \AD \subseteq \mathsf{AntiVIS}$: we have 
that $\VIS? \rcomp \AD \subseteq \VIS? \rcomp \AD \rcomp \VIS? = \mathsf{AntiVIS}$. 
Inequation \SavisR\; can be proved similarly. 
\end{description}
\qed

\myparagraph{Proof of Theorem \ref{thm:depgraphs}\eqref{thm:psi}}
Let $\Delta_{\PSI} = \{\delta_{\PSI_0}\} \cup \{ \delta_{\PSI(x)} \}_{x \in \Obj}$. 
Recall that 
\begin{align*}
\delta_{\PSI_0}: & \G \mapsto (\RF_{\G} \cup \VO_{\G})^{+}\\
\delta_{\PSI(x)} : & \G \mapsto ((\RF_{\G} \cup \VO_{\G})^{\ast} \rcomp \AD(x))^{+}.
\end{align*}
We need to show that $\modelof(\Sigma_{\PSI}) = \modelof(\Delta_{\PSI})$:  
for any execution $\aexec \in \aeset(\Sigma_{\PSI})$, $\graphof(\aexec) \in \graphs(\Delta_{\PSI})$, 
and for any $\G \in \graphs(\Delta_{\PSI})$, there exists an execution $\aexec \in \aeset(\Sigma_{\PSI})$ 
such that $\graphof(\aexec) = \G$. 

We prove this result in several step. First, define 
\[\delta'_{\PSI} : \G \mapsto (\RF_{\G} \cup \VO_{\G})^{+} 
\cup \bigcup_{x \in \Obj} \left( \circled{\WTr_x} \rcomp (\RF_{\G} \cup \VO_{\G})^{\ast} \rcomp \AD_{\G}(x)\right)^{+}.
\]
We prove that $\modelof(\Sigma_{\PSI}) = \modelof(\{\delta'_{\PSI}\})$.
By Theorem \ref{thm:PSI.acyclic} we have that, for any $\aexec \in \aeset(\PSI)$, the relation 
$\delta'_{\PSI}(\G)$ is irreflexive, hence $\modelof(\Sigma_{\PSI}) \subseteq \modelof(\{\delta'_{\PSI}\})$. 
Let then $\G \in \modelof(\delta'_{\PSI})$, that is the relation $\delta'_{\PSI}(\G)$ is irreflexive. By Proposition \ref{prop:psi.correct} 
we have that $(X_V = \_, X_A = \delta'_{\PSI}(\G), X_N = \_)$ is a solution to $\System_{\PSI}(\G)$, and by 
Theorem \ref{thm:completeness} it follows that there exists a relation $\aexec \in \aeset(\Sigma_{\PSI})$ such 
that $\graphof(\aexec) = \G$. That is, $\modelof(\{\delta'_{\PSI}\}) \subseteq \modelof(\Sigma_{\PSI})$. 

Next, for any object $x \in \Obj$, define $\delta'_{\PSI(x)}(\G) = (\circled{\WTr_x} \rcomp (\RF_{\G} \cup \VO_{\G})^{\ast} \rcomp \AD_{\G}(x))^{+}$. 
It is immediate to observe that $\modelof(\{\delta'_{\PSI}\}) = \modelof(\{\delta_{\PSI_0}\} \cup \{ \delta'_{\PSI(x)} \mid x \in \Obj\})$. 
In fact, for any $\G \in \graphs$, we have that $\delta'_{\PSI}(\G) = 
\delta_{\PSI_0}(\G) \cup \bigcup_{x \in \Obj} \delta'_{\PSI(x)}(\G)$, 
hence $\delta'_{\PSI}(\G) \cap \Id = \emptyset$ if and only if 
$\delta_{\PSI_0}(\G) \cap \Id = \emptyset$, and $\delta'_{\PSI}(x)(\G) \cap \Id = \emptyset$. 
At this point we have that $ \modelof(\Sigma_{\PSI}) = \modelof(\{\delta'_{\PSI}\}) = \modelof(\{\delta_{\PSI_0}\} \cup \{ \delta'_{\PSI(x) \mid x \in \Obj}\})$.

As a last step, we show that for each dependency graph $\G$ and object $x$, the relation $\delta'_{\PSI(x)}(\G)$ is irreflexive 
if and only if the relation $\delta_{\PSI(x)}(\G)$ is irreflexive, where we recall that $\delta_{\PSI(x)}(\G) = ((\RF_{\G} \cup \VO_{\G})^{\ast} \rcomp \AD_{\G}(x))^{+}$. 
An immediate consequence of this fact is that $\modelof(\Sigma_{\PSI}) = \modelof(\{\delta_{\PSI_0}\} \cup \{\delta_{\PSI(x)} \mid x \in \Obj\}) = 
\modelof(\Delta_{\PSI})$, which is exactly what we want to prove.

Note that $\delta'_{\PSI(x)}(\G) = (\circled{\WTr_x} \rcomp (\RF_{\G} \cup \VO_{\G})^{\ast} \rcomp \AD_{\G}(x))^{+} 
\subseteq )(\RF_{\G} \cup \VO_{\G})^{\ast} \rcomp \AD_{\G}(x))^{+} = \delta_{\PSI(x)}(\G)$: if 
$\delta_{\PSI(x)}(\G)$ is irreflexive, then so if $\delta'_{\PSI(x)}(\G)$. 
Finally, suppose that $\delta'_{\PSI(x)}(\G) \cap \Id \subseteq \emptyset$. That is, 
$(\circled{\WTr_x} \rcomp (\RF_{\G} \cup \VO_{\G})^{\ast} \rcomp \AD_{\G}(x))^{+} \cap \Id \subseteq \emptyset$. 
We apply the following Theorem from Kleene Algebra: for any relations $R_1, R_2 \subseteq \T_\G \times \T_\G$, 
$(R_1 \rcomp R_2)^{+} = R_1 \rcomp (R_2 \rcomp R_1)^{\ast} \rcomp  R_2$. This leads to the following: 
\[
\left( \circled{\WTr_x} \rcomp ((\RF_{\G} \cup \VO_{\G})^{\ast} \rcomp \AD(x) \rcomp \circled{\WTr_x})^{\ast} \rcomp 
((\RF_{\G} \cup \VO_{\G})^{\ast} \rcomp \AD(x) ) \right) \cap \Id \subseteq \emptyset
\]
Also, by Proposition \ref{prop:irrefl.change}, the latter can be rewritten as follows: 
\[ 
\left(((\RF_{\G} \cup \VO_{\G})^{\ast} \rcomp \AD_{\G}(x) \rcomp \circled{\WTr_x})^{\ast} \rcomp 
(\RF_{\G} \cup \VO_{\G})^{\ast} \rcomp \AD_{\G}(x) \rcomp \circled{\WTr_x}) \right) \cap \Id \subseteq \emptyset
\]
which can be simplified into
\[ 
((\RF_{\G} \cup \VO_{\G})^{\ast} \rcomp \AD_{\G}(x) \rcomp \circled{\WTr_x})^{+} \cap \Id \subseteq \emptyset.
\]
As a last step, note that $\AD_{\G}(x) \rcomp \circled{\WTr_x} \subseteq \AD_{\G}(x)$, hence we have 
\[ 
((\RF_{\G} \cup \VO_{\G})^{\ast} \rcomp \AD(x))^{+} \cap \Id \subseteq \emptyset 
\]
which is exactly $\delta_{\PSI(x)}(\G) \cap \Id \subseteq \emptyset$. 
\qed

\subsection{Incompleteness for Arbitrary x-specifications of Consistency Models} 
One could ask whether Theorem \ref{thm:completeness} holds for non-simple 
x-specifications $\Sigma$, where $\System_{\Sigma}(\G)$ is defined by 
including inequations of the form \Saxiom, \Scoaxiom, for each consistency 
guarantee $(\rho, \pi) \in \Sigma$. Unfortunately, this is not the case.
Consider the x-specification $\Sigma = \{(\rho_\Id, \rho_{\SI}), 
(\rho_{S}, \rho_{S})\}$, and let $\G$ be the dependency graph 
depicted to the right. Recall that transactions with a double border 
are marked as serialisable.    
\setlength{\intextsep}{0pt}
\begin{mywrapfigure}[9]{r}{0.5\textwidth}
\begin{tikzpicture}[every node/.style={transform shape}, font=\small]
\node (Wy2) {$\WR\; y: 2$};
\path(Wy2.center) + (0,-0.5) node (Rx0) {$\RD\; x: 0$};
\path(Rx0.center) + (0,-1.5) node (Wx1) {$\WR\; x: 1$};
\path(Wx1.center) + (0,-0.5) node (Wz1) {$\WR\; z: 1$};
\path(Wz1.center) + (3.3,0) node (Wz2) {$\WR \; z: 2$};
\path(Wz2.center) + (0,0.5) node (Rv0) {$\RD \;v : 0$}; 
\path(Rv0.center) + (0,1.5) node (Wv1) {$\WR\; v: 1$};
\path(Wv1.center) + (0, 0.5) node (Wy1) {$\WR\; y: 1$};
%
%
\begin{pgfonlayer}{background}
\node(t1) [background, fit=(Wy2) (Rx0), inner sep=0.1cm] {};
\node(t11) [background, fit=(Wy2) (Rx0), inner sep=0.2cm] {};
\node(t2) [background, fit=(Wx1) (Wz1), inner sep=0.1cm] {};
\node(t22) [background, fit=(Wx1) (Wz1), inner sep=0.1cm] {};
\node(t3) [background, fit=(Wz2) (Rv0), inner sep=0.1cm] {};
\node(t33) [background, fit=(Wz2) (Rv0), inner sep=0.2cm] {};
\node(t4) [background, fit=(Wv1) (Wy1), inner sep=0.1cm] {};
\node(t44) [background, fit=(Wv1) (Wy1), inner sep=0.1cm] {};
\path(t1.west) + (-0.5,0) node[font=\normalsize] (T1) {$T_1$};
\path(t2.west) + (-0.5,0) node[font=\normalsize] (T2) {$T_2$};
\path(t3.east) + (0.5,0) node[font=\normalsize] (T3) {$T_3$};
\path(t4.east) + (0.5, 0) node[font=\normalsize] (T4) {$T_4$};
%
\path[->] 
  (t11.south) edge node[left]   {$\AD(x)$} (t22.north)
  (t22.east) edge node[below] {$\VO(z)$} (t33.west)
  (t33.north) edge node[right] {$\AD(v)$} (t44.south) 
  (t44.west) edge node[above] {$\VO(y)$} (t11.east);
\end{pgfonlayer}
\end{tikzpicture}
\end{mywrapfigure} 
We omitted from $\G$ a transaction $T_0$
which writes the value $0$ 
for objects $x,v$, and which is seen by $T_1, T_3$. 
For the dependency graph $\G$, the least solution of 
$\System_{\Sigma}(\G)$ is $(X_V = \_, X_A = \AR_0, X_N = \_)$, 
where $\AR_0 = \{(T_2, T_3), (T_4, T_1)\} \cup \{(T_0, T_i)\}_{i=1}^{4}$. 
That is, $\AR_0$ is acyclic. However, there exists no abstract execution 
$\aexec \in \aeset(\Sigma)$ such that $\graphof(\aexec) = \G$. In 
fact, if such $\aexec$ existed, then $T_1$ and $T_3$ should 
be related by $\AR_{\aexec}$. However, it cannot be $T_1 \xrightarrow{\AR_{\aexec}} T_3$: 
the axiom of the consistency guarantee $( \rho_S, \rho_S)$, 
$\circled{\SerTX} \rcomp \AR_{\aexec} \rcomp \circled{\SerTX} \subseteq \VIS_{\aexec}$,  
would imply $T_1 \xrightarrow{\VIS_{\aexec}} T_3$; 
together with $T_3 \xrightarrow{\AD_{\aexec}} T_4$ 
and the co-axiom induced by $(\rho_{\Id}, \rho_{\SI})$, 
$(\VIS_\aexec \rcomp \overline{\VIS^{-1}_{\aexec}}) \setminus \Id \subseteq 
\AR_{\aexec}$, 
this would mean that $T_1 \xrightarrow{\AR_{\aexec}} T_4$. 
But we also 
have $T_4 \xrightarrow{\AR_{\aexec}} T_1$, hence a contradiction. 
Similarly, we can prove  $\neg(T_3 \xrightarrow{\AR_{\aexec}} T_1)$.

\else 
\fi

\end{document}